\documentclass[letterpaper,11pt]{article}
\pdfoutput=1

\usepackage{jheppub}
\usepackage{graphicx}
\usepackage{amsmath}
\usepackage{amssymb}
\usepackage{mathrsfs}
\usepackage{bm}
\usepackage{braket}
\usepackage{slashed}
\usepackage{xspace}
\usepackage[dvipsnames]{xcolor}
\usepackage{mdwlist}
\usepackage[normalem]{ulem}
\usepackage[utf8]{inputenc}

\newcommand{\eq}[1]{eq.~\eqref{eq:#1}}
\newcommand{\eqs}[2]{eqs.~\eqref{eq:#1} and \eqref{eq:#2}}
\renewcommand{\sec}[1]{sec.~\ref{sec:#1}}
\newcommand{\secs}[2]{secs.~\ref{sec:#1} and \ref{sec:#2}}
\newcommand{\fig}[1]{fig.~\ref{fig:#1}}

\def\mybox#1{\vbox{\kern0.1cm\hbox{#1}\kern0.1cm}}

\newcommand{\ord}[1]{{\mathcal O}(#1)}

\newcommand{\nn}{\nonumber}

\newcommand{\df}{\mathrm{d}}
\newcommand{\img}{\mathrm{i}}
\newcommand{\sdt}{\!\cdot\!}
\newcommand{\tr}{\mathrm{tr}}
\newcommand{\lra}{\leftrightarrow}

\newcommand{\al}{\alpha}
\newcommand{\bt}{\beta}
\newcommand{\ga}{\gamma}
\newcommand{\Ga}{\Gamma}
\newcommand{\de}{\delta}

\newcommand{\eps}{\epsilon}

\newcommand{\si}{\sigma}

\newcommand{\cL}{{\mathcal L}}
\newcommand{\cS}{{\mathcal S}}

\newcommand{\nslash}{n\!\!\!\slash}
\newcommand{\bnslash}{\bar{n}\!\!\!\slash}
\newcommand{\bn}{{\bar{n}}}

\def\hyp{\mathsf{y}}

\def\C{\mathcal{C}}
\def\W{\mathcal{W}}

\newcommand{\Ecm}{E_{\rm cm}}
\newcommand{\MSbar}{\ensuremath{\overline{\text{MS}}}}

\allowdisplaybreaks[2]

\preprint{\vbox{\hbox{Nikhef 2017-067}}}

\title{Electroweak Logarithms in Inclusive Cross Sections}

\author[a]{Aneesh V.~Manohar,}
\author[b,c]{Wouter J.~Waalewijn}

\affiliation[a]{Department of Physics, University of California, San Diego, 9500 Gilman Drive, La Jolla, CA 92093, USA}
\affiliation[b]{Institute for Theoretical Physics Amsterdam and Delta Institute for Theoretical Physics, University of Amsterdam, Science Park 904, 1098 XH Amsterdam, The Netherlands}
\affiliation[c]{Nikhef, Theory Group, Science Park 105, 1098 XG, Amsterdam, The Netherlands}

\abstract{We develop the framework to perform all-orders resummation of electroweak logarithms of $Q/M$ for inclusive scattering processes at energies $Q$ much above the electroweak scale $M$. We calculate all ingredients needed at next-to-leading logarithmic (NLL) order and provide an explicit recipe to implement this for $2 \to 2$ processes. PDF evolution including electroweak corrections, which lead to Sudakov double logarithms, is computed. If only the invariant mass of the final state is measured, all electroweak logarithms can be resummed by the PDF evolution, at least to LL. However, simply identifying a lepton in the final state requires the corresponding fragmentation function and introduces  angular dependence through the exchange of soft gauge bosons. Furthermore, we show the importance of polarization effects for gauge bosons, due to the chiral nature of SU(2) --- even the gluon distribution in an unpolarized proton becomes polarized at high scales due to electroweak effects. We justify our approach with a factorization analysis, finding that the objects entering the factorization theorem do not need to be $SU(2) \times U(1)$ gauge singlets, even though we perform the factorization and resummation in the symmetric phase. We also discuss a range of extensions, including jets and how to calculate the EW logarithms when you are fully exclusive in the central (detector) region and fully inclusive in the forward (beam) regions.}
\begin{document}
\maketitle

%%%%%%%%%%%%%%%%%%%%%%%%%%%%%%%%%%%%%%%%%%%%%%%%%%%%%%%%%%%%%%%%%%%%%%%%%%%%%%%%
\section{Introduction}
\label{sec:intro}
%%%%%%%%%%%%%%%%%%%%%%%%%%%%%%%%%%%%%%%%%%%%%%%%%%%%%%%%%%%%%%%%%%%%%%%%%%%%%%%%

At LHC energies, the effect of electroweak (EW) corrections on the cross section can be significant ($\sim 10$\%). These are dominated by EW Sudakov double logarithms, 
%%%
\begin{align}
  \si  = \si_0\, \sum_{m \leq 2n} c_{nm}\, \al_w^n \ln^m \frac{Q}{M}
\,,\end{align}
%%%
where $\si_0$ is the Born cross section, $\al_w$ is a weak coupling constant ($\alpha_2$ or $\alpha_1$), $Q$ denotes the hard scale (typically taken to be the partonic center of mass energy $\sqrt{\hat s}$) and $M$ is an electroweak scale such as $M_W$, $M_Z$, $m_H$, $m_t$, which we consider to be of the same parametric size.\footnote{Of course there are also mixed QCD-EW corrections, and we will consider their interplay in the analysis.} The energy dependence of EW corrections makes it important to include them when searching for new physics in tails of distributions. It also highlights that these effects are indispensable for cross section predictions at a FCC, where EW logarithms are truly large \cite{Mangano:2016jyj}, and make order one corrections to the cross section. EW Sudakov logarithms also play an important role in calculations of WIMP dark matter, see e.g.~refs.~\cite{Ciafaloni:2010ti,Baumgart:2014vma, Bauer:2014ula, Ovanesyan:2014fwa, Baumgart:2017nsr}.

\begin{figure}
\centering
 \includegraphics[width=0.6\textwidth]{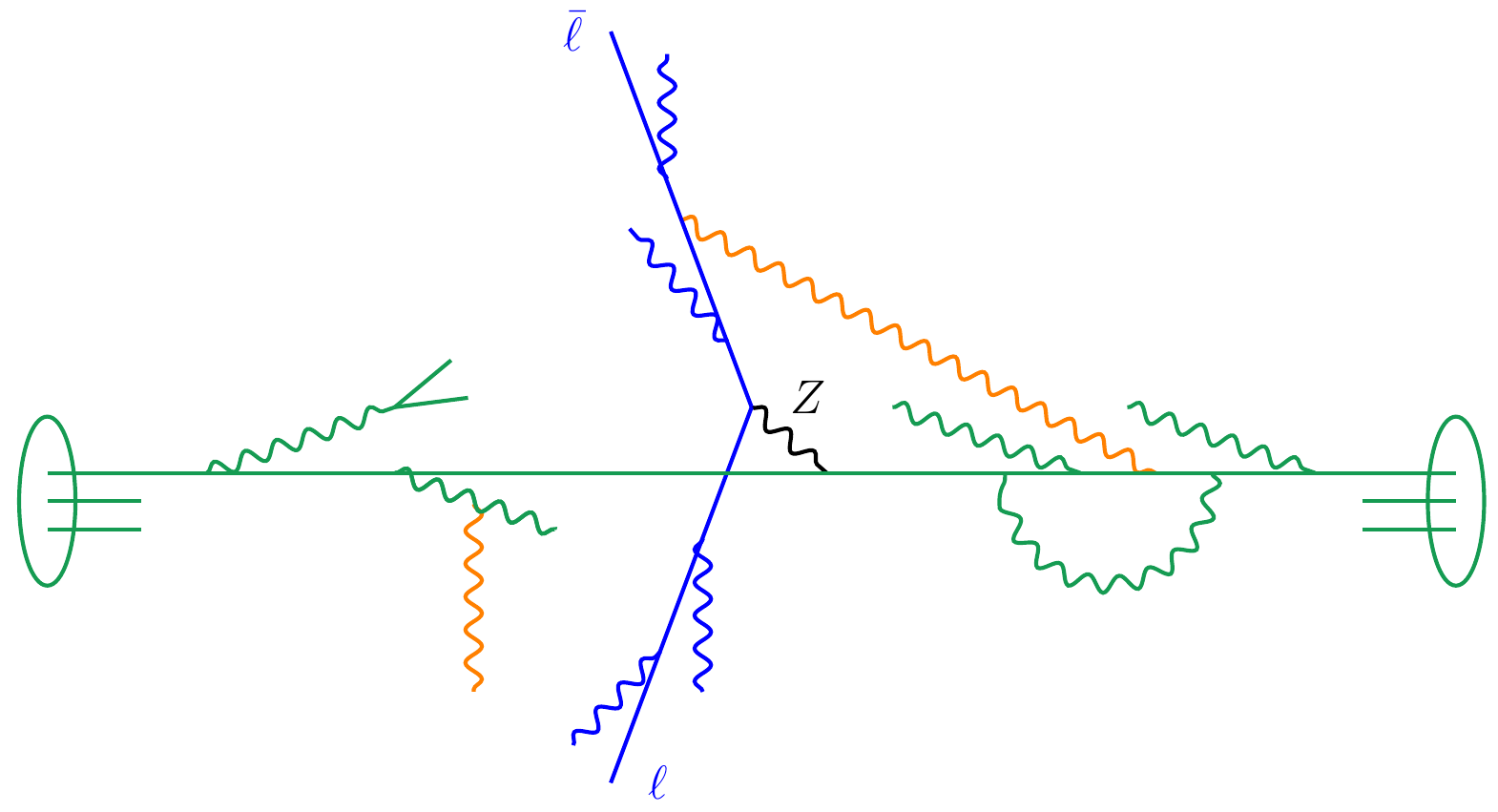}
 \caption{EW corrections to Drell-Yan production: the parton from each proton (blob) emits  {\color{ForestGreen} initial-state radiation} before participating in the hard scattering ($Z$ exchange). The outgoing leptons produce {\color{blue} final-state radiation}. These collinear effects are described by the DGLAP evolution of the corresponding PDFs and FFs. Surprisingly, {\color{orange}soft radiation} between different collinear directions matters because the incoming and outgoing particles are not $SU(2)$ singlets, and this also modifies the DGLAP evolution.}
\label{fig:factorization}
\end{figure}

Most studies of EW logarithms focus on virtual effects. The underlying assumption is that one is fully exclusive, i.e.~all real EW radiation is resolved by detectors. This is not unreasonable because the $W$ and $Z$ boson are massive, and can be tagged experimentally via their decay products. Electroweak Sudakov logarithms were first studied in refs.~\cite{Ciafaloni:1998xg,Kuhn:1999de,Fadin:1999bq,Beenakker:2000kb,melles1}, a recipe for the next-to-leading order (NLO) corrections was presented in refs.~\cite{Denner:2000jv,Denner:2001gw,Denner:2006jr} and the two-loop logarithms for a four-fermion process were obtained in refs.~\cite{Kuhn:2001hz,Jantzen:2005az}. Refs.~\cite{Chiu:2007yn,Chiu:2007dg} developed a resummation framework using Soft-Collinear Effective Theory (SCET)~\cite{Bauer:2000ew, Bauer:2000yr, Bauer:2001ct, Bauer:2001yt}, obtaining results at next-to-leading-logarithmic (NLL) plus NLO accuracy. The effect of real radiation can be important, and was studied at NLO in e.g.~refs.~\cite{Baur:2006sn,Bell:2010gi,Stirling:2012ak,Manohar:2014vxa}. 

In this paper we start from the opposite extreme, considering inclusive processes. One example is Drell-Yan, $pp \to \ell \bar \ell X$, where $X$ is unconstrained, illustrated in \fig{factorization}. When the lepton pair has a small transverse momentum or has a large invariant mass (threshold limit), the cross section contains large double logarithms. Here we will not focus on these regions of phase space, so the QCD corrections do not involve double logarithms. Nevertheless, because the proton is not an electroweak singlet, EW double logarithms remain present~\cite{Ciafaloni:2000df,Ciafaloni:2000rp}, which is one of the salient features of our analysis. In this paper, we develop a framework to perform the resummation of EW logarithms using a factorization theorem that is valid to all-orders in perturbation theory. Important ingredients for resummation are the collinear splitting functions, which were determined at leading order in refs.~\cite{Ciafaloni:2001mu,Ciafaloni:2005fm}. These have been implemented into a parton shower~\cite{Christiansen:2014kba,Krauss:2014yaa,Chen:2016wkt} and used to resum initial-state radiation by including them in the evolution of parton distribution functions (PDFs)~\cite{Bauer:2017isx,Bauer:2017bnh}. Our calculation gives the same result for the Sudakov double logarithms as ref.~\cite{Bauer:2017isx}, but we also consider final-state radiation and extend to NLL. Interestingly, we will see that the splitting functions alone are not enough to account for all the EW logarithms, and soft anomalous dimensions must also be included. We also show the importance of polarization effects for gauge bosons, which are a consequence of the chiral nature of $SU(2)$ and the helicity dependence of splitting functions, and were missed in earlier studies.

We achieve resummation using an effective field theory analysis, in the spirit of refs.~\cite{Chiu:2007yn,Chiu:2007dg}. First the hard scattering is integrated out at the scale $Q$, matching onto an effective field theory in the symmetric phase of $SU(2) \times U(1)$. We then factorize the cross section and use the renormalization group evolution to evolve to the low scale $M$, thereby resumming EW logarithms. Only at the low scale $M$ do we switch to the broken phase.  Anomalous dimensions are related to ultraviolet divergences and do not depend on symmetry breaking, which is an infrared effect. The collinear initial- and final-state radiation will be resummed using the DGLAP evolution~\cite{Gribov:1972ri,Altarelli:1977zs,Dokshitzer:1977sg} of the corresponding PDFs and fragmentation functions (FFs). Surprisingly, for the nonsinglet PDFs there is also a sensitivity to soft radiation. This introduces rapidity divergences, and we use the rapidity renormalization group~\cite{Chiu:2011qc,Chiu:2012ir} to resum the corresponding single logarithms of $Q/M$. We calculate all ingredients necessary for resummation at NLL and provide an explicit recipe on how to implement them for $2 \to 2$ processes in the appendix.

We end the paper by discussing a range of generalizations:
\begin{itemize*}
\item Resummation beyond NLL.
\item Other processes.
\item Kinematic hierarchies which arise when not all of the Mandelstam invariants are of order $Q$.
\item Jets identified (inclusively) using a jet algorithm
\item Less inclusive processes where radiation within the range of the detectors is observed, but radiation near the beam axis is not.
\end{itemize*}

The outline of our paper is as follows. Our factorization analysis, which splits the cross section into collinear and soft parts, is described in \sec{framework}. 
The renormalization group equations for the collinear sector are given in \sec{RGE_C}, and for the soft sector in \ref{sec:RGE_S}. The matching onto the broken phase of the gauge theory is presented in \sec{low_matching}. The evolution from the hard scale $Q$ to the electroweak scale $M$ accomplishes the resummation of electroweak logarithms of $Q/M$, as discussed in \sec{resummation}. In \sec{comparison} we show how our results compare to electroweak resummation for PDFs in the literature. We discuss the generalizations listed above in \sec{extensions}, and conclude in \sec{conclusions}. For readers mostly interested in the final results, we provide a recipe to include electroweak resummation at NLL accuracy in appendix \ref{app:recipe}. In appendix~\ref{app:xsec}, we give examples of the possible PDF combinations which enter the production of a heavy particle in quark-antiquark annihilation.

%%%%%%%%%%%%%%%%%%%%%%%%%%%%%%%%%%%%%%%%%%%%%%%%%%%%%%%%%%%%%%%%%%%%%%%%%%%%%%%%
\section{Factorization}
\label{sec:framework}
%%%%%%%%%%%%%%%%%%%%%%%%%%%%%%%%%%%%%%%%%%%%%%%%%%%%%%%%%%%%%%%%%%%%%%%%%%%%%%%%

\begin{figure}
\begin{center}
\includegraphics[width=0.35\textwidth]{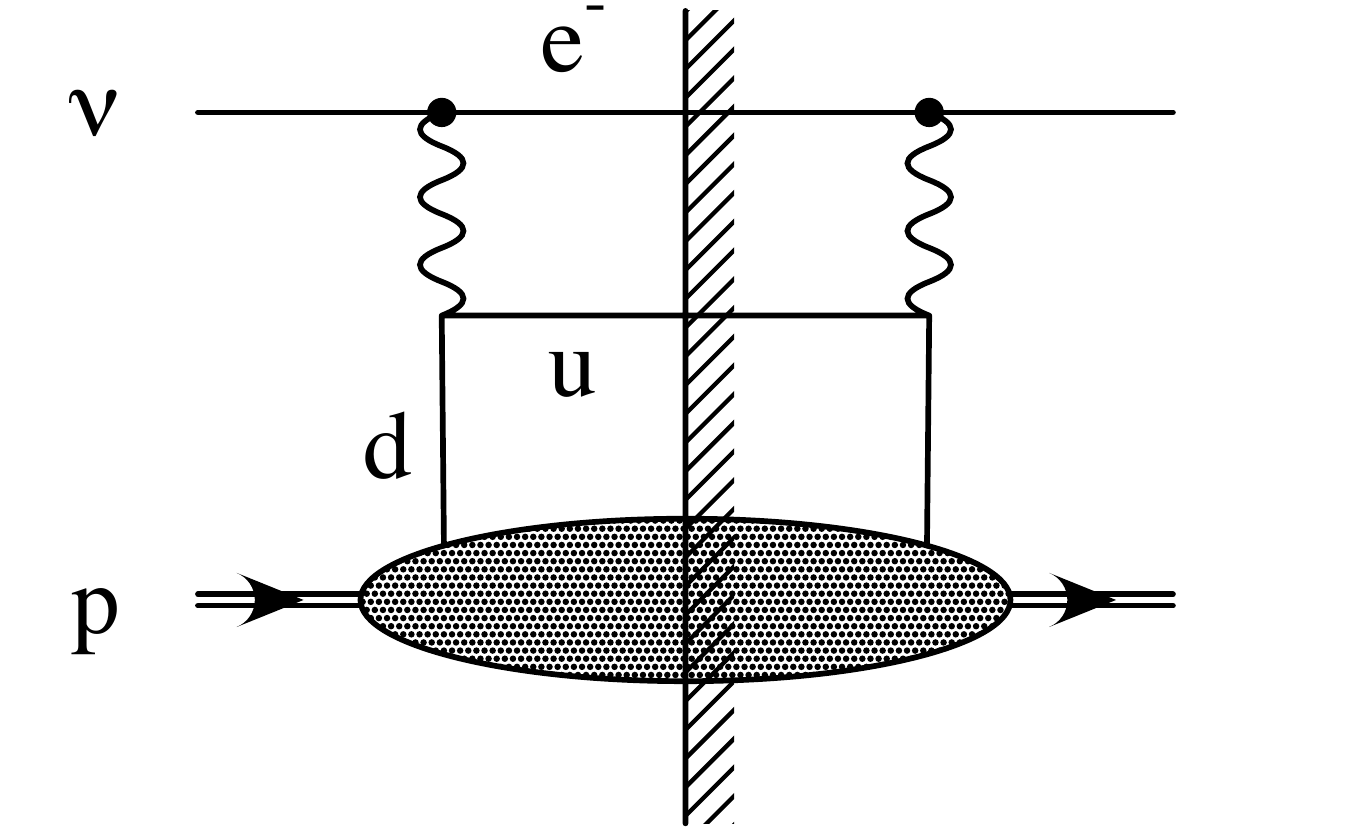}
\end{center}
\vspace{-2ex}

\caption{\label{fig:nu_dis} Tree-level diagram that contributes to deep-inelastic neutrino scattering.}
\end{figure}

In this section we present our framework for resumming electroweak logarithms in inclusive cross sections, considering as an example deep-inelastic neutrino scattering $\nu p \to \ell X$, for which a tree-level diagram is shown in \fig{nu_dis}. We start, in \sec{high_matching},  by integrating out the short-distance scattering at the hard scale $Q$. Here we can work in the symmetric phase of the gauge theory, since $Q \gg M \sim M_W, M_Z, m_h, m_t$. The scattering amplitude can be factored into a coefficient and hard scattering operator. We discuss the factorization of the hard-scattering operator into collinear and soft operators in \sec{factorization}, which allows one to sum collinear and soft logarithms using RGEs. The gauge and spin indices are disentangled in \sec{gauge_spin}, allowing one to write the scattering amplitude for any process in terms of a standard basis of collinear and soft operators. 

In previous work on electroweak resummation using SCET in refs.~\cite{Chiu:2007dg,Chiu:2007yn,Chiu:2008vv,Chiu:2009ft,Chiu:2009mg,Chiu:2009yx,Chiu:2011qc,Chiu:2012ir,Fuhrer:2010eu,Fuhrer:2010vi}, the final state was assumed to consist of particles and jets with masses smaller than the electroweak scale such that only virtual electroweak corrections needed to be taken into account. This allows for the entire analysis from $Q$ down to $M$ to be carried out at the amplitude level. In this paper, we are interested in inclusive cross sections where we sum over final states with masses larger than the electroweak scale (e.g.~in semi-inclusive cross sections), so we square the amplitude and factorize the cross section above the electroweak scale.

%===============================================================================
\subsection{Matching at the hard scale}
\label{sec:high_matching}
%===============================================================================

The sample process we study is given by lepton-quark scattering, and the hard scattering operators that contribute are 
%%%
\begin{align} \label{eq:hard_ops}
  O^{(3)}_{\ell q} &= (\bar \ell_1 \ga^\mu t^a \ell_2)\, (\bar q_3 \ga_\mu t^a q_4)
  \,, &
  O_{\ell q} &= (\bar \ell_1 \ga^\mu \ell_2)\, (\bar q_3 \ga_\mu q_4)
  \,, \nn \\
  O_{\ell u} &= (\bar \ell_1 \ga^\mu \ell_2)\, (\bar u_3 \ga_\mu u_4)
  \,,  &
  O_{\ell d} &= (\bar \ell_1 \ga^\mu \ell_2) \, (\bar d_3 \ga_\mu d_4)
  \,, \nn \\
  O_{eq} &= (\bar e_1 \ga^\mu e_2)\, (\bar q_3 \ga_\mu q_4)
  \,, \nn \\
  O_{eu} &= (\bar e_1 \ga^\mu e_2)\, (\bar u_3 \ga_\mu u_4)
  \,, &
  O_{ed} &= (\bar e_1 \ga^\mu e_2) \, (\bar d_3 \ga_\mu d_4)
  \,,\end{align}
%%%
at leading power in $M/Q \ll 1$. The electroweak doublet fields $\ell = (e_L, \nu_L)$ and $q = (u_L, d_L)$ are left-handed,  the electroweak singlet fields $e = e_R$, $u = u_R$ and $d=d_R$ are right-handed, and $t^a$ are the $SU(2)$ generators. For quark-quark scattering, one can also have operators such as $ (\bar q_1 \ga^\mu T^A q_2)\, (\bar q_3 \ga_\mu T^A q_4)$ which involve the color generators $T^A$, or $ (\bar q_1 \ga^\mu t^a  T^A q_2)\, (\bar q_3 \ga_\mu t^a T^A q_4)$ which involve both weak and color generators.

The subscripts on the fields indicate their momentum, e.g.~$\bar e_1$ has momentum $p_1$. This is important because the hard-matching coefficients $\mathcal{H}_i$ depend on $p_i$,
%%%
\begin{align} \label{eq:hard_L}
  \cL_{\rm hard} = \sum_i \bigg(\prod_k \int\! \frac{\df^4 p_k}{(2\pi)^4} \bigg)\, (2\pi)^4 \de^4\Big(\sum_m p_m\Big)\, \mathcal{H}_i(\{p_k\})\, O_i(\{p_k\})
\,.\end{align}
%%%
We will use the convention that all momenta are incoming. Thus an outgoing particle has momentum $p$ with $p^0<0$. This convention avoids certain minus signs between incoming and outgoing particles in subsequent results, and allows us to treat both with a unified notation. The field $\bar e_1$ contributes to processes with either an outgoing right-handed electron or an incoming left-handed positron, and the two are distinguished by the sign of $p_1^0$.

At tree-level, $O^{(3)}_{\ell q}$ is generated by $SU(2)$ gauge boson exchange. The other operators in \eq{hard_ops}, which we denote by $O_{AB}$ with $A = \ell, e$ and $B=q, u, d$, are due to the exchange of a $U(1)$ gauge boson. This leads to the matching coefficients 
%%%
\begin{align} \label{eq:H_tree}
  \mathcal{H}^{(3)}_{\ell q} = \frac{\img g_2^2}{2 p_1 \sdt p_2}  
  \,, \qquad
  \mathcal{H}_{AB} = \frac{\img g_1^2\, \hyp_A\, \hyp_B}{2 p_1 \sdt p_2}
\,,\end{align}
%%%
where $g_2$ and $g_1$ are the $SU(2)$ and $U(1)$ couplings, and $\hyp_A$ and $\hyp_B$ are the $U(1)$ hypercharge of the fields $A$ and $B$. Since $p_1 \cdot p_2 \sim Q^2 \gg M^2$, gauge boson masses in the propagator are power suppressed, and have been dropped. For example, for neutrino-proton scattering via
$\nu q  \to e^- X$, the hard-scattering at tree level is given by
%%%
\begin{align}\label{eq:DY}
\sum_i \mathcal{H}_i O_i &= \frac{\img g_2^2}{2 p_1 \sdt p_2}\, O^{(3)}_{\ell q} + \frac{\img g_1^2 }{2 p_1 \sdt p_2} \left[\hyp_\ell\, \hyp_q\, O_{\ell q} +
\hyp_\ell\, \hyp_u\, O_{\ell u} + \hyp_\ell\, \hyp_d\, O_{\ell d}  \right] \,.
\end{align}
%%%
Because we will carry out the collinear and soft evolution to the scale $Q$ of the hard scattering, we do not have to calculate the matching coefficients $\mathcal{H}_i$ at one loop to achieve resummation at NLL accuracy.

After integrating out the hard gauge boson to obtain \eq{hard_L}, only collinear and soft fluctuations of the fields remain. These can be described using SCET, where the Lagrangian $\cL_{\rm SCET}$ encodes the dynamics of the collinear and soft fields. For our analysis, we do not need all the technical details of SCET, so we present the discussion in terms of soft and collinear corrections, which should be accessible to a wider audience.\footnote{That is, we use pseudo-SCET analogous to pseudocode in computer science.}

We will make frequent use of the following light-like vectors for incoming particles,
%%%
\begin{align}
n_i = (1, \mathbf{n}_i)
\,, \qquad 
\bar n_i = (1, - \mathbf{n}_i)
\,,\end{align}
%%%
where the unit vector $\mathbf{n}_i$ points along the direction of $\mathbf{p}_i$. For outgoing particles with energy $E_j$ and momentum $\mathbf{p}_j$, our convention is that $p_j=(-E_j,-\mathbf{p}_j)$, and 
%%%
\begin{align}\label{eq:3.5}
n_j = (-1, -\mathbf{n}_j)
\,, \qquad 
\bar n_j = (-1,  \mathbf{n}_j)
\,,\end{align}
%%%
where $\mathbf{n}_j$ is a unit vector in the direction $\mathbf{p}_j$.

The matching in \eq{hard_L} removes fluctuations of virtuality $\gtrsim Q$, and the full gauge invariance of the theory is replaced by collinear gauge invariance for each collinear direction, as well as soft gauge invariance~\cite{Bauer:2001yt}. Each field in \eq{hard_ops} corresponds to a distinct collinear direction, so it must be (made) collinearly gauge invariant by itself. This is accomplished by including collinear Wilson lines in the definitions of fields~\cite{Bauer:2001ct}. By including soft Wilson lines, the interactions between collinear and soft fields can be removed from the Lagrangian~\cite{Bauer:2001yt}, and included in the hard scattering operator.
For example, the incoming field $q_4$ in \eq{hard_ops} is short-hand for a collinear fermion field $\psi_{4}$ (typically denoted by $\xi_{n_4}$ in SCET) combined with a collinear Wilson line $\W$ in the $\bn_4$ direction and a soft Wilson line $\cS$ in the $n_4$ direction,
(using the covariant derivative convention $D_\mu=\partial_\mu + i g A_\mu$)
%%%
\begin{align} \label{eq:Wilson}
 q_4(x) &= S_{4} \int \df^4x\, e^{\img p_1 \cdot x}\, \W_{4}^\dagger(x) \psi_{4}(x)
 \,, \nn \\
 \W_{4}(x) &= {\rm P} \exp\bigg\{-\img \int_{\infty}^0\!\df s\, \bn_4 \sdt \big[g_3 A_{n_4}(x+s \bar n_4) + g_2 W_{n_4}(x+s \bar n_4) + g_1 \hyp_q B_{n_4}(x+s \bar n_4) \big] \bigg\}\,, \nn \\
 \cS_{4} &= {\rm P} \exp\bigg\{-\img \int_{-\infty}^0\!\df s\, 
  n_4 \sdt \big[g_3 A_s(s\, n_4) + g_2 W_s(s\, n_4) + g_1 \hyp_q B_s(s\, n_4) \big]\bigg\}\,.
\end{align}
%%%
$A_{n_4}$, $W_{n_4}$ and $B_{n_4}$ denote $SU(3)$, $SU(2)$ and $U(1)$  gauge fields whose momenta are collinear to the $n_4$ direction, and $A_s, W_s, B_s$ denote soft gauge fields.\footnote{The $SU(2)$ gauge field $W$ should not be confused with $\W$, a collinear Wilson line.}
The Wilson lines $\W$ and $\cS$ depend on the gauge representation of the particle.
The soft Wilson line integral is over the worldline of the particle. For incoming particles, the soft Wilson line integral is from $t=-\infty$ to $t=0$.
With our sign convention, \eq{Wilson} also holds for outgoing particles, and the minus sign in \eq{3.5} converts the integral from $t=-\infty$ to $t=0$ into one from $t=0$ to $t=\infty$. The direction of the soft Wilson line affects the sign of  $\img 0$ terms in the eikonal propagators, and the sign of scattering phases.

The interaction of the collinear fields $\W_{4}$ and $\psi_{4}$ in \eq{Wilson}, which is given by the full QCD interaction for particles in the $n_4$ direction~\cite{Goerke:2017ioi}, leads to the production of a jet of particles in the $n_4$ direction, with invariant mass much smaller than $Q$. The soft Wilson line $\cS_{4}$  sums the emission of soft radiation from the collinear fields, so the collinear fields no longer interact with soft fields in this picture. To avoid additional notation, in the remainder of the paper $q_4$ will denote the collinear part of the right-hand side of \eq{Wilson}, $\W_4^\dagger \psi_4 \to q_4$, so that $q_4$ in \eq{hard_ops} is now denoted by $q_4 \to S_4 q_4$.

%===============================================================================
\subsection{Factorization into collinear and soft}
\label{sec:factorization}
%===============================================================================

Eq.~\eqref{eq:Wilson} factors the operator describing the hard interaction in \eq{hard_ops} into different collinear sectors and a soft sector that no longer interact.
The cross section is given by taking the matrix element of the hard scattering in \eq{hard_L} with initial and final-state particles, squaring, and including the phase-space integration, flux factor and measurement. This is largely an exercise in bookkeeping, where most of the complications arise from the phase space, and leads to the usual factorization theorems for hard scattering processes in QCD.

Schematically, the cross section for $\nu p \to e^- X$ is given by
%%%
\begin{align}\label{eq:3.8}
\sigma \sim
\sum_X \braket{\nu p | \cL_{\text{hard}} | e^- X}\braket{e^- X | \cL_{\text{hard}} | \nu p}\,.
\end{align}
%%%
The fields in $\cL_{\text{hard}}$ are the product of soft and collinear terms (see \eqs{hard_L}{Wilson}), and the matrix element in \eq{3.8} is factored into the product of soft and collinear matrix elements, by writing $\ket{e^-X}$ as the product $\ket{X_s} \otimes \ket{X_{c,1}} \otimes \ket{X_{c,2}} \ldots$ of soft particles and collinear particles in different collinear sectors in the final state. For $\nu p \to e^- X$, there are four sectors given by the directions of $\nu$, $p$, $e^-$ and the outgoing jet produced by the struck quark. Using only the $O_{\ell q}^\dagger O_{\ell q}$ contribution to \eq{3.8} as an example, the relevant matrix element is 
%%%
\begin{align} \label{eq:factor}
&(\ga^\mu_{\bt_1\ga_1}  \ga_{\mu, \bt_2 \ga_2}  \ga^\nu_{\bt_3\ga_3}  \ga_{\nu,\bt_4\ga_4})
\Bigl[ \sum_{X_1}  \langle 0 | \ell_{1',\de_3} |X_1  \rangle \langle X_1| \bar \ell_{1,\al_1} | 0 \rangle \Bigr] 
 \Bigl[\langle \nu| \bar \ell_{2',\al_3} \ell_{2,\de_1} | \nu \rangle\Bigr] \nn \\
& \Bigl[ \sum_{X_3}  
  \langle 0 | q_{3',\de_4} |X_3  \rangle \langle X_3| \bar q_{3,\al_2} | 0 \rangle\Bigr]
 \Bigl[  \langle p | \bar q_{4',\al_4}  q_{4,\de_2} | p \rangle\Bigr] \nn \\
& \Bigl[ \sum_{X_s} \langle 0 | \cS_{1,\ga_3,\de_3} \cS_{2,\al_3 \bt_3}^\dagger
  \cS_{3,\ga_4,\de_4} \cS_{4,\al_4 \bt_4}^\dagger
   |X_s  \rangle \langle X_s| \cS_{1,\al_1,\bt_1}^\dagger \cS_{2,\ga_1 \de_1}
  \cS_{3,\al_2 \bt_2}^\dagger \cS_{4,\ga_2 \de_2}
   | 0 \rangle \Bigr]
\,.\end{align}
%%%
Here the indices $\al_1, \dots, \de_4$ include both spin and gauge indices. Since the hard interaction \eq{hard_L} has a sum over the momenta of the colliding partons weighted with a hard coefficient $H(\{p_k\})$, the labels on the fields in $O^\dagger$ have been distinguished from those in $O$ by a prime.  Eventually these will be equal because of momentum conservation in the matrix elements.

Eq.~\eqref{eq:factor} has factored the total cross section into collinear sectors corresponding to the incoming proton and neutrino, outgoing lepton (in $X_1$) and jet (in $X_3$), and the soft sector. This factorization is what  enables the resummation of logarithms of $Q/M$, by separating the ingredients at different invariant mass and rapidity scales, as discussed in \sec{resummation}. In the next section we  show how to disentangle the gauge/spin indices for all combinations of hard-scattering operators. Since we only probe the hard scattering kinematics, the collinear matrix elements will correspond to parton distributions functions for incoming directions and fragmentation functions for outgoing directions, as is the case in QCD factorization for inclusive cross sections. (We avoid kinematic limits, such as small transverse momentum, which would require a transverse momentum dependent parton distribution or fragmentation function.) What is perhaps surprising is the appearance of a soft function, since it would seem that soft radiation is not directly probed by the measurement, i.e.~we are not in a kinematic limit that makes the measurement sensitive to soft radiation. For the QCD corrections, color conservation forces the hadronic matrix elements of quark operators to be diagonal in color, leading to color contractions of indices on the soft Wilson lines. Because the observables we consider do not directly probe the soft radiation, we can perform the sum over $\ket{X_s}$ and then find that the soft matrix element is the identity by using $\cS_i^\dagger \cS_i=1$. The fundamental difference in the electroweak case is that electroweak symmetry is broken, so the  matrix elements do not have to be diagonal in electroweak indices,
%%%
\begin{align} \label{eq:soft}
 \langle 0 | \cS_{1,\ga_3,\de_3} \cS_{2,\al_3 \bt_3}^\dagger
  \cS_{3,\ga_4,\de_4} \cS_{4,\al_4 \bt_4}^\dagger \cS_{1,\al_1,\bt_1}^\dagger \cS_{2,\ga_1 \de_1}
  \cS_{3,\al_2 \bt_2}^\dagger \cS_{4,\ga_2 \de_2}
   | 0 \rangle 
\,.\end{align}
%%%

The proton matrix element of $q_4$ in \eq{factor} gives the sum of left-handed $u$ and $d$ quark PDFs in the proton.  
Similarly, the neutrino matrix element corresponds to a lepton PDF in the neutrino. Because the neutrino does not have QCD or QED interactions, this PDF is a delta function at tree-level at the electroweak scale. The matrix element of $q_3$ reduces to a quark jet function, after summing on $X_3$. The matrix element involving $\ell_1$ would reduce to a lepton (electroweak) jet function if one sums over all $X_1$. However, in DIS the energy and direction of the outgoing electron are measured, so one sums over $\ket{X_1}=\ket{e(p_e), X}$ where $p_e$ is measured. This corresponds to a fragmentation function, as it only probes the energy $p_e^0$ (the electron is collinear to the field $\ell_1$). On the other hand, the soft function is sensitive to the direction of $p_e$ but not its energy. Thus we have factorized the cross section into collinear and soft pieces which can be studied independently.

There is a subtlety in \eq{factor}. The soft Wilson lines have been written as $\cS$ or $\cS^\dagger$ depending on whether they arose from the field $\psi$ or $\overline \psi$. This keeps track of the gauge indices in the Wilson lines. The Wilson lines in $O_H$ give the scattering amplitude, whereas those in $O_H^\dagger$ give the complex conjugate of the amplitude. Thus the Wilson lines from $O_H$ are time-ordered, whereas those from $O_H^\dagger$ are anti-time-ordered. In \eq{factor}, $S_1^\dagger$, $S_2$, $S_3^\dagger$ and $S_4$ are time-ordered, whereas $S_1$, $S_2^\dagger$, $S_3$ and $S_4^\dagger$ are anti-time-ordered. We will not carefully keep track of this in our notation, because our calculation of the anomalous dimension in \sec{RGE_S} shows that this is irrelevant.

%===============================================================================
\subsection{Disentangling gauge and spin indices}
\label{sec:gauge_spin}
%===============================================================================

The next step is to disentangle the spin and gauge indices on the fermion fields in the product of two operators $O^\dagger O$, which enter the factorization formula \eq{factor}. This can be achieved by using the relations  
%%%
\begin{align} \label{eq:disentangle}
\bar \ell_{i\al} \ell^j_\bt &= \frac{1}{2 N_w}\, \delta^j{}_i (P_L \nslash)_{\bt \al}\, \bar \ell \tfrac{\bnslash}{2} \ell + (t^a)^j{}_i  (P_L \nslash)_{\bt\al}\, \bar \ell \tfrac{\bnslash}{2} t^a \ell
\,, \nn \\
\bar e_{\al} e_{\bt} &= \tfrac{1}{2} (P_R \nslash)_{\bt \al}\, \bar e_{} \tfrac{\bnslash}{2} e 
\,. \end{align}
%%%
Here $\alpha,\beta$ are spinor indices,  $i,j$ are $SU(2)$ gauge indices, and $N_w=2$. The lepton fields are treated as massless and assumed to correspond to the same collinear direction $n$. There are similar relations for the quarks. Eventually, we will take the proton matrix element of the quark operators. Since color is an unbroken gauge symmetry, and the proton is a color singlet state, matrix elements of color non-singlet operators in the proton vanish. We therefore drop these from the outset. 
 
We start with the most complicated case, namely $O^{(3)\dagger}_{\ell q} O^{(3)}_{\ell q}$:
%%%
\begin{align} \label{eq:O1_fact}
O^{(3)\dagger}_{\ell q} O^{(3)}_{\ell q}&= 
(\bar \ell_2 \cS_2^\dagger \ga^\nu t^b \cS_1 \ell_1)\,(\bar q_4 \cS_4^\dagger \ga_\nu t^b \cS_3 q_3)\,
  (\bar \ell_1 \cS_1^\dagger \ga^\mu t^a \cS_2 \ell_2)\, (\bar q_3 \cS_3^\dagger \ga_\mu t^a \cS_4 q_4)
\,.
\end{align}
%%%
We can use \eq{disentangle} to combine $\bar \ell_1$ and $\ell_1$ into a bilinear, $\bar \ell_2$ and $\ell_2$ into a bilinear, etc., and drop color non-singlet operators to obtain
%%%
\begin{align} \label{eq:2.13}
O^{(3)\dagger}_{\ell q} O^{(3)}_{\ell q}
  &= 
    (n_1 \sdt n_3)(n_2 \sdt n_4) \bigg[\frac{1}{N_c N_w^4}\, \C_{\ell_1}\, \C_{\ell_2}\, \C_{q_3}\, \C_{q_4}\, \tr(t^a t^b)\, \tr(t^a t^b) 
    \\ & \quad
 + \frac{4}{N_c N_w^2} \, \C_{\ell_1}^c\, \C_{\ell_2}^d\, \C_{q_3}\, \C_{q_4}\,
  \tr(\cS_1 t^c \cS_1^\dagger t^a \cS_2 t^d \cS_2^\dagger t^b)\, \tr(t^a t^b) 
  \nn \\ & \quad
 + \frac{4}{N_c N_w^2} \, \C_{\ell_1}\, \C_{\ell_2}\, \C^c_{q_3}\, \C^d_{q_4}\,
  \tr(\cS_3 t^c \cS_3^\dagger t^a \cS_4 t^d \cS_4^\dagger t^b)\, \tr(t^a t^b) 
  \nn \\ & \quad
 + \frac{4}{N_c N_w^2} \, \C_{\ell_1}^c\, \C_{\ell_2}\, \C_{q_3}^d\, \C_{q_4}\,
  \tr(\cS_1 t^c \cS_1^\dagger t^a t^b)\, \tr(\cS_3 t^d \cS_3^\dagger t^a t^b) + (3\text{ more})
  \nn \\ & \quad
 + \frac{8}{N_c N_w} \, \C_{\ell_1}^c\, \C_{\ell_2}^d\, \C_{q_3}^e\, \C_{q_4}\,
  \tr(\cS_1 t^c \cS_1^\dagger t^a \cS_2 t^d \cS_2^\dagger t^b)\, \tr(\cS_3 t^e \cS_3^\dagger t^a t^b) + (3\text{ more})
  \nn \\ & \quad
 + \frac{16}{N_c} \, \C_{\ell_1}^c\, \C_{\ell_2}^d\, \C_{q_3}^e\, \C_{q_4}^f\,
  \tr(\cS_1 t^c \cS_1^\dagger t^a \cS_2 t^d \cS_2^\dagger t^b)\, \tr(\cS_3 t^e \cS_3^\dagger t^a \cS_4 t^f \cS_4^\dagger t^b) \bigg]
\,.\nn\end{align}
%%%
Here we introduce the abbreviation
%%%
\begin{align} \label{eq:coll_abb}
  \C_{\ell_1} = \bar \ell_1 \tfrac{\bnslash_1}{2} \ell_1
  \,, \qquad
  \C_{\ell_1}^c = \bar \ell_1 \tfrac{\bnslash_1}{2} t^c \ell_1  
\,, \qquad
\dots
\,,\end{align}
%%%
where the superscript distinguishes the gauge group representations of the collinear operator: $\C_{\ell_1}$ is a weak singlet, and $\C_{\ell_1}^c$ is a weak triplet.

We simplify the soft operators using the $SU(N_w)$ completeness relation
%%%
\begin{align}
(t^a)^\alpha{}_\beta\,(t^a)^\gamma{}_\delta = \frac12 \de^\alpha{}_\delta\,\de^\gamma{}_\beta - \frac{1}{2N_w} \de^\alpha{}_\beta\, \de^\gamma{}_\delta\,.
\end{align}
%%%
The relevant identities are
%%%
\begin{align} \label{eq:soft_id}
  \tr(t^a t^b)\, \tr(t^a t^b) &= \tfrac{1}{4}(N_w^2-1)
  \,, \\
  \tr(\cS_1 t^c \cS_1^\dagger t^a \cS_2 t^d \cS_2^\dagger t^b)\, \tr(t^a t^b)
  &= - \tfrac{1}{4N_w}\,  \cS_{12}^{cd}
  \,,\nn \\ 
  \tr(\cS_1 t^c \cS_1^\dagger t^a t^b)\, \tr(\cS_3 t^d \cS_3^\dagger t^a t^b) 
  &= -\tfrac{1}{2N_w}\, \cS_{13}^{cd}
  \,, \nn \\
\tr(\cS_1 t^c \cS_1^\dagger t^a \cS_2 t^d \cS_2^\dagger t^b)\, \tr(\cS_3 t^e \cS_3^\dagger t^a t^b) 
&= -\tfrac{1}{4N_w} (\cS_{123}^{cde} + \cS_{132}^{ced})
\,,\nn \\ \quad
\tr(\cS_1 t^c \cS_1^\dagger t^a \cS_2 t^d \cS_2^\dagger t^b)\, \tr(\cS_3 t^e \cS_3^\dagger t^a \cS_4 t^f \cS_4^\dagger t^b) 
&= \tfrac{1}{4N_w^2} \cS_{12}^{cd} \cS_{34}^{ef}
\!+\! \tfrac14 \cS_{14}^{cf} \cS_{23}^{de}
\!-\!\tfrac{1}{4N_w} (\cS_{1234}^{cdef} \!+\! \cS_{1432}^{cfed} )
\nn \\
&\!\!\!\!\stackrel{N_w=2}{=}  -\tfrac{1}{16} \cS_{12}^{cd} \cS_{34}^{ef}
+\tfrac{1}{8} 
 \cS_{13}^{ce} \cS_{24}^{df} 
+ \tfrac{1}{8}  \cS_{14}^{cf} \cS_{23}^{de}
\,,\nn\end{align}
%%%
where the last one only holds for $SU(2)$.
Here we used $\cS_i^\dagger \cS_i = 1$ and introduced the shorthand notation
%%%
\begin{align} \label{eq:soft_ops}
  \cS_{12}^{cd} &= \tr(\cS_1 t^c \cS_1^\dagger \cS_2 t^d \cS_2^\dagger) 
 \,, \nn \\
   \cS_{123}^{cde} &=  \tr(\cS_1 t^c \cS_1^\dagger\, \cS_2 t^d \cS_2^\dagger\,  \cS_3 t^e \cS_3^\dagger )
  \,, \nn \\
   \cS_{1234}^{cdef} &= \tr(\cS_1 t^c \cS_1^\dagger\,  \cS_2 t^d \cS_2^\dagger\,  \cS_3 t^e \cS_3^\dagger\, \cS_4 t^f \cS_4^\dagger)  
\,.\end{align}
%%%
For $N_w=2$, the last relation in \eq{soft_id} was simplified using
%%%
\begin{align}
   \cS_{1234}^{cdef} \stackrel{N_w=2}{=} \tfrac12\big( \cS_{12}^{cd} \cS_{34}^{ef} - \cS_{13}^{ce} \cS_{24}^{df} + \cS_{14}^{cf} \cS_{23}^{de} \big)
\,.\nn\end{align}
%%%
Using the above relations gives
%%%
\begin{align} 
O^{(3)\dagger}_{\ell q} O^{(3)}_{\ell q}
  &= 
    (n_1 \sdt n_3)(n_2 \sdt n_4) \bigg[\frac{N_w^2-1}{4N_c N_w^4}\, \C_{\ell_1}\, \C_{\ell_2}\, \C_{q_3}\, \C_{q_4}\, 
    \\ & \quad
 -    \frac{1}{N_c N_w^3} \, \C_{\ell_1}^c\, \C_{\ell_2}^d\, \C_{q_3}\, \C_{q_4}\,
\cS_{12}^{cd}
 - \frac{1}{N_c N_w^3} \, \C_{\ell_1}\, \C_{\ell_2}\, \C^c_{q_3}\, \C^d_{q_4}\, \cS_{34}^{cd}
   \nn \\ & \quad
 - \frac{2}{N_c N_w^3} \, \C_{\ell_1}^c\, \C_{\ell_2}\, \C_{q_3}^d\, \C_{q_4}\, \cS_{13}^{cd} + (3\text{ more})
  \nn \\ & \quad
 - \frac{2}{N_c N_w^2} \, \C_{\ell_1}^c\, \C_{\ell_2}^d\, \C_{q_3}^e\, \C_{q_4}\, (\cS_{123}^{cde} + \cS_{132}^{ced}) + (3\text{ more})
  \nn \\ & \quad
 + \frac{1}{N_c}\, \C_{\ell_1}^c\, \C_{\ell_2}^d\, \C_{q_3}^e\, \C_{q_4}^f\,
 ( - \cS_{12}^{cd} \cS_{34}^{ef}
+2  \cS_{13}^{ce} \cS_{24}^{df} 
+ 2 \cS_{14}^{cf} \cS_{23}^{de}) \bigg]
\,.\nn\end{align}
%%%

We reiterate that a color-adjoint collinear operator of the form $\C_{q_4}^A = \bar q_4 \tfrac{\bnslash_4}{2} T^A q_4$, where $T^A$ is a color generator, would never have been considered in QCD. Although it could in principle be kept in intermediate steps of the calculation, it would be dropped at the end because its proton matrix element vanishes, since the proton is a color-singlet state. However, the proton is not an electroweak singlet and gives a nonzero matrix element for
the $SU(2)$ adjoint operator $\C_{q_4}^a = \bar q_4 \tfrac{\bnslash_4}{2} t^a q_4$, where $t^a$ is an $SU(2)$ generator, see \sec{low_matching}. Related to this, we note that only the $SU(2)$ Wilson lines survive in the soft operators in \eq{soft_ops}, since the colored Wilson lines are paired with colored operators which have vanishing proton matrix elements. The new features in the remaining discussion therefore center on $SU(2)$. For $SU(3)$ and $U(1)$ we have the standard PDF and fragmentation function evolution for the collinear operators and we have no soft operators.

Next we consider the interference contribution $O_{\ell q}^\dagger O^{(3)}_{\ell q}$ (and its conjugate $O_{\ell q}^{(3)\dagger} O_{\ell q}$), which can be obtained from \eq{O1_fact} by dropping the $t^b$'s  
%%%
\begin{align} 
O_{\ell q}^\dagger O^{(3)}_{\ell q} &= 
  (\bar \ell_2 \cS_2^\dagger \ga^\nu  \cS_1 \ell_1)\, (\bar q_4 \cS_4^\dagger \ga_\nu  \cS_3 q_3)\,
  (\bar \ell_1 \cS_1^\dagger \ga^\mu t^a \cS_2 \ell_2)\, (\bar q_3 \cS_3^\dagger \ga_\mu t^a \cS_4 q_4)
  \nn \\
  &= 
   (n_1 \sdt n_3)(n_2 \sdt n_4) \bigg[ \frac{4}{N_c N_w^2} \, \C_{\ell_1}^c\, \C_{\ell_2}\, \C_{q_3}^d\, \C_{q_4}\,
  \tr(\cS_1 t^c \cS_1^\dagger t^a )\, \tr(\cS_3 t^d \cS_3^\dagger t^a ) + (3\text{ more})
  \nn \\ & \quad
 + \frac{8}{N_c N_w} \, \C_{\ell_1}^c\, \C_{\ell_2}^d\, \C_{q_3}^e\, \C_{q_4}\,
  \tr(\cS_1 t^c \cS_1^\dagger t^a \cS_2 t^d \cS_2^\dagger )\, \tr(\cS_3 t^e \cS_3^\dagger t^a ) + (3\text{ more})
  \nn \\ & \quad
 + \frac{16}{N_c} \, \C_{\ell_1}^c\, \C_{\ell_2}^d\, \C_{q_3}^e\, \C_{q_4}^f\,
  \tr(\cS_1 t^c \cS_1^\dagger t^a \cS_2 t^d \cS_2^\dagger )\, \tr(\cS_3 t^e \cS_3^\dagger t^a \cS_4 t^f \cS_4^\dagger ) \bigg]
\,.\end{align}
%%%
This can be simplified using
%%%
\begin{align}
  \tr(\cS_1 t^c \cS_1^\dagger t^a)\, \tr(\cS_3 t^d \cS_3^\dagger t^a)
  &= \tfrac12 \cS_{13}^{cd}
  \,, \nn \\
  \tr(\cS_1 t^c \cS_1^\dagger t^a \cS_2 t^d \cS_2^\dagger )\, \tr(\cS_3 t^e \cS_3^\dagger t^a )  
  &= \tfrac12 \cS_{132}^{ced}
  \,, \nn \\
\tr(\cS_1 t^c \cS_1^\dagger t^a \cS_2 t^d \cS_2^\dagger )\, \tr(\cS_3 t^e \cS_3^\dagger t^a \cS_4 t^f \cS_4^\dagger )
 &= \tfrac{1}{2} \cS_{1432}^{cfed} 
 - \tfrac{1}{2N} \cS_{12}^{cd} \cS_{34}^{ef}
 \nn \\
 &\!\!\!\!\stackrel{N_w=2}{=} 
  -\tfrac{1}{4} \cS_{13}^{ce} \cS_{24}^{df}
+ \tfrac{1}{4}  \cS_{14}^{cf} \cS_{23}^{de}
\,,\end{align}
%%%
to get
%%%
\begin{align} 
O_{\ell q}^\dagger O^{(3)}_{\ell q}   &= 
   (n_1 \sdt n_3)(n_2 \sdt n_4) \bigg[ \frac{2}{N_c N_w^2} \, \C_{\ell_1}^c\, \C_{\ell_2}\, \C_{q_3}^d\, \C_{q_4}\, \cS_{13}^{cd}+ (3\text{ more})
  \nn \\ & \quad
 + \frac{4}{N_c N_w} \, \C_{\ell_1}^c\, \C_{\ell_2}^d\, \C_{q_3}^e\, \C_{q_4}\, \cS_{132}^{ced}
 + (3\text{ more})
 \nn \\ & \quad
 + \frac{4}{N_c}\, \C_{\ell_1}^c\, \C_{\ell_2}^d\, \C_{q_3}^e\, \C_{q_4}^f\,
( - \cS_{13}^{ce} \cS_{24}^{df}+ \cS_{14}^{cf} \cS_{23}^{de} ) \bigg]
\,.\end{align}
%%%

The expressions for $O_{\ell q}^\dagger O_{\ell q}$ can directly be obtained from \eq{O1_fact}, dropping $t^a$ \emph{and} $t^b$,  
%%%
\begin{align} \label{eq:O2_fact}
  O_{\ell q}^\dagger O_{\ell q} &= 
 ( \bar \ell_2 \cS_2^\dagger \ga^\nu  \cS_1 \ell_1)\, (\bar q_4 \cS_4^\dagger \ga_\nu  \cS_3 q_3)\,
  (\bar \ell_1 \cS_1^\dagger \ga^\mu  \cS_2 \ell_2)\, (\bar q_3 \cS_3^\dagger \ga_\mu  \cS_4 q_4)
  \nn \\
  &=
  (n_1 \sdt n_3)(n_2 \sdt n_4) \bigg[ \frac{1}{N_cN_w^2}\, \C_{\ell_1}\, \C_{\ell_2}\, \C_{q_3}\, \C_{q_4}
 + \frac{4}{N_cN_w} \, \C_{\ell_1}^c\, \C_{\ell_2}^d\, \C_{q_3}\, \C_{q_4}\,
 \cS_{12}^{cd} + (1\text{ more})
 \nn \\ & \quad
  + \frac{16}{N_c}\, \C_{\ell_1}^c\, \C_{\ell_2}^d\, \C_{q_3}^e\, \C_{q_4}^f\,
 \cS_{12}^{cd} \cS_{34}^{ef} \bigg]
\,.\end{align}
%%%

For $O_{\ell u}^\dagger O_{\ell u}$ there is a further simplification compared to \eq{O2_fact} because the $SU(2)$ doublet $q$ is replaced by the singlet $u$, 
%%%
\begin{align} \label{eq:O3_fact}
  O_{\ell u}^\dagger O_{\ell u} &= 
  (\bar \ell_2 \cS_2^\dagger \ga^\nu  \cS_1 \ell_1)\, (\bar u_4 \cS_4^\dagger \ga_\nu  \cS_3 u_3)\,
  (\bar \ell_1 \cS_1^\dagger \ga^\mu  \cS_2 \ell_2)\, (\bar u_3 \cS_3^\dagger \ga_\mu  \cS_4 u_4)
  \nn \\
  &=
  (n_1 \sdt n_3)(n_2 \sdt n_4) \bigg[ \frac{1}{N_cN_w}\, \C_{\ell_1}\, \C_{\ell_2}\, \C_{u_3}\, \C_{u_4}
  +\frac{4}{N_c}\, \C_{\ell_1}^c\, \C_{\ell_2}^d\, \C_{u_3}\, \C_{u_4}\,
  \cS_{12}^{cd} \bigg]
\,.\end{align}
%%%
The expression for $O_{\ell d}^\dagger O_{\ell d}$ can directly be obtained from \eq{O3_fact} by replacing $u \to d$, and the expression for $O_{eq}^\dagger O_{eq}$ follows from interchanging $q \lra \ell$ and $e \lra u$.
Finally, for $O_{eu}^\dagger O_{eu}$  
%%%
\begin{align}
  O_{eu}^\dagger O_{eu} &= 
  (\bar e_2 \cS_2^\dagger \ga^\nu  \cS_1 e_1)\, (\bar u_4 \cS_4^\dagger \ga_\nu  \cS_3 u_3)\,
  (\bar e_1 \cS_1^\dagger \ga^\mu  \cS_2 e_2)\, (\bar u_3 \cS_3^\dagger \ga_\mu  \cS_4 u_4)
  \nn \\
  &=
  (n_1 \sdt n_3)(n_2 \sdt n_4) \frac{1}{N_c}\, \C_{e_1}\, \C_{e_2}\, \C_{u_3}\, \C_{u_4}
,\end{align}
%%%
and similarly for $O_{ed}^\dagger O_{ed}$.

The above identities can be used to write the factorized cross section \eq{factor} as a product of collinear and soft factors, which we now study separately.
The collinear factors are the usual PDFs. The soft factors do not arise in QCD factorization theorems, but are present in electroweak cross sections.
In appendix~\ref{app:xsec}, we give examples of the possible PDF combinations which enter the production of a heavy particle in quark-antiquark annihilation, and show that they are all given in terms of the singlet and triplet PDFs.

%%%%%%%%%%%%%%%%%%%%%%%%%%%%%%%%%%%%%%%%%%%%%%%%%%%%%%%%%%%%%%%%%%%%%%%%%%%%%%%%
\section{Collinear evolution}
\label{sec:RGE_C}
%%%%%%%%%%%%%%%%%%%%%%%%%%%%%%%%%%%%%%%%%%%%%%%%%%%%%%%%%%%%%%%%%%%%%%%%%%%%%%%%

In this section we determine the renormalization group (RG) evolution of the collinear operators entering the factorized cross section. The splitting functions for $z < 1$ agree with those computed in ref.~\cite{Bauer:2017isx}.
We begin with the collinear operators corresponding to the incoming particles, discussing the fragmentation case in \sec{RGE_FF}.
The collinear operators that enter the cross section can be written in terms of the usual PDF operators. In QCD processes,  PDF operators are singlets under the $SU(3)$ gauge group, and there are no rapidity divergences, as these cancel between real and virtual graphs. In the electroweak case, the factorization formula has terms involving the product of collinear and soft operators which are not separately gauge singlets, as we have seen in the previous section. There are rapidity divergences in the collinear and soft sectors. One can see this must be true in the collinear sector by noting that real and virtual graphs have different group theory factors for gauge non-singlet PDFs.

Before discussing the electroweak case, we first review QCD, for which the standard definitions for the unpolarized PDF quark operator is~\cite{Collins:1981uw}
%%%
\begin{align}\label{eq:cs}
O_Q(r^-) &= \frac{1 }{ 4 \pi} \int\! \df \xi\, e^{-i \xi r^-}
[ \bar Q(\bar n \xi )\, \W (\bar n \xi )]\ \slashed{\bar n}\ [\W^\dagger (0) \,  Q(0 ) ]
\,. \end{align}
%%%
Here the Wilson line $\W$ is defined in \eq{Wilson} and the null vectors are
%%%
\begin{align}
n^\mu=(1,\mathbf{\hat n})\,, \qquad \bar n^\mu=(1,-\mathbf{\hat n})
\,.\end{align}
%%%
Note that the operator product in \eq{cs} is an ordinary product, not a time-ordered product. One can insert a complete set of states between $\W$ and $\W^\dagger$, allowing us to evaluate matrix elements of the PDF operator using cut Feynman rules. The PDF operators in \eq{cs} are written using standard QCD notation. In terms of collinear fields introduced in \eq{Wilson}, the quark PDF operator is given by $[W^\dagger Q] \to Q$, since the collinear Wilson line was included in $Q$.  
The quark PDF is given by the matrix elements of this operators in a target state $T$ of momentum $p$,
%%%
\begin{align}\label{eq:pdf}
f_{Q/T}(r^-/p^-,\mu) &\equiv \braket{T,p| O_Q(r^-) |T,p}
\,,\end{align}
%%%
where $p^- = \bn \sdt p$ and the operators are renormalized in the \MSbar\ scheme. The polarized quark PDF $f_{\Delta Q}$ is given by replacing $\slashed{\bar n}$ by $\slashed{\bar n}\gamma_5$. In terms of $f_{Q_+}$ and $f_{Q_-}$, the distributions of quarks with helicity $h=+1/2$ and $h=-1/2$, the unpolarized and polarized PDFs are $f_Q = f_{Q_+}+f_{Q_-}$ and $f_{\Delta Q} = f_{Q_+}-f_{Q_-}$.

We need to generalize \eq{cs}, \eq{pdf} to include PDFs which are not gauge singlets. The gauge indices in a fermion bilinear $\bar Q \, Q$ can be combined into a gauge singlet or adjoint. We thus define two different fermion operators 
%%%
\begin{align} \label{eq:3.33}
O^{(1)}_Q(r^-) &= \frac{1 }{ 4 \pi} \int \df \xi\, e^{-i \xi r^-}\,
[ \bar Q(\bar n \xi ) \, \W (\bar n \xi )]_i \ \slashed{\bar n}\
\, \delta^i_j\,  [\W^\dagger (0) \,  Q(0 ) ]^j
\,,  \nn \\
O^{(\text{adj},a)}_Q(r^-) &= \frac{1 }{ 4 \pi} \int \df \xi\, e^{-i \xi r^-}\, [ \bar Q(\bar n \xi )\, \W (\bar n \xi )]_i\ \slashed{\bar n}\
 [t^a]^i{}_j\,  [\W^\dagger (0) \,  Q(0 ) ]^j
\,, \end{align}
%%%
where $i,j,k,l$ are gauge indices in the fundamental representation.
Note that it is the gauge indices at the $x=\infty$ end of the Wilson line which are combined into a singlet or adjoint.
The anti-fermion and anti-scalar PDFs are given by $CP$ conjugation, where $P$ is implemented by reflection in a plane containing the direction of the proton $n$. This amounts to exchanging $\bar Q \leftrightarrow Q$ and letting $t^a \to -(t^a)^T$ in the adjoint PDFs, 
%%%
\begin{align} \label{eq:3.33f}
O^{(1)}_{\bar Q}(r^-) &= \frac{1 }{ 4 \pi} \int \df \xi\, e^{-i \xi r^-}\,
\tr  \left\{ [\W^\dagger (\bar n \xi ) \,  Q(\bar n \xi ) ]^j
\, \delta^i_j\,   [ \bar Q( 0 ) \, \W ( 0 )]_i  \, \slashed{\bar n}\right\}
\,,  \nn \\
O^{(\text{adj},a)}_{\bar Q}(r^-) &= \frac{1 }{ 4 \pi} \int \df \xi\, e^{-i \xi r^-}\, 
\tr \left\{  [\W^\dagger (\bar n \xi) \,  Q( \bar n \xi ) ]^j  [-(t^a)^T]_j{}^i\,  [ \bar Q( 0 )\, \W ( 0 )]_i\ \slashed{\bar n}\right\}
\,, \end{align}
%%%
which can also be written
%%%
\begin{align} \label{eq:3.33g}
O^{(1)}_{\bar Q}(r^-) &=- \frac{1 }{ 4 \pi} \int \df \xi\, e^{-i \xi r^-}\,
[ \bar Q(0 ) \, \W (0 )]_i \ \slashed{\bar n}\
\, \delta^i_j\,  [\W^\dagger (\bar n \xi) \,  Q(\bar n \xi ) ]^j
\,,  \nn \\
O^{(\text{adj},a)}_{\bar Q}(r^-) &= \frac{1 }{ 4 \pi} \int \df \xi\, e^{-i \xi r^-}\, [ \bar Q(0 )\, \W (0)]_i\ \slashed{\bar n}\
 [t^a]^i{}_j\,  [\W^\dagger (\bar n \xi) \,  Q(\bar n \xi ) ]^j
\,, \end{align}
%%%
on anticommuting the fermion fields. The $Q$ and $\bar Q$ PDFs have been defined in the conventional way, such that for example the deep-inelastic structure functions are proportional to $f_{Q}(x,\mu)+f_{\bar Q}(x,\mu)$. If the singlet PDF is treated as the PDF for quark number, then the total baryon number is given by the difference of the $Q$ and $\bar Q$ PDFs with the sign convention of \eq{3.33g}. The triplet PDFs have an $SU(2)$ charge, and this requires the additional minus sign, so that the total $SU(2)$ charge is given by the sum of  $Q$ and $\bar Q$ triplet PDFs. 
The matrix elements in a target $T$ of momentum $p$ define the singlet and adjoint PDFs
%%%
\begin{align}\label{eq:3.7}
f^{(1)}_{Q/T}(r^-/p^-,\mu) &\equiv \braket{T,p| O^{(1)}_Q(r^-) |T,p}, & f^{(1)}_{\bar Q/T}(r^-/p^-,\mu) &\equiv \braket{T,p| O^{(1)}_{\bar Q}(r^-) |T,p}, \nn \\
f^{(\text{adj},a)}_{Q/T}(r^-/p^-,\mu) &\equiv \braket{T,p| O^{(\text{adj},a)}_Q(r^-) |T,p} , &
f^{(\text{adj},a)}_{\bar Q/T}(r^-/p^-,\mu) &\equiv \braket{T,p| O^{(\text{adj},a)}_{\bar Q}(r^-) |T,p}\,.
\end{align}
%%%
The singlet PDF for QCD is the same as the usual PDF. The adjoint PDF vanishes for QCD, but not in the electroweak case.

The generalization of eqs.~\eqref{eq:3.33}, \eqref{eq:3.33f} and \eqref{eq:3.7} to the electroweak case is straightforward. Since the weak interactions are chiral, it is necessary to use polarized PDFs. One defines fermion and antifermion PDFs using \eqs{3.33}{3.33f} with the Wilson line in the appropriate $SU(3) \times SU(2) \times U(1)$ representation, and the field $Q$ replaced by the chiral fields $q,l,u,d,e$. For left-handed fields $q$ and $l$, this gives the distribution of particles with helicity $h=-1/2$, and for right-handed fields, the distribution of particles with helicity $h=+1/2$. In terms of the usual polarized and unpolarized quark distributions, the  PDF in \eq{cs} with $Q \to u$ corresponds to $(f_u+f_{\Delta u})/2=f_{u_+}$, and with $Q \to q$ corresponds to $(f_u-f_{\Delta u})/2+(f_d-f_{\Delta d})/2=f_{u_-}+f_{d_-}$, etc.\ where the $\pm$ subscript denotes the helicity (not chirality) of the quark. For antiquarks, the relation between helicity and chirality is reversed, so the antiquark PDFs for $Q\to u$ and $Q \to q$ give $(f_{\bar u}-f_{\Delta \bar u})/2=f_{\bar u_-}$ and $(f_{\bar u}+f_{\Delta \bar u})/2+(f_{\bar d} + f_{\Delta \bar d})/2=f_{\bar u_+}+f_{\bar d_+}$, respectively. In the unbroken theory, the PDFs are defined using chiral SM fields, and we will use the notation $f_u$, $f_q$, $f_{\bar u}$, $f_{\bar q}$ to denote these four PDFs, dropping the helicity label.
In the electroweak sector, we will also denote singlet and adjoint PDFs using  superscripts $(I=0)$ and $(I=1)$.

We also need the unpolarized and polarized gluon PDF operators defined by~\cite{Collins:1981uw,Manohar:1990kr,Manohar:1990jx}
%%%
\begin{align}\label{eq:pol}
O_G(r^-) &= -\frac{1 }{ 2 \pi r^-} \int\! \df  \xi\, e^{-i \xi r^-}
 \bn_\mu [G^{\mu \lambda }(\bar n \xi)\, \W (\bar n \xi )]\ \bn_\nu [\W^\dagger (0) \,G^{\nu}{}_\lambda(0)]\,, \nn \\
O_{\Delta G}(r^-) &= \frac{i}{ 2 \pi r^-} \int\! \df  \xi\, e^{-i \xi r^-}
 \bn_\mu [G^{\mu \lambda }(\bar n \xi)\, \W (\bar n \xi )]\ \bn_\nu [\W^\dagger (0) \, {\widetilde G}^{\nu}{}_\lambda(0)]\,,
\end{align}
%%%
($\widetilde G_{\alpha \beta} = \frac12 \epsilon_{\alpha \beta \ga \de} G^{\ga \de}$ with $\epsilon_{0123}=+1$)
whose matrix elements give the PDFs $f_{G_+} + f_{G_-}$ and $f_{G_+} - f_{G_-}$. It is more convenient to use $f_{G_+}$ and $f_{G_-}$, the distribution of helicity $h=1$ and $h=-1$ gauge bosons, given by the sum and difference of the equations in \eq{pol}.\footnote{
$  \bn_\mu G^{\mu \lambda }(\bar n \xi)\, \bn_\nu {\widetilde G}^{\nu}{}_\lambda(0) = - \bn_\mu {\widetilde G}^{\mu \lambda }(\bar n \xi)\, \bn_\nu {G}^{\nu}{}_\lambda(0)  $ and $  \bn_\mu G^{\mu \lambda }(\bar n \xi)\, \bn_\nu G^{\nu}{}_\lambda(0) = \bn_\mu {\widetilde G}^{\mu \lambda }(\bar n \xi)\, \bn_\nu {\widetilde G}^{\nu}{}_\lambda(0) $ so the other two possibilities for replacing $G^{\mu \lambda} \to \widetilde G^{\mu \lambda}$ do not lead to new PDFs.}  There are also double helicity-flip gauge PDFs which are leading twist~\cite{Manohar:1988pp,Jaffe:1989xy}. These correspond to a transition where a helicity $h=\pm 1$ gauge boson is emitted and a helicity $h=\mp 1$ gauge boson is absorbed. Since gauge boson helicity changes by two, there has to be a corresponding change in helicity of the target hadron; as a result these operators only contribute to scattering off targets with spin $\ge 1$, and we neglect them here. They occur in the factorization theorem through box graphs.

The transverse $W$ PDF is given by replacing the gluon field-strength tensor by the $SU(2)$ field-strength, and using a Wilson line in the adjoint of $SU(2)$. The transverse $B$ PDF is given by using $B_{\mu \nu}$, and no Wilson line is required since a $U(1)$ field-strength is a gauge singlet.  
The gauge operator involves two adjoint fields, and 
%%%
\begin{align}\label{eq:reps}
\text{adj} \otimes \text{adj} &= [1+ \text{adj} + \bar a a + \bar s s]_S + [\text{adj} + \bar a s + \bar s a]_A\,,
\end{align}
%%%
where the first four representations are in the symmetric product of the two adjoints, and the last three are in the antisymmetric product. The representation $\bar a a$ is a traceless tensor $t^{ab}_{cd}$ antisymmetric in its lower and in its upper indices, $\bar a s$ is a traceless tensor $t^{ab}_{cd}$ antisymmetric in its lower and 
symmetric in its upper indices, etc. For the special case of $SU(3)$, $\bar ss=\mathbf{27}$, $\bar a s=\mathbf{10}$,  $\bar s a = \mathbf{\overline{10}}$, and the $\bar a a$ does not exist. For the special case of $SU(2)$, $\bar ss$ is the isospin $I=2$ representation, $\bar a s$, $\bar s a$, $\bar a a$ do not exist, and $\text{adj}_S$ does not exist since the $d$-symbol vanishes. Further details on the group theory can be found in app.~A of ref.~\cite{Dashen:1994qi}.

The various gauge operators are given by ($G_{\mu \lambda}$ denotes a generic field-strength tensor)
%%%
\begin{align}\label{eq:3.36}
O_G^{(R,c)} (r^-) &= -\frac{1 }{ 2 \pi r^-} \int \! \df  \xi\, e^{-i \xi r^-}\,
\bn_\mu [G^{\mu \lambda }(\bar n \xi)\, \W (\bar n \xi )]_a 
 \mathscr{C}_{ab}^{(R,c)}\,   \bn_\nu [\W^\dagger (0) \,G^{\nu}{}_\lambda(0)]_b \,,
\end{align}
%%%
where $a,b,c$ are gauge indices in the adjoint representation (upper vs.~lower indices do not matter, since the adjoint representation is real).
$\mathscr{C}_{ab}^{(R,c)}$ is a Clebsch-Gordan coefficient for combining the two adjoints into state $c$ of representation $R$ given in \eq{reps}.
The Clebsch-Gordan coefficients for the singlet, and the symmetric and antisymmetric adjoints are
%%%
\begin{align}
\mathscr{C}_{ab}^{(1)} &= \delta_{ab}\,, &
\mathscr{C}_{ab}^{(\text{adj}_S,c)} &= d_{abc}\,, &
\mathscr{C}_{ab}^{(\text{adj}_A,c)} &= -\img f_{abc} \,.
\end{align}
%%%
The Clebsch-Gordan coefficients for $\bar aa$, $\bar s s$, $\bar a s$ and $\bar s a$ are given in ref.~\cite{Dashen:1994qi}. We also have the corresponding polarized PDFs given by replacing $G^{\nu}{}_\lambda(0)$ by $- \img\, {\widetilde G}^{\nu}{}_\lambda(0)$  in \eq{3.36}. 

For the $SU(2) \times U(1)$ case, there are some additional PDFs. There are two more isotriplet gauge PDFs, 
%%%
\begin{align}\label{eq:3.36a}
O_{W\!B}^{(I=1,c)} (r^-) &= -\frac{1 }{ 2 \pi r^-} \int \! \df  \xi\, e^{-i \xi r^-}\,
\bn_\mu [W^{\mu \lambda }(\bar n \xi)\, \W (\bar n \xi )]_c \,   \bn_\nu [ B^{\nu}{}_\lambda(0)] \,,\nn \\
O_{BW}^{(I=1,c)} (r^-) &= -\frac{1 }{ 2 \pi r^-} \int \! \df  \xi\, e^{-i \xi r^-}\,
\bn_\mu [B^{\mu \lambda }(\bar n \xi)] \,   \bn_\nu [\W^\dagger (0) \,W^{\nu}{}_\lambda(0)]_c \,.
\end{align}
%%%
A Wilson line is not needed for the $U(1)$ field-strength tensor, since it is a gauge singlet. Taking the Hermitian conjugate gives
$[ O_{W\!B}^{(I=1,c)} (r^-) ]^\dagger = O_{BW}^{(I=1,c)} (r^-)$. We also have the polarized versions of \eq{3.36}, $O_{\Delta W B}$ and $O_{\Delta B W}$.

For the massive electroweak gauge bosons, \eq{3.36} and its polarized version only give the PDFs for $h=\pm 1$. We also need the PDFs for $h=0$ longitudinally polarized gauge bosons. As we now discuss, these PDFs can not be written as light-cone Fourier transforms of operators involving the field-strength tensor. For a massive $W$ moving in the $+z$ direction, i.e.~$\hat n = (0,0,1)$, its momentum and polarization are
%%%
\begin{align}
p^\mu &=(E,0,0,p), & \epsilon_+ &= -\frac{1}{\sqrt 2}(0,1,i,0),  & \epsilon_- &= \frac{1}{\sqrt 2}(0,1,-i,0), &
\epsilon_0 &= \frac{1}{M_W}(p,0,0,E).
\end{align}
%%%
The field strength tensor annihilates a $W$ with amplitude
%%%
\begin{align}
\mathcal{A}^{\mu \nu} = -i(p^\mu \epsilon^\nu - p^\nu \epsilon^\mu).
\end{align}
%%%
For helicity $h=\pm1$ with polarization vectors $\epsilon_{\pm}$, $\mathcal{A}^{\mu \nu}$ has components of order $E$. For $h=0$,
%%%
\begin{align}
\epsilon_0 &=\frac{1}{M_W} p^\mu - \frac{M_W}{E+p} \bar n^\mu,
\end{align}
%%%
and
%%%
\begin{align}
\mathcal{A}^{\mu \nu} = i  \frac{M_W^2}{E+p} (p^\mu \bar n^\nu - p^\nu \bar n^\mu).
\end{align}
%%%
Since the $p^\mu/M_W$ term does not contribute to $\mathcal{A}^{\mu \nu}$,  $\mathcal{A}^{\mu \nu}$ has components of order $M_W/E$, which are suppressed by $M_W^2/E^2$ relative to transverse polarization, and so are not leading twist.

The longitudinal gauge boson PDFs are given in terms of scalar PDFs, using the Goldstone boson equivalence theorem~\cite{chanowitz,bohmbook}.
We use the PDF operators
%%%
\begin{align}\label{eq:cs2}
O_H^{(I=0)}(r^-) &= \frac{r^- }{ 2 \pi} \int\! \df \xi\, e^{-i \xi r^-}
[ H^\dagger (\bar n \xi )\, \W (\bar n \xi )]\ [\W^\dagger (0) \,  H(0 ) ] \,, \nn \\
O_H^{(I=1,a)}(r^-) &= \frac{r^- }{ 2 \pi} \int\! \df \xi\, e^{-i \xi r^-}
[ H^\dagger (\bar n \xi )\,  \W (\bar n \xi )]\ t^a\ [\W^\dagger (0) \,  H(0 ) ] \,.
\end{align}
%%%
for the Higgs field, which is given by
%%%
\begin{align}
H &= \begin{pmatrix}
H^+  \\
H^0 \end{pmatrix} = \frac{1}{\sqrt 2}\! \begin{pmatrix}
\varphi^2 + \img \varphi^1 \\
v + h - \img \varphi^3 \end{pmatrix}\,,
\end{align}
%%%
in the unbroken and broken phase, respectively. Here $h$ is the physical Higgs particle, and the unphysical scalars $\varphi^3,\varphi^\pm=(\varphi^1 \mp \img \varphi^2)/\sqrt{2}$ in the Higgs multiplet can be related to longitudinal electroweak gauge bosons $Z_L, W^\pm_L$ using the Goldstone boson equivalence theorem~\cite{chanowitz,bohmbook}. The $CP$-conjugate $\bar H$ PDFs are given by $H \leftrightarrow \bar H$ and $t^a \to (-t^a)^T$.

There are two additional Higgs PDFs $O_{\widetilde HH}^{(I=0)} $, $O_{\widetilde H H}^{(I=1,c)} $ given by replacing $H^\dagger$ in \eq{cs2} by $\widetilde H^\dagger$, where
%%%
\begin{align}\label{eq:wide}
\widetilde H_j &= \epsilon_{jk} H^{\dagger\, k} =
 \begin{pmatrix}
\bar H^0  \\
-H^- \end{pmatrix} .
\end{align}
%%%
The $CP$-conjugate PDFs are given by swapping $H \leftrightarrow \widetilde H$, and $t^a \to (-t^a)^T$. The operator $O_{\widetilde HH}^{(I=0)}$ breaks electromagnetism and doesn't contribute to factorization formulae. $O_{\widetilde H H}^{(I=1,c)}$, $O_{H \widetilde H}^{(I=1,c)}$ do, and are included in our analysis. They have $Y=\pm 1$, and can occur in the factorization theorem in pairs. The $\Delta Q=0$ components  have non-zero proton matrix elements.
Taking Hermitian conjugates gives $[ O_{\widetilde HH}^{(I=1,c)}(r^-)]^\dagger = -O_{H \widetilde H}^{(I=1,c)} (r^-)$.

%===============================================================================
\subsection{Anomalous dimensions}
%===============================================================================

We will first briefly review rapidity divergences, and the rapidity regulator of ref.~\cite{Chiu:2011qc,Chiu:2012ir}, that we use to treat them. Rapidity divergences arise in e.g.~transverse momentum dependent factorization theorems, where the emission of a single soft gluon involves an integral over its rapidity with a rapidity-independent (i.e.~constant) integrand. To bring this divergence under control, the $\eta$ regulator of ref.~\cite{Chiu:2011qc,Chiu:2012ir} explicitly breaks boost invariance. The resulting $1/\eta$ poles cancel between the collinear and soft operators. These poles lead to renormalization group equations involving the rapidity renormalization scale $\nu$. Just as the $\mu$-evolution can be used to resum invariant mass logarithms, the $\nu$-evolution can be used to resum rapidity logarithms, which arise because the collinear and soft operators have different natural rapidity scales. This will be discussed in more detail in \sec{resummation}, see in particular \fig{nu}. In our case we do not measure the transverse momentum of the gauge boson, but instead have to account for the gauge boson mass.

The RG equations of the collinear operators take the form
%%%%
\begin{align} \label{eq:C_RGE}
\frac{\df}{\df \ln \mu}\, O_i(r^-,\mu,\nu) &= \sum_{j} \int_0^1\! \frac{\df z}{z}\,    \gamma_{\mu,ij}(z,r^-,\mu,\nu)\, O_j\Big(\frac{r^-}{z},\mu,\nu\Big)
\,,\nn \\
\frac{\df}{\df \ln \nu}\, O_i(r^-,\mu,\nu) &=  \ga_{\nu,i}(\mu,\nu)\, O_i(r^-,\mu,\nu)
\,.\end{align}
%%%\
The lower limit on $z$ turns into $r^-/p^-$ when its matrix element is taken in a state with momentum $p$. The anomalous dimension $\gamma_\mu$ depends not just on $z$ but also on $r^-$ because of rapidity divergences. The collinear operators mix under $\mu$ evolution, but are multiplicatively renormalized under $\nu$ evolution. 
The convolution in \eq{C_RGE} will be abbreviated by $\otimes$. 
Since we limit ourselves to one-loop results, we find it convenient to use the notation
%%%
\begin{align}
\gamma &\equiv \frac{\alpha}{\pi} \, \hat \gamma\, .
\end{align}
%%%
in intermediate expressions. We now compute the anomalous dimensions $\gamma_\mu$ and $\gamma_\nu$. 

The indices $i$ and $j$ in \eq{C_RGE} run over all the fermion and gauge boson PDFs. We will use the helicity basis $Q_+$, $Q_-$, $G_+$, $G_-$ rather than the more conventional basis $Q$, $\Delta Q$, $G$, $\Delta G$ used in QCD. In QCD, there is no $\nu$-evolution and parity invariance implies that the $\mu$-evolution kernels satisfy
%%%
\begin{align}
 \gamma_{\mu,Q_+ Q_+} & = \gamma_{\mu,Q_- Q_-}\,, &
 \gamma_{\mu,Q_+ Q_-} & = \gamma_{\mu,Q_- Q_+}\,, &
 \gamma_{\mu,G_+ G_+} & = \gamma_{\mu,G_- G_-}\,, &
 \gamma_{\mu,G_+ G_-} & = \gamma_{\mu,G_- G_+}\,, \nn \\
 \gamma_{\mu,Q_+ G_+} & = \gamma_{\mu,Q_- G_-}\,, &
 \gamma_{\mu,Q_+ G_-} & = \gamma_{\mu,Q_- G_+}\,, &
 \gamma_{\mu,G_+ Q_+} & = \gamma_{\mu,G_- Q_-}\,, &
 \gamma_{\mu,G_+ Q_-} & = \gamma_{\mu,G_- Q_+}\,, 
\end{align}
%%%
which allows one to write evolution equations which mix $\{ f_Q,\, f_G\}$, and separately mix $\{f_{\Delta Q}, f_{\Delta G}\}$ using the kernels
%%%
\begin{align}
 \gamma_{\mu,QQ} &=  \gamma_{\mu,Q_+Q_+}  +  \gamma_{\mu,Q_+Q_-}\,, &
 \gamma_{\mu,\Delta Q \Delta Q} &=  \gamma_{\mu,Q_+Q_+}  -  \gamma_{\mu,Q_+Q_-} \nn  \\
 \gamma_{\mu,QG} &=  \gamma_{\mu,Q_+G_+}  +  \gamma_{\mu,Q_+G_-}  &
 \gamma_{\mu,\Delta Q\Delta G} &=  \gamma_{\mu,Q_+G_+}  -  \gamma_{\mu,Q_+G_-}  \nn \\
 \gamma_{\mu,GQ} &=  \gamma_{\mu,G_+Q_+}  +  \gamma_{\mu,G_+Q_-} &
 \gamma_{\mu,\Delta G\Delta Q} &=  \gamma_{\mu,G_+Q_+}  -  \gamma_{\mu,G_+Q_-} \nn \\
 \gamma_{\mu,G G} &=  \gamma_{\mu,G_+G_+}  +  \gamma_{\mu,G_+G_-}  &
 \gamma_{\mu,\Delta G\Delta G} &=  \gamma_{\mu,G_+G_+}  -  \gamma_{\mu,G_+G_-}. 
\end{align}
%%%
This simplification is not possible in the electroweak sector, since parity is not a good symmetry. We therefore write our results using the helicity basis
$ \gamma_{Q_+Q_+}$, $ \gamma_{Q_+ Q_-}$, etc.

%===============================================================================
\subsection{$\gamma_{QQ}$ and $\gamma_{HH}$}
%===============================================================================

\begingroup
\renewcommand{\arraystretch}{1.5}
\setlength{\arraycolsep}{3pt}
\begin{table*}
\centering
\begin{eqnarray*}
\begin{array}{c|c|c}
\hline\hline
\text{Graph} &\hat \gamma_\mu &\hat \gamma_\nu \\
 \hline
\begin{minipage}{2cm} \mybox{\includegraphics[width=2cm]{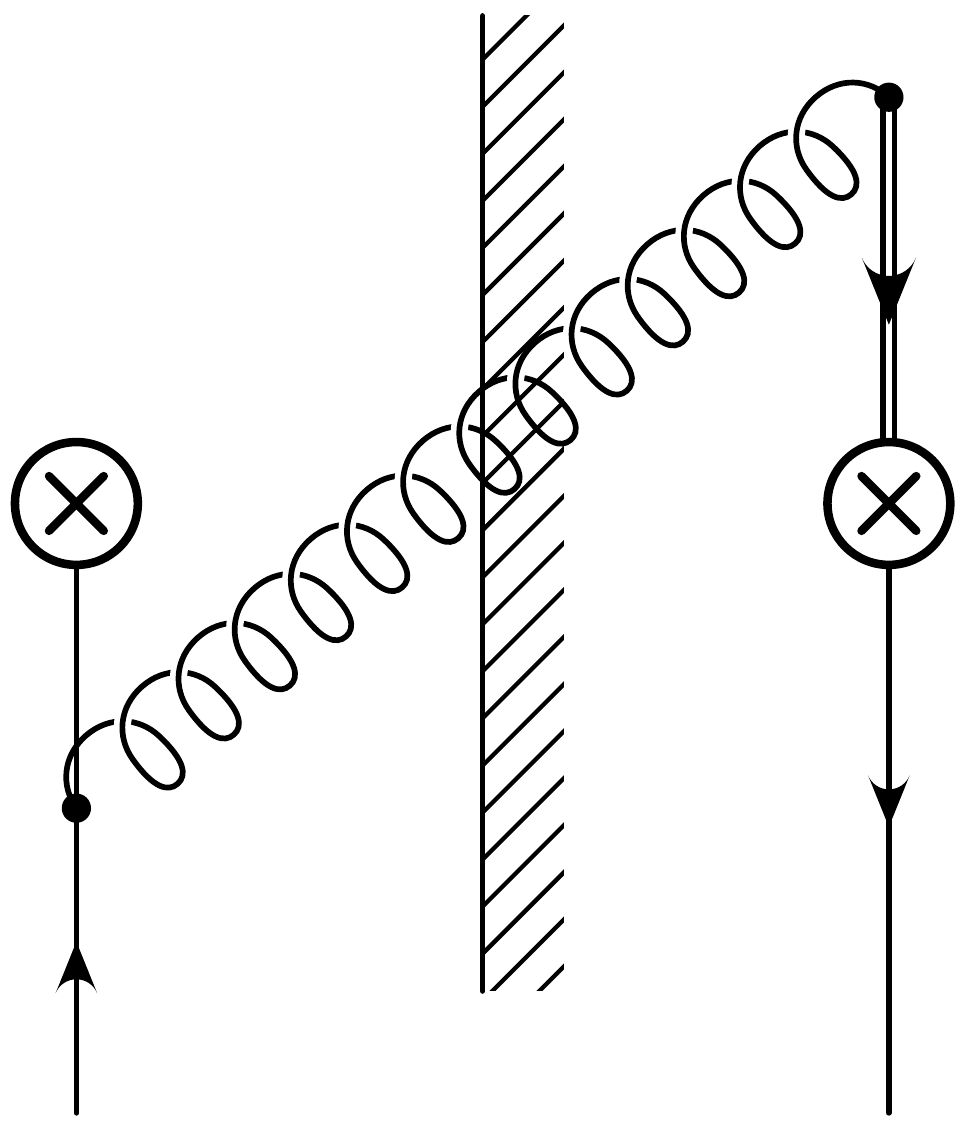}} \end{minipage}
&   \frac{2}{(1-z)}_{\!+} -2- 2\,\delta(1-z) \ln \frac{\nu}{\bar n \cdot r} & -  \ln \frac{\mu^2}{M^2}  \\  
\begin{minipage}{2cm} \mybox{\includegraphics[width=2cm]{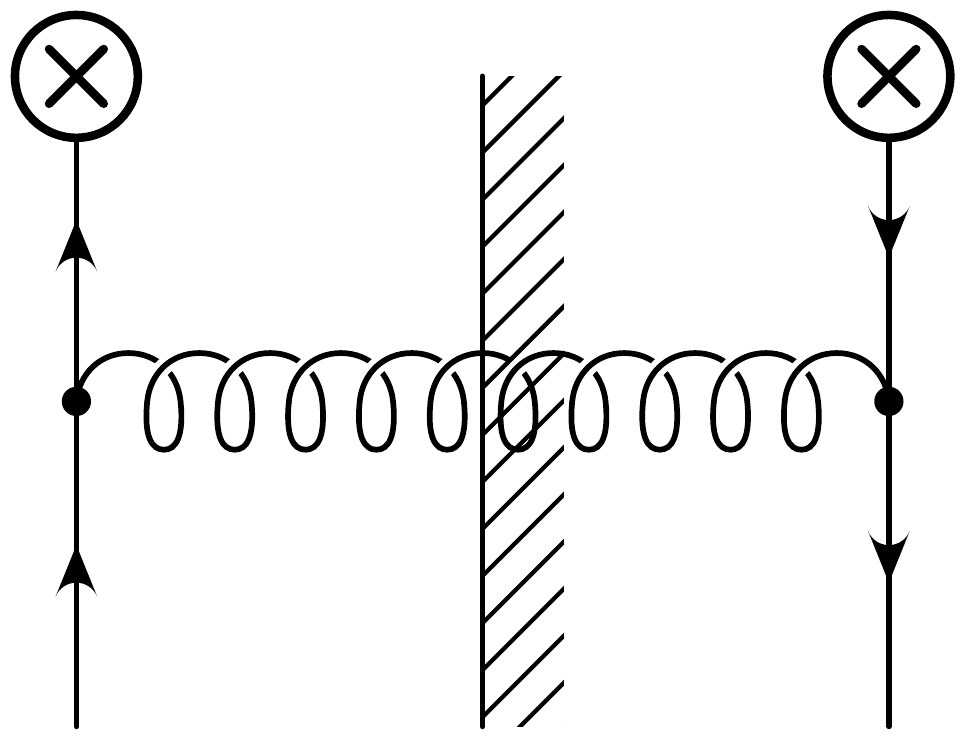}} \end{minipage}
  &  1-z  & 0  \\  \hline 
  \text{Total}_1 & \frac{2}{(1-z)}_{\!+} -z-1- 2\,\delta(1-z) \ln \frac{\nu}{\bar n \cdot r} & - \ln \frac{\mu^2}{M^2}   \\ \hline
\begin{minipage}{2cm} \mybox{\includegraphics[width=2cm]{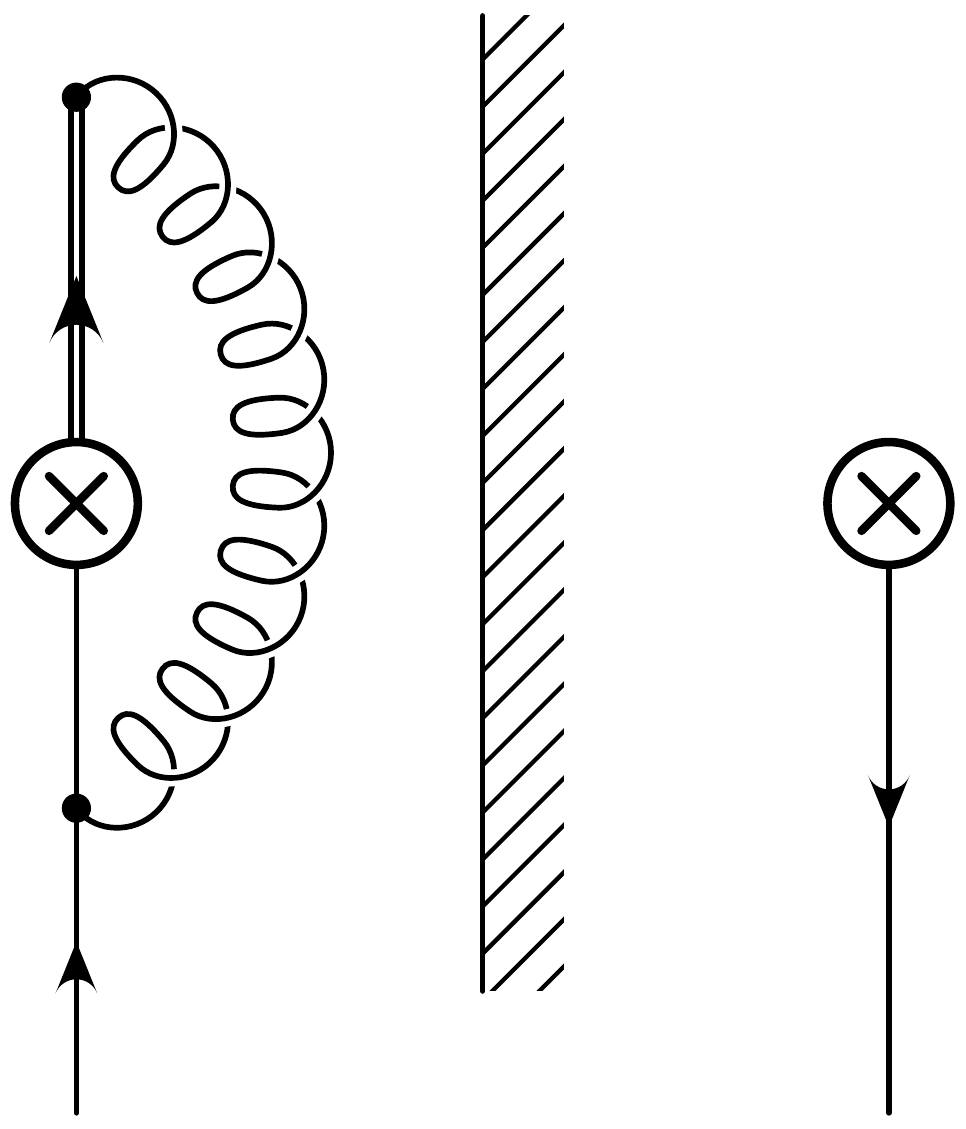}} \end{minipage}
 &2 \bigl( \ln \frac{\nu}{\bar n \cdot r} +1\bigr)\delta(1-z)  &  \ln \frac{\mu^2}{M^2}  \\ 
\begin{minipage}{1.8cm} \mybox{\includegraphics[width=1.8cm]{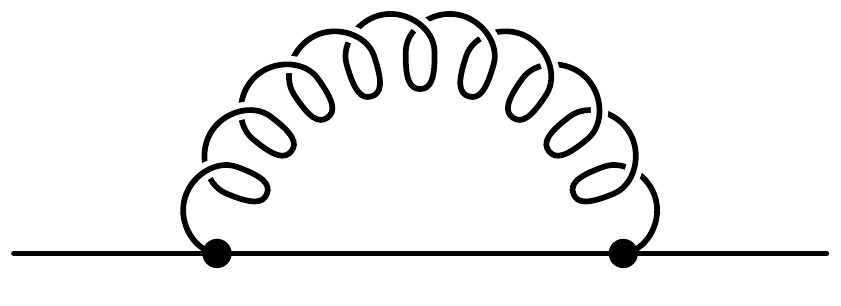}} \end{minipage}
&  -\frac12\delta(1-z) & 0  \\
\hline
  \text{Total}_2 & \bigl(2\, \ln \frac{\nu}{\bar n \cdot r} +\frac 32 \bigr)  \delta(1-z) &  \ln \frac{\mu^2}{M^2}   \\ \hline \hline
\end{array}
\end{eqnarray*}
\caption{One-loop diagrams for the renormalization of fermion collinear operators. The columns show the graph and contribution to the $\mu$ and $\nu$ anomalous dimension. The combinatoric factor of $2$ for the first and third graphs, and $-1$ for the wavefunction graph has been included. Subsets of the graphs have been summed to give $\text{Total}_1$ and $\text{Total}_2$. For the singlet fermion PDF, $\text{Total}_1$ and $\text{Total}_2$ have group theory factor $c_F$. For the adjoint PDF, $\text{Total}_1$ has group theory factor $c_F-c_A/2$ and $\text{Total}_2$ has group theory factor $c_F$.}
\label{tab:coll_fermion}
\end{table*}
\endgroup

The one-loop gauge diagrams and resulting contributions to the $QQ$ anomalous dimensions are shown in table~\ref{tab:coll_fermion}. Their calculation was performed in section V of ref.~\cite{Manohar:2012jr}, using dimensional regularization for the UV divergences and the rapidity renormalization group~\cite{Chiu:2011qc,Chiu:2012ir} to treat the rapidity divergences. The graphs are divided into two sets, with the sum of each set given in the table. The value of individual diagrams depends on the gauge choice, but $\text{Total}_1$ and $\text{Total}_2$ remain the same. The color factor $c_{QQ}(R)$ for $\text{Total}_1$ depends on the PDF representation, whereas the color factor for $\text{Total}_2$ is $c_F$.

The one-loop graphs in Table~\ref{tab:coll_fermion} conserve fermion helicity, implying that the helicity mixing terms $\gamma_{Q_+Q_-}$ and $\gamma_{Q_-Q_+}$ vanish at this order. The expressions for the diagrams hold for both $\gamma_{Q_+Q_+}$ for $Q=u,d,e$ and $\gamma_{Q_-Q_-}$ for $Q=q,l$, and lead to the following anomalous dimensions
%%%
\begin{align} \label{eq:ga_C}
\hat \gamma_{\mu,QQ}^{(R)}  &= c_{QQ}(R) \Big( \frac{2}{(1-z)}_{\!+} -z-1 \Big)
+  \frac 32 c_F \delta(1-z)  + 2\left[ c_F- c_{QQ}(R)\right]  \ln \frac{\nu}{\bar n \cdot r}\, \delta(1-z) 
\,,\nn \\
&= c_{QQ}(R) \widetilde P_{QQ}(z)
+ \left[c_F- c_{QQ}(R)\right]\Big( 2  \ln \frac{\nu}{\bar n \cdot r} + \frac 32 \Big) \delta(1-z) 
\,,\nn \\
\hat \gamma_{\nu,Q}^{(R)}  &= \left[c_F - c_{QQ}(R)\right]  \ln \frac{\mu^2}{M^2} 
\,,\end{align}
%%%
with
%%%
\begin{align} \label{eq:ga_C_ap}
\widetilde P_{QQ}(z) &=   \frac{2}{(1-z)}_{\!+} -z-1  +  \frac 32\, \delta(1-z)  
\,, \end{align}
%%%
the usual Altarelli-Parisi evolution kernel.
The group theory factor $c_{QQ}(R)$ is 
%%%
\begin{align}\label{eq:3.41}
c_{QQ}(1) &= c_F & c_{QQ}(\text{adj}) &= c_F - \frac12 c_A\,.
\end{align}
%%%
As stated above, the expressions for the anomalous dimensions in \eq{ga_C} hold for both $Q_+Q_+$ and $Q_-Q_-$. However, there are differences in the anomalous dimensions due to group theory factors, since the gauge quantum numbers of SM fields depend on chirality (and hence helicity).
The above results also hold for the $\bar Q$ PDF.
The $\ln \nu/(\bar n \cdot r)$ in \eq{ga_C} arises due to the rapidity divergence. With our conventions $\bar n \cdot r=2 E$, where $E$ is the energy of the outgoing parton.
The $QQ$ anomalous dimension for the singlet PDF reproduces the standard result~\cite{Altarelli:1977zs,Dokshitzer:1977sg}. The rapidity divergences cancel in this case to yield $\gamma_\nu=0$. 

In ref.~\cite{Manohar:2012jr}, the gauge boson mass $M$ that appears in these expressions played the role of infrared regulator, and dropped out in the final result. Here the gauge boson mass $M$ is physical and does not drop out. As we will discuss in \sec{mixing}, the only gauge boson mass that enters for $SU(2) \times U(1)$ is $M=M_W$,  with the exception of $O_{\widetilde HH}$, where also $M=M_Z$ contributes.

\begingroup
\renewcommand{\arraystretch}{1.5}
\setlength{\arraycolsep}{3pt}
\begin{table*}
\centering
\begin{eqnarray*}
\begin{array}{c|c|c}
\hline\hline
\text{Graph} &\hat \gamma_\mu &\hat \gamma_\nu \\
 \hline
\begin{minipage}{2cm} \mybox{\includegraphics[width=2cm]{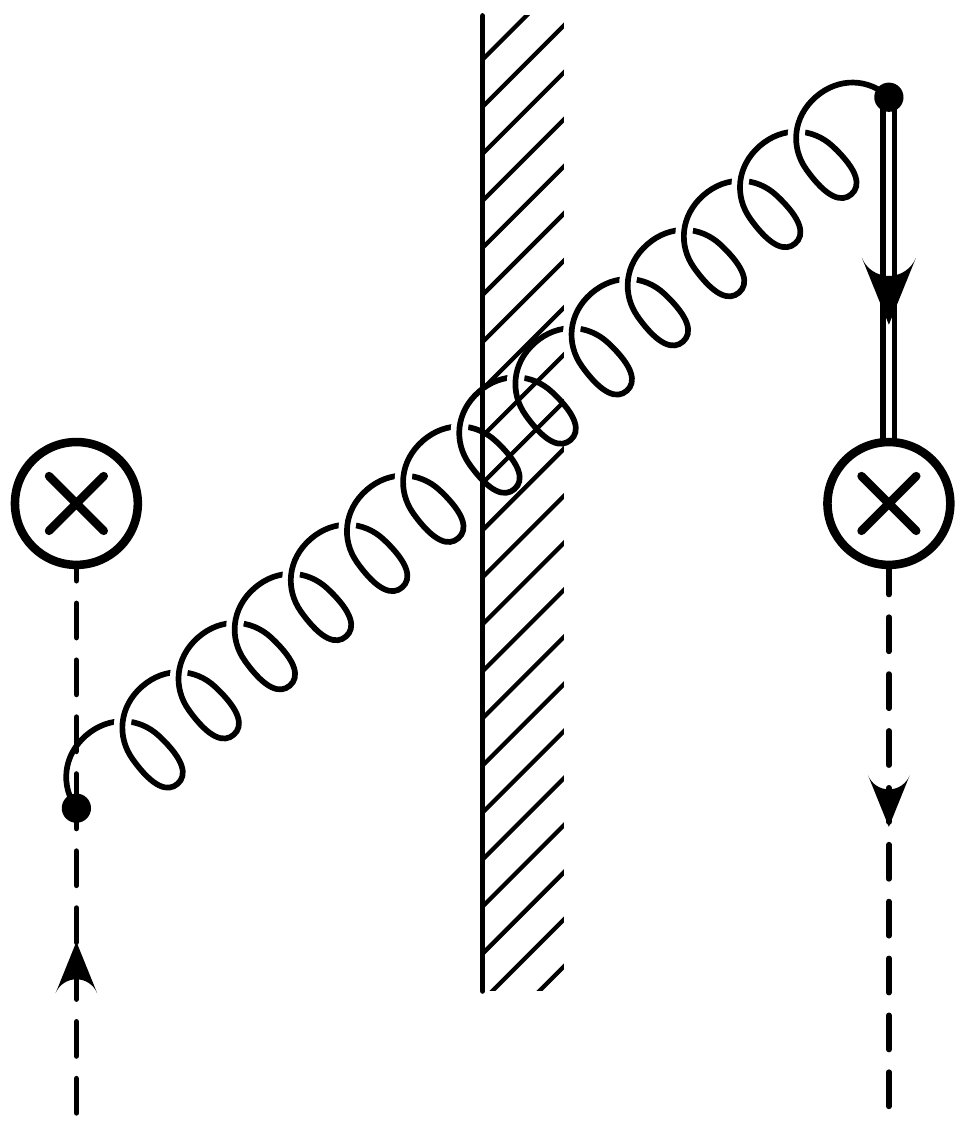}} \end{minipage}
&   \frac{2}{(1-z)_+}-z - 2  - 2  \delta(1-z) \ln \frac{\nu}{\bn \cdot r} & -  \ln \frac{\mu^2}{M^2}  \\  
\begin{minipage}{2cm} \mybox{\includegraphics[width=2cm]{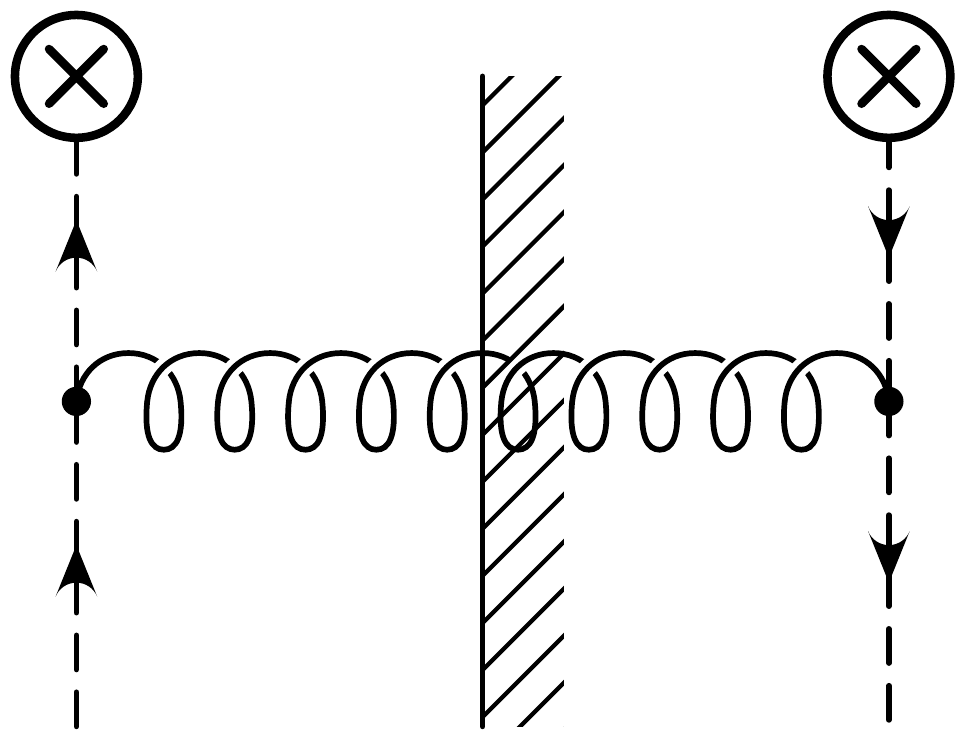}} \end{minipage}
  &  z  & 0  \\  \hline 
  \text{Total}_1 & \frac{2}{(1-z)_+}- 2  - 2 \delta(1-z) \ln \frac{\nu}{\bn \cdot r} & - \ln \frac{\mu^2}{M^2}   \\ \hline
\begin{minipage}{2cm} \mybox{\includegraphics[width=2cm]{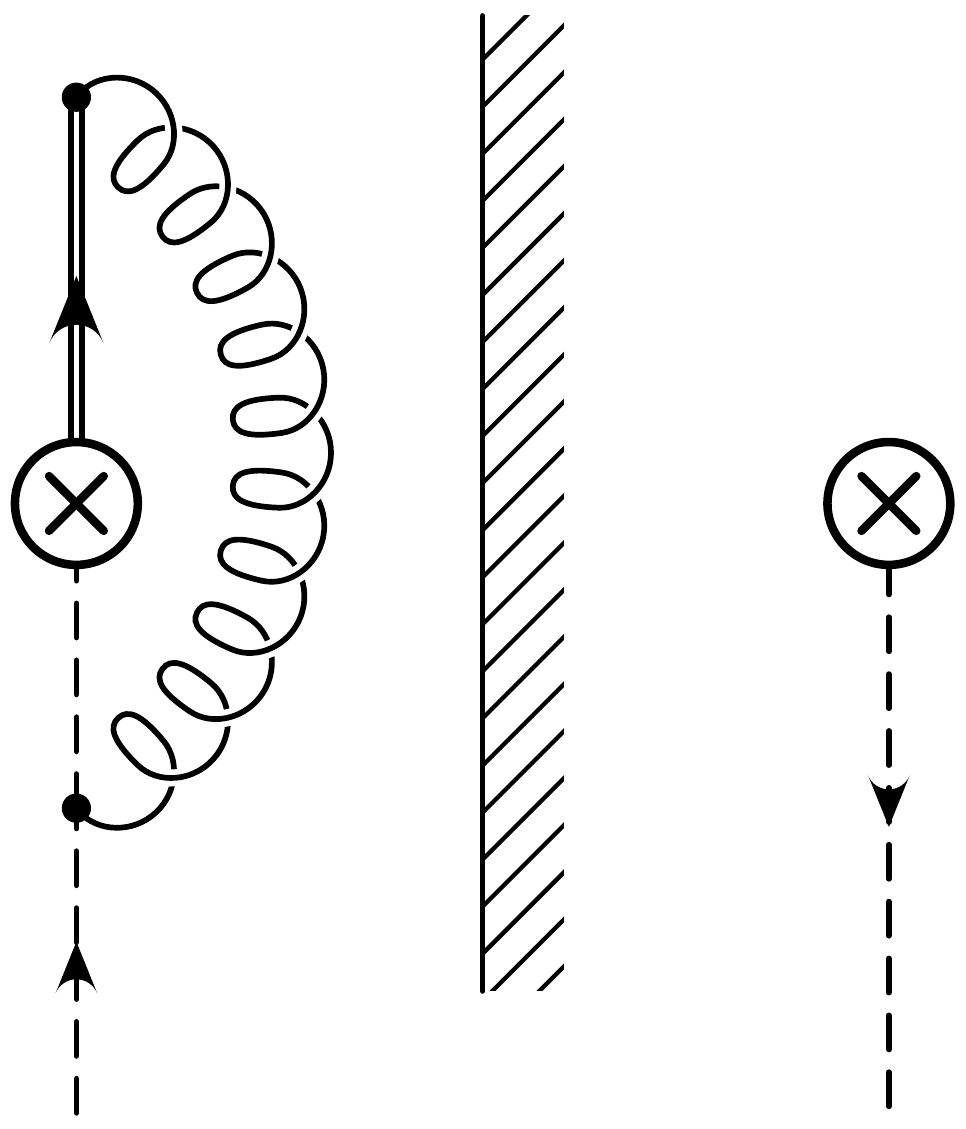}} \end{minipage}
 &  \bigl(2 \ln \frac{\nu}{\bn \cdot r} +1\bigr)\delta(1-z)   &  \ln \frac{\mu^2}{M^2}  \\ 
\begin{minipage}{1.8cm} \mybox{\includegraphics[width=1.8cm]{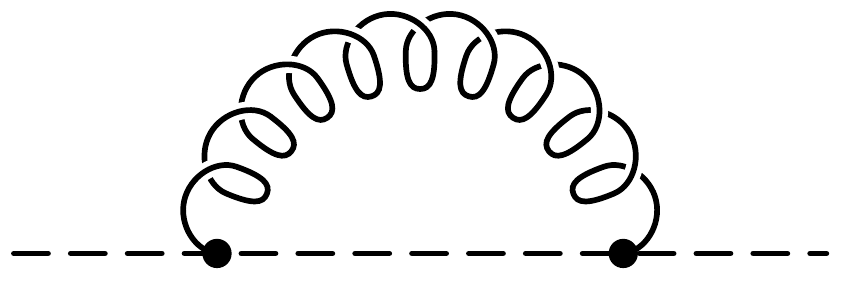}} \end{minipage}
&  \delta(1-z) & 0  \\ \hline
  \text{Total}_2 & \bigl(  2  \ln \frac{\nu}{\bn \cdot r} + 2 \bigr) \delta(1-z) &  \ln \frac{\mu^2}{M^2}   \\ \hline \hline
\end{array}
\end{eqnarray*}
\caption{One-loop diagrams for the renormalization of scalar collinear operators. Subsets of the graphs have been summed to give $\text{Total}_1$ and $\text{Total}_2$. For the singlet scalar PDF, $\text{Total}_1$ and $\text{Total}_2$ have group theory factor $c_F$. For the adjoint PDF, $\text{Total}_1$ has group theory factor $c_F-c_A/2$ and $\text{Total}_2$ has group theory factor $c_F$.}
\label{tab:6}
\end{table*}
\endgroup

An almost identical analysis holds for the mixing of scalar (i.e.\ Higgs) and gauge PDFs. The graphs are shown in table~\ref{tab:6}, and give
%%%
\begin{align} \label{eq:3.17}
\hat \gamma_{\mu,HH}^{(R)}  &= c_{HH}(R) \widetilde P_{HH}(z)
+ \left[c_F- c_{HH}(R)\right]\Big(2  \ln \frac{\nu}{\bar n \cdot r} + 2\Big) \delta(1-z) 
\,,\nn \\
\hat \gamma_{\nu,H}^{(R)}  &= \left[c_F - c_{HH}(R)\right]  \ln \frac{\mu^2}{M^2} 
\,,\end{align}
%%%
with
%%%
\begin{align}
\widetilde P_{HH}(z) &=   \frac{2}{(1-z)}_{\!+} -2  +  2\, \delta(1-z) 
\,. \end{align}
%%%
The group theory factor $c_{HH}(R)$ for scalars is the same as $c_{QQ}(R)$ for fermions. The scalar results also hold for the $\bar H$ PDF.

\begingroup
\renewcommand{\arraystretch}{1.5}
\setlength{\arraycolsep}{3pt}
\begin{table*}
\centering
\begin{eqnarray*}
\begin{array}{c|c|c}
\hline\hline
\text{Graph} &\hat \gamma_\mu &\hat \gamma_\nu \\
 \hline
\begin{minipage}{2cm} \mybox{\includegraphics[width=2cm]{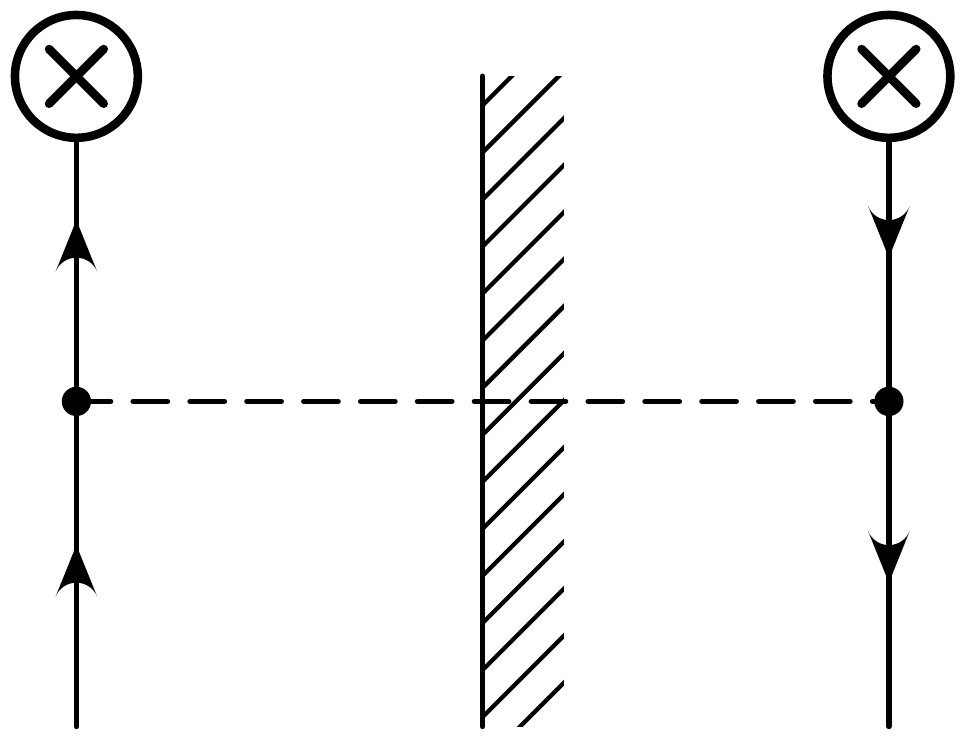}} \end{minipage}
  &  \frac12(1-z)  & 0  \\   \hline
\begin{minipage}{1.8cm} \mybox{\includegraphics[width=1.8cm]{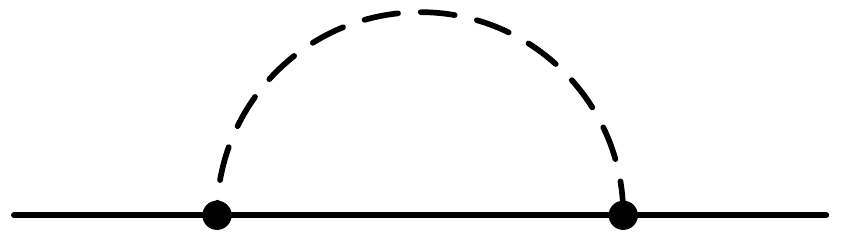}} \end{minipage}
&  -\frac14\delta(1-z) & 0  \\
\hline
\end{array}
\end{eqnarray*}
\caption{One-loop Yukawa diagrams for the renormalization of fermion collinear operators. The Yukawa factors multiplying the graphs are given in \eqs{QQ_yuk}{HH_yuk}.}
\label{tab:2}
\end{table*}
\endgroup

The Yukawa diagrams which contribute to the fermion  anomalous dimensions are shown in table~\ref{tab:2}. We will use the convention
%%%
\begin{align}
\mathcal{L}_Y &= - H^{\dagger j} \overline d_r\, [Y_d]_{rs}\, q_{j,s} - \widetilde H^{\dagger j} \overline u_r\, [Y_u]_{rs}\, q_{j,s} - H^{\dagger j} \overline e_r\, [Y_e]_{rs} \,  \ell_{j,s} + \hbox{h.c.}
\label{eq:sm}
\end{align}
%%%
for the Yukawa couplings, where $j$ is an $SU(2)$ index, $r,s$ are flavor indices, and $\widetilde H$ is given in \eq{wide}.
The Lagrangian in \eq{sm} is written in the weak eigenstate basis. 
The Yukawa interaction \eq{sm} is gauge invariant only if the weak gauge group is $SU(2)$, so the Yukawa contributions given below are only valid in this case. In the Standard Model, one can pick a flavor basis in which
%%%
\begin{align}\label{eq:3.21}
\frac{v}{\sqrt 2} Y_e &= M_e, &
\frac{v}{\sqrt 2} Y_u &= M_u,  &
\frac{v}{\sqrt 2} Y_d &= M_d V^\dagger,
\end{align}
%%%
where $M_e=\text{diag}(m_e,m_\mu,m_\tau)$, $M_u=\text{diag}(m_u,m_c,m_t)$, $M_d=\text{diag}(m_d,m_s,m_b)$, and
$V$ is the CKM mixing matrix. Since the only heavy fermion is the top quark, one can, to a very good approximation let
%%%
\begin{align}\label{eq:approx}
Y_e &\to 0, & Y_d &\to 0, & Y_u &\to \text{diag}(0,0,\sqrt{2}\, m_t/v)\,,
\end{align}
%%%
in the anomalous dimensions. We will, however, retain the full Yukawa dependence for the moment.

The scalar exchange contribution to the evolution kernel is not diagonal in flavor. Letting $O_{Q,r,s}$ denote the $Q$ PDF with fields $\bar Q_r$ and $Q_s$ in \eq{3.33}, where $r,s$ are flavor (generation) indices, the Yukawa contribution to the evolution equations are:
%%%
\begin{align} \label{eq:QQ_yuk}
\mu \frac{\df}{\df \mu}\ O^{(I=0)}_{d,r,s} &= \frac{1}{4\pi^2} \biggl[  \frac12 (1-z) [Y_d^\dagger]_{vr} [Y_d]_{sw} \otimes O^{(I=0)}_{q,v,w}  -\frac14 \delta(1-z) [Y_d Y_d^\dagger]_{vr}\otimes  O^{(I=0)}_{d,v.s} \nn \\
& -  \frac14 \delta(1-z) [Y_d Y_d^\dagger]_{sv} \otimes O^{(I=0)}_{d,r.v} \biggr] \,, \nn \\
\mu \frac{\df}{\df \mu}\ O^{(I=0)}_{u,r,s} &= \frac{1}{4\pi^2} \biggl[  \frac12 (1-z) [Y_u^\dagger]_{vr} [Y_u]_{sw} \otimes O^{(I=0)}_{q,v,w}  -\frac14 \delta(1-z) [Y_u Y_u^\dagger]_{vr} \otimes O^{(I=0)}_{u,v.s} \nn \\
& -  \frac14 \delta(1-z) [Y_u Y_u^\dagger]_{sv}\otimes  \otimes O^{(I=0)}_{u,r.v} \biggr] \,, \nn \\
\mu \frac{\df}{\df \mu}\ O^{(I=0)}_{e,r,s} &= \frac{1}{4\pi^2} \biggl[  \frac12 (1-z) [Y_e^\dagger]_{vr} [Y_e]_{sw} \otimes O^{(I=0)}_{\ell,v,w}  -\frac14 \delta(1-z) [Y_e Y_e^\dagger]_{vr} \otimes O^{(I=0)}_{e,v.s} \nn \\
& -  \frac14 \delta(1-z) [Y_e Y_e^\dagger]_{sv} \otimes O^{(I=0)}_{e,r.v} \biggr] \,, \nn \\
\mu \frac{\df}{\df \mu}\ O^{(I=0)}_{q,r,s} &= \frac{1}{4\pi^2} \biggl[ (1-z) [Y_d]_{vr} [Y_d^\dagger]_{sw} \otimes O^{(I=0)}_{d,v,w} +
(1-z) [Y_u]_{vr} [Y_u^\dagger]_{sw} \otimes O^{(I=0)}_{u,v,w}  \nn \\
&  -\frac18 \delta(1-z) [Y_d^\dagger Y_d + Y_u^\dagger Y_u]_{vr} \otimes O^{(I=0)}_{q,v.s} 
 -  \frac18 \delta(1-z) [Y_d^\dagger Y_d +Y_u^\dagger Y_u]_{sv} \otimes O^{(I=0)}_{q,r.v} \biggr] \,, \nn \\
\mu \frac{\df}{\df \mu}\  O^{(I=0)}_{\ell,r,s} &= \frac{1}{4\pi^2} \biggl[ (1-z) [Y_e]_{vr} [Y_e^\dagger]_{sw} \otimes O^{(I=0)}_{e,v,w} 
  -\frac18 \delta(1-z) [Y_e^\dagger Y_e ]_{vr} \otimes O^{(I=0)}_{\ell,v.s}  \nn \\
& -  \frac18 \delta(1-z) [Y_e^\dagger Y_e]_{sv} \otimes O^{(I=0)}_{\ell,r,v} \biggr] \,, \nn \\
 \mu \frac{\df}{\df \mu}\ O^{(I=1)}_{q,r,s} &= \frac{1}{4\pi^2} \biggl[   -\frac18 \delta(1-z) [Y_d^\dagger Y_d + Y_u^\dagger Y_u]_{vr} \otimes O^{(I=1)}_{q,v.s} 
 -  \frac18 \delta(1-z) [Y_d^\dagger Y_d +Y_u^\dagger Y_u]_{sv} \otimes O^{(I=1)}_{q,r,v} \biggr] \,,\nn \\
 \mu \frac{\df}{\df \mu}\  O^{(I=1)}_{\ell,r,s} &= \frac{1}{4\pi^2} \biggl[
  -\frac18 \delta(1-z) [Y_e^\dagger Y_e ]_{vr} \otimes O^{(I=1)}_{\ell,v.s} 
 -  \frac18 \delta(1-z) [Y_e^\dagger Y_e]_{sv} \otimes  O^{(I=1)}_{\ell,r.v} \biggr] \,.
\end{align}
%%%
The antifermion evolution equations are given by $CP$ conjugation, i.e.\ by replacing $d,r,s \leftrightarrow \bar d,s,r$, etc.\ on both sides of the equation. The factorization theorem leads to collinear operators with $r=s$, which can mix with $r \not= s$ operators under evolution. In the final results, we will use \eq{approx}, which greatly simplifies the results. Most terms vanish, and the Yukawa evolution is flavor diagonal and only contributes to the third generation.

Yukawa couplings give an additional contribution to Higgs wavefunction renormalization, and hence an additional term to the $HH$ anomalous dimension
%%%
\begin{align} \label{eq:HH_yuk}
\gamma_{\mu,HH} &= -\frac1{8\pi^2} \tr \big[ N_c Y_u^\dagger Y_u + N_c Y_d^\dagger Y_d + Y_e^\dagger Y_e \big] \delta(1-z)\,,
\end{align}
%%%
which must be added to \eq{3.17}.

%===============================================================================
\subsection{$\gamma_{GG}$}
%===============================================================================

\begingroup
\renewcommand{\arraystretch}{1.5}
\setlength{\arraycolsep}{3pt}
\begin{table*}
\centering
\begin{eqnarray*}
\begin{array}{c|c|c|c|c}
\hline\hline 
 \text{Graph} & \multicolumn{2}{c|}{P_{G_+ G_+}} & \multicolumn{2}{c}{P_{G_+ G_-}} \\
 \hline
 & \hat \gamma_\mu & \hat \gamma_\nu & \hat \gamma_\mu & \hat \gamma_\nu \\
\hline
\begin{minipage}{2cm} \mybox{\includegraphics[width=2cm]{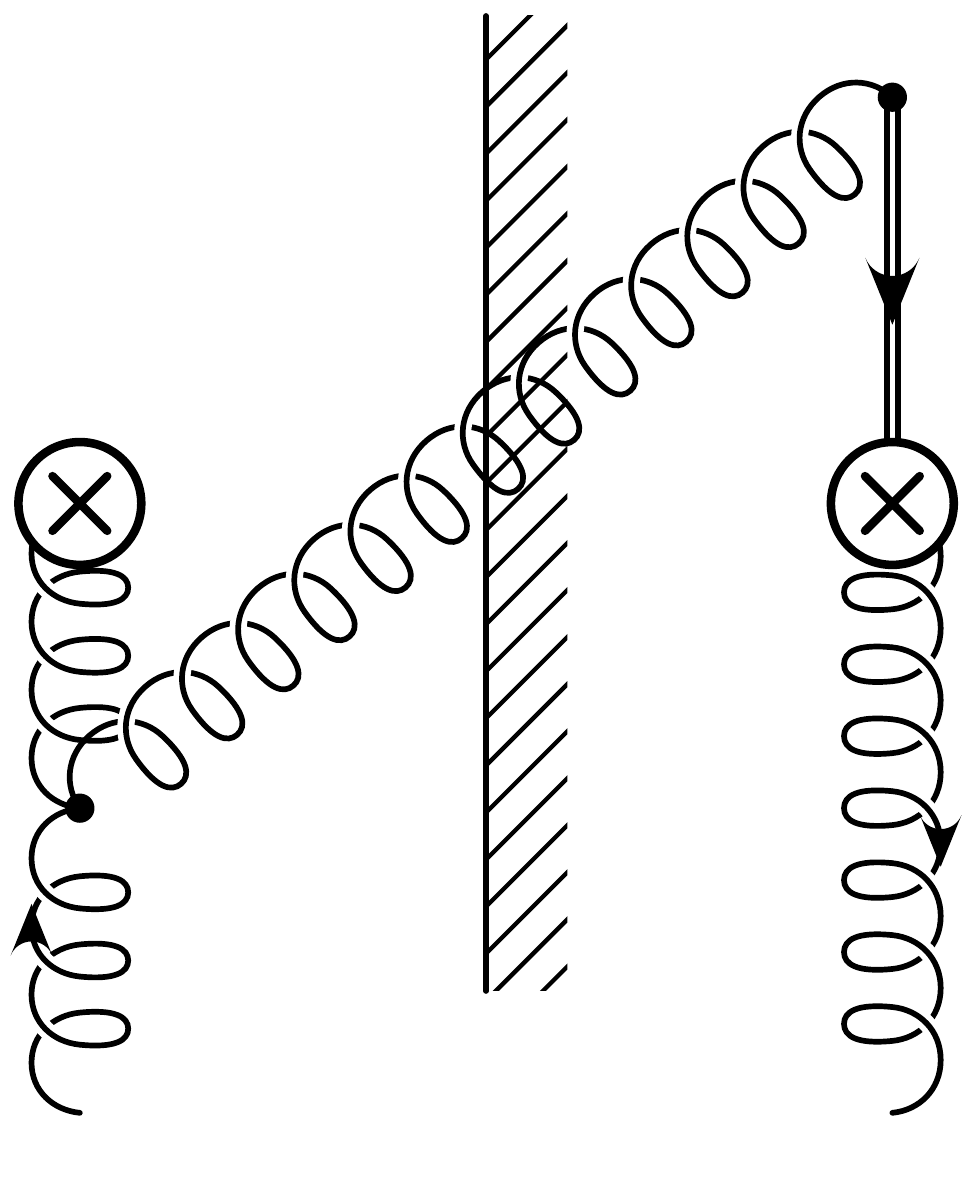}} \end{minipage} &  \frac{2}{(1-z)_+}-1- 2 \ln \frac{\nu}{\bar n \cdot r }\,\delta(1-z)  & -\ln \frac{\mu^2}{M^2} & 0 & 0  \\
\begin{minipage}{2cm} \mybox{\includegraphics[width=2cm]{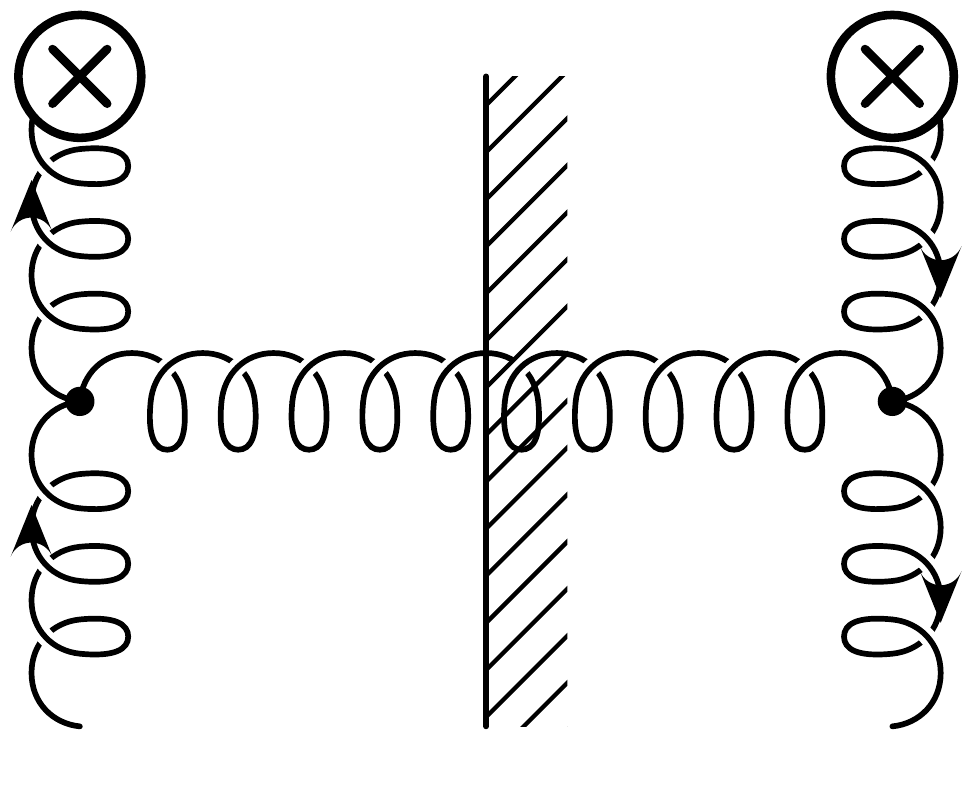}}  \end{minipage} & \frac{1}{z}+1-z^2  & 0 & \frac{(1-z)^3}{z}  & 0  \\
\begin{minipage}{2cm} \mybox{\includegraphics[width=2cm]{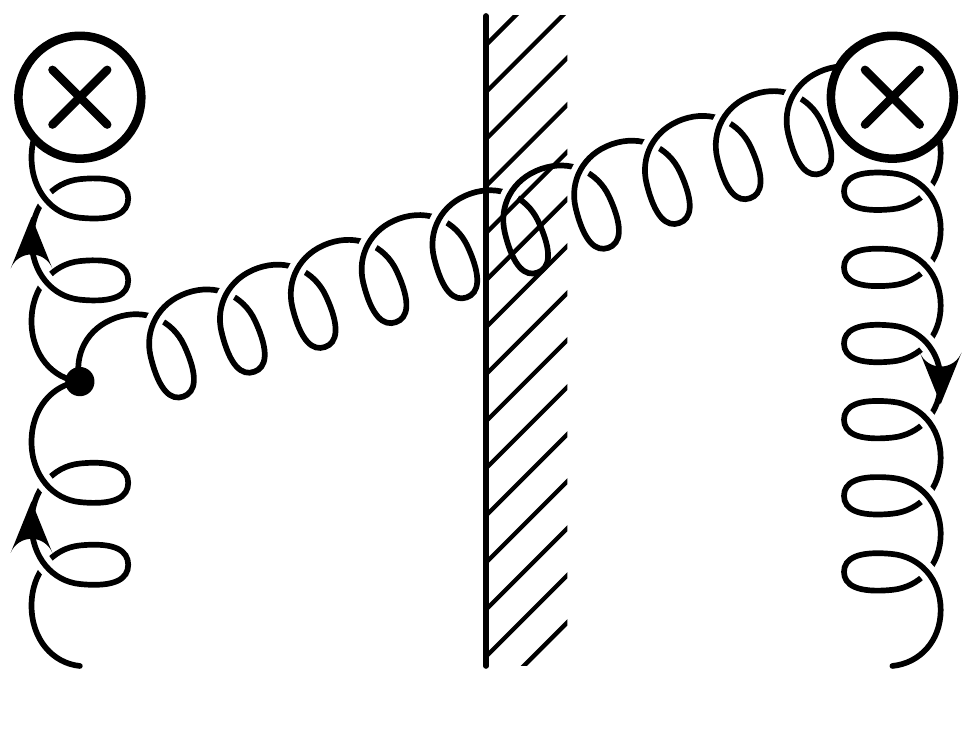}} \end{minipage} & -1-z & 0 & 0 & 0 \\
\hline
\text{Total}_1 & \frac{2}{(1-z)_+} +\frac{1}{z}-1-z-z^2 - 2 \ln \frac{\nu}{\bar n \cdot r }\,\delta(1-z) & -\ln \frac{\mu^2}{M^2} &   \frac{(1-z)^3}{z}  & 0  \\  \hline
\begin{minipage}{2cm} \mybox{\includegraphics[width=2cm]{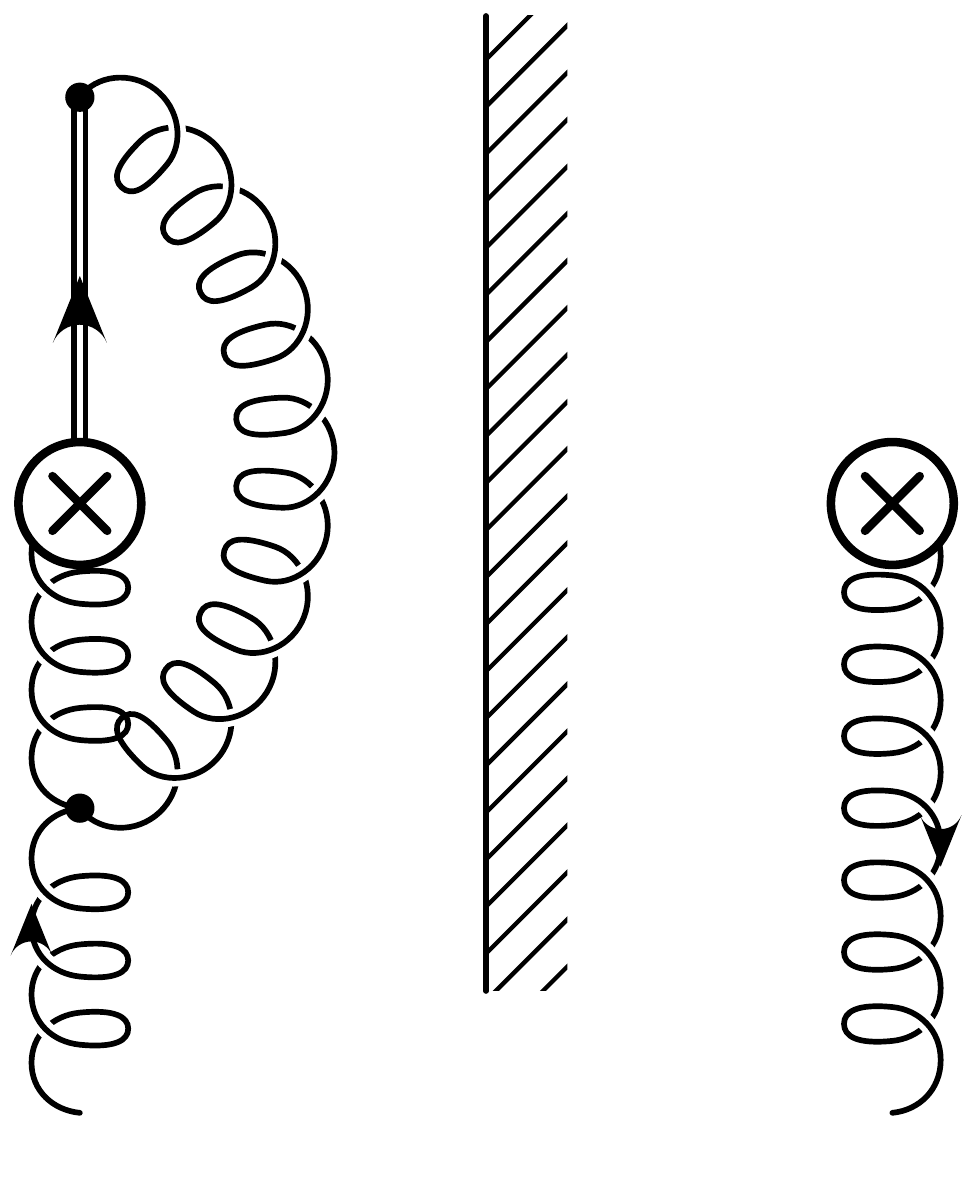}} \end{minipage} & c_A\big( 2 \ln  \frac{\nu}{\bar n \cdot r}+\frac52\big) \delta(1-z)  & c_A  \ln \frac{\mu^2}{M^2} & 0 & 0  \\
\begin{minipage}{2cm} \mybox{\includegraphics[width=2cm]{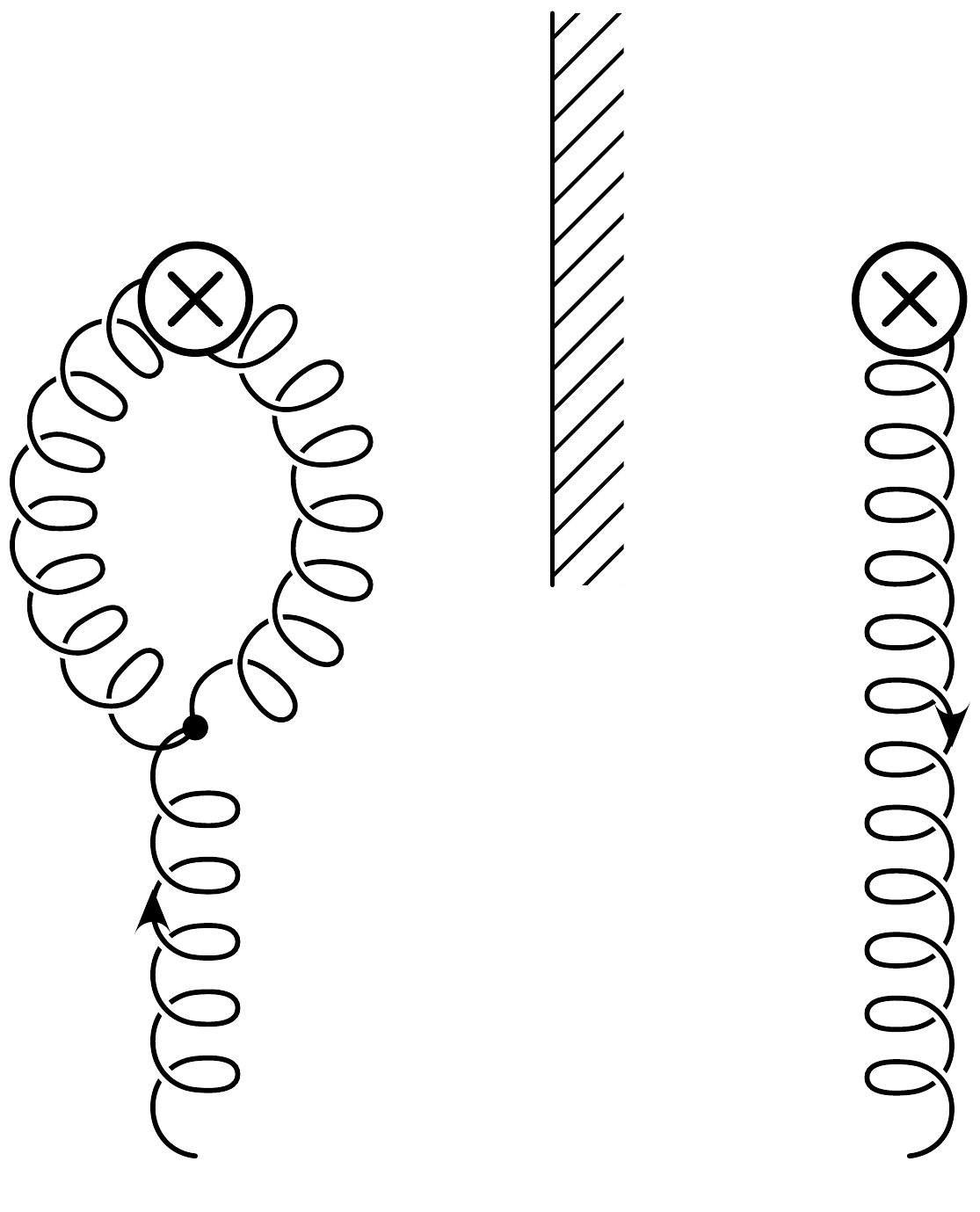}}
\end{minipage} & -\frac32 c_A \delta(1-z) & 0 & 0  & 0  \\
\begin{minipage}{2cm} \mybox{\includegraphics[width=2cm]{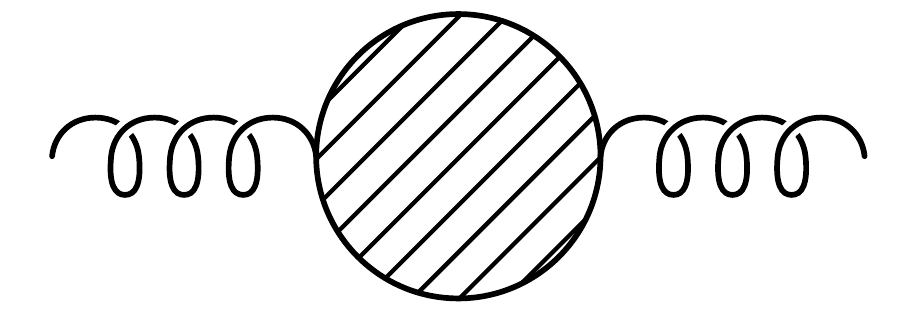}} \end{minipage} & \big(\frac{b_0}{2} - c_A\big)\delta(1-z) & 0  & 0 & 0   \\
\hline
\text{Total}_2 &
\big(\frac{b_0}{2}+2 c_A \ln \frac{\nu}{\bar n \cdot r} \big)\delta(1-z) & c_A \ln \frac{\mu^2}{M^2} &
0 & 0  \\
\hline\hline
\end{array}
\end{eqnarray*}
\caption{One-loop diagrams for the renormalization of collinear gauge boson operators. The columns show the graph and contribution to the $\mu$ and $\nu$ anomalous dimension for $P_{G_+G_+}$, $P_{G_+G_-}$. At one-loop,  $P_{G_+G_+}=P_{G_-G_-}$ and $P_{G_+G_-}=P_{G_-G_+}$. Combinatoric factors have been included. Subsets of the graphs have been summed to give $\text{Total}_1$ and $\text{Total}_2$. $\text{Total}_2$ has group theory factor $1$ in all cases, since its group theory factors are already included in the table. $\text{Total}_1$ has group theory factors given in \eq{3.43}.} 
\label{tab:coll_gauge}
\end{table*}
\endgroup

The one-loop diagrams contributing to the evolution of gauge-boson collinear operators are listed in table \ref{tab:coll_gauge} for $P_{G_+G_+}$ and $P_{G_+G_-}$. At one-loop order $P_{G_+G_+}=P_{G_-G_-}$ and $P_{G_+G_-}=P_{G_-G_+}$, but these relations do not hold at higher order.
As for  fermions, the individual graphs depend on the gauge, but $\text{Total}_1$ and $\text{Total}_2$ do not.
The anomalous dimensions are
%%%
\begin{align} \label{eq:ga_C_W}
 \hat \gamma_{\mu,G_+G_+}^{(R)}   &=c_{GG}(R)  \left[\frac{2}{(1-z)_+} + \frac{1}{z}-1-z-z^2\right]
 + \left[\frac{b_0}{2}+2 \left(c_A-c_{GG}(R) \right) \ln \frac{\nu}{\bar n \cdot r} \right]\delta(1-z)  \nn \\
   &=c_{GG}(R)  \widetilde P_{G_+G_+}(z)
 + \left[\frac{b_0}{2}+2 \left(c_A-c_{GG}(R) \right) \ln  \frac{\nu}{\bar n \cdot r} \right]\delta(1-z) \,, \nn \\
 \hat \gamma_{\mu,G_+G_-}^{(R)}   &= c_{GG}(R) \frac{(1-z)^3}{z} =c_{GG}(R)  \widetilde P_{G_+G_-}(z)\,, \nn \\
\hat \gamma_{\nu,G_+G_+}^{(R)}  &= \left[ c_A-c_{GG}(R) \right] \ln \frac{\mu^2}{M^2}\delta(1-z) \,, \nn \\
\hat \gamma_{\nu,G_+G_-}^{(R)}  &= 0
\,,\end{align}
%%%
%%%
where
\begin{align}
 \widetilde P_{G_+G_+}(z)    &=\frac{2}{(1-z)_+} + \frac{1}{z}-1-z-z^2 \,,\nn \\
 \widetilde P_{G_+G_-}(z)    &= \frac{(1-z)^3}{z} \,,
\end{align}
%%%
are the helicity components of the usual Altarelli-Parisi kernels excluding the $\delta(1-z)$ term,
and $b_0$ is the first term in the $\beta$-function,
%%%
\begin{align}
\mu \frac{\df g}{\df \mu} &=- \frac{g^3}{16\pi^2}\, b_0 + \ord{g^5}\,.
\end{align}
%%%
The group theory factors are
%%%
\begin{align}\label{eq:3.43}
c_{GG}(1) &= c_A\,, & c_{GG}(\text{adj}_S) &= \frac12 c_A\,, & c_{GG}(\text{adj}_A) &= \frac12 c_A\,, \nn \\
 c_{GG}(\bar a s) &= 0\,,
& c_{GG}(\bar sa ) &= 0\,, & c_{GG}(\bar a a) &= 1\,, & c_{GG}(\bar s s) &= -1\,.
\end{align}
%%%
For the singlet PDF, $c_{GG}(1)=c_A$, \eq{ga_C_W} reduces to the standard result~\cite{Altarelli:1977zs}, and $\gamma_\nu$ vanishes.
For the present analysis, we  need the results for gauge group $SU(2)$, so the only representations in \eq{3.43} which occur are $1$,
$\text{adj}_A$ and $\bar ss$, which are the $SU(2)$ singlet, triplet, and  quintet representation with weak isospin $I=0,1,2$.

For the $W\!B$ and $BW$ PDFs, there are only virtual corrections which are diagonal in helicity,
%%%
\begin{align}
\gamma_{\mu,W\!B}  &= \gamma_{\mu,BW}  = \frac{\alpha_2}{\pi} \left[\frac{b_{0,2}}{4}+2 \ln \frac{\nu}{\bar n \cdot r} \right]\delta(1-z) +
\frac{\alpha_1}{\pi} \left[\frac{b_{0,1}}{4} \right]\delta(1-z)\, , \nn \\
\gamma_{\nu,W\!B}  &= \gamma_{\nu,BW} = \frac{\alpha_2}{\pi} \ln \frac{\mu^2}{M^2} \,, 
\end{align}
%%%
where $b_{0,2}$ is $b_0$ for the $SU(2)$ gauge group, and similarly for $b_{0,3}$ and $b_{0,1}$.
The $W\!B$ PDFs do not mix with the triplet $W$ PDF at one-loop, since $W$ and $B$ bosons do not interact at tree level.

%===============================================================================
\subsection{$\gamma_{QG}$ and $\gamma_{HG}$}
%===============================================================================

\begingroup
\begin{table*}
\renewcommand{\arraystretch}{1.5}
\setlength{\arraycolsep}{3pt}
\centering
\begin{align*}
\begin{array}{c|c|c|c|c}
\hline\hline
 \text{Graph} & \multicolumn{2}{c|}{P_{Q_+ G_+}} & \multicolumn{2}{c}{P_{Q_+ G_-}} \\
 \hline
 & \hat \gamma_\mu & \hat \gamma_\nu & \hat \gamma_\mu & \hat \gamma_\nu \\
\hline
\begin{minipage}{2cm} \mybox{\includegraphics[width=2cm]{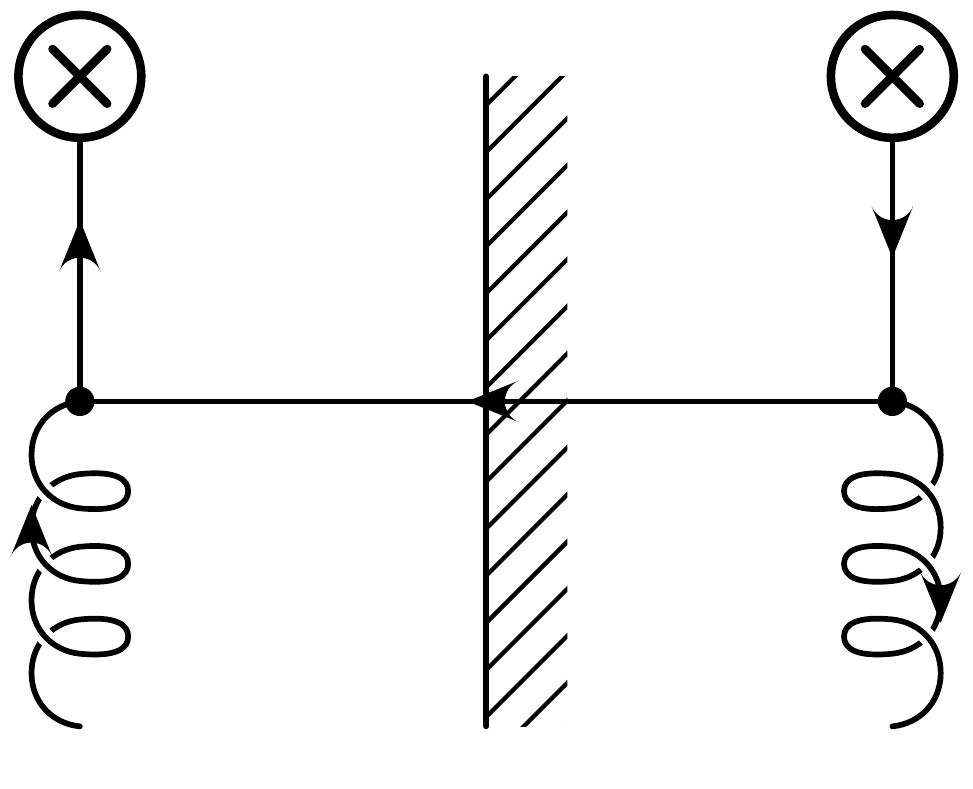}} \end{minipage} & z^2  & 0 & (1-z)^2  & 0 
\\ \hline\hline
\end{array}
\end{align*}
\caption{\label{tab:QG} One-loop diagrams contributing to $QG$ mixing. The columns show the graph and contribution to the $\mu$ and $\nu$ anomalous dimension. The group theory factors are given in \eqs{3.48}{barQG}.}
\end{table*}
\endgroup

Fermion and gauge boson PDFs can mix, and the $QG$ element of the mixing matrix is due to the graph shown in table~\ref{tab:QG}. The graph has no rapidity divergence so (for all four choices of signs)
%%%
\begin{align}\label{eq:3.47}
\hat \gamma_{\mu,Q_{\pm}G_{\pm}}^{(R)}   &= c_{QG}(R)\, \widetilde P_{Q_{\pm} G_{\pm}}(z) , &
\widetilde P_{Q_+G_+}(z)  &=  z^2 , &
\widetilde P_{Q_+G_-}(z)  &=  (1-z)^2 , 
\end{align}
%%%
where $c_{QG}(R)$ is the group theory factor, and  $\widetilde P_{Q_{\pm} G_{\pm}}(z)$  are the usual Altarelli-Parisi kernels. At one-loop, $\widetilde P_{Q_+G_+}(z)=\widetilde P_{Q_-G_-}(z)$ and $\widetilde P_{Q_+G_-}(z)=\widetilde P_{Q_-G_+}(z)$, but the anomalous dimensions do not satisfy these equalities because the group theory factors for $Q_+$ and $Q_-$ differ.
Fermion and gluon PDFs which mix must have the same gauge representation, so the only mixing is in the singlet and adjoint sectors, for which
%%%
\begin{align}\label{eq:3.48}
c_{QG}(1) &= t_F
\,, \qquad 
c_{QG}(\text{adj}_S) = \frac12 t_F
\,, \qquad 
c_{QG}(\text{adj}_A) = \frac12 t_F
\,.\end{align}
%%%
where $t_F=1/2$ is the index of the fundamental representation.
For the antifermion PDF, 
%%%
\begin{align}\label{eq:barQG}
c_{\bar QG}(1) &= t_F
\,, \qquad 
c_{\bar QG}(\text{adj}_S) = -\frac12 t_F
\,, \qquad 
c_{\bar QG}(\text{adj}_A) = \frac12 t_F
\,.\end{align}
%%%
The triplet quark PDF can mix with the $W\!B$ and $BW$ PDFs, with anomalous dimensions
\begin{align}\label{eq:3.30}
\gamma_{\mu, Q\pm\,W\!B\pm} &= \gamma_{\mu, Q\pm\,BW\pm}  = \frac{g_1 g_2}{4 \pi^2}\, \hyp_Q\,\widetilde P_{Q\pm G\pm}(z),  \nn \\
\gamma_{\mu, \bar Q\pm\,W\!B\pm} &= \gamma_{\mu, \bar Q\pm\,BW\pm}  =- \frac{g_1 g_2}{4 \pi^2}\, \hyp_Q\, \widetilde P_{Q\pm G\pm}(z)  .
\end{align}
%%%

\begingroup
\begin{table*}
\renewcommand{\arraystretch}{1.5}
\setlength{\arraycolsep}{3pt}
\centering
\begin{minipage}{0.4\textwidth}
\begin{align*}
\begin{array}{c|c|c}
\hline\hline
 \text{Graph} & \multicolumn{2}{c}{P_{HG_+}={P_{HG_-}}}  \\
 \hline
 & \hat \gamma_\mu & \hat \gamma_\nu  \\
\hline
\begin{minipage}{2cm} \mybox{\includegraphics[width=2cm]{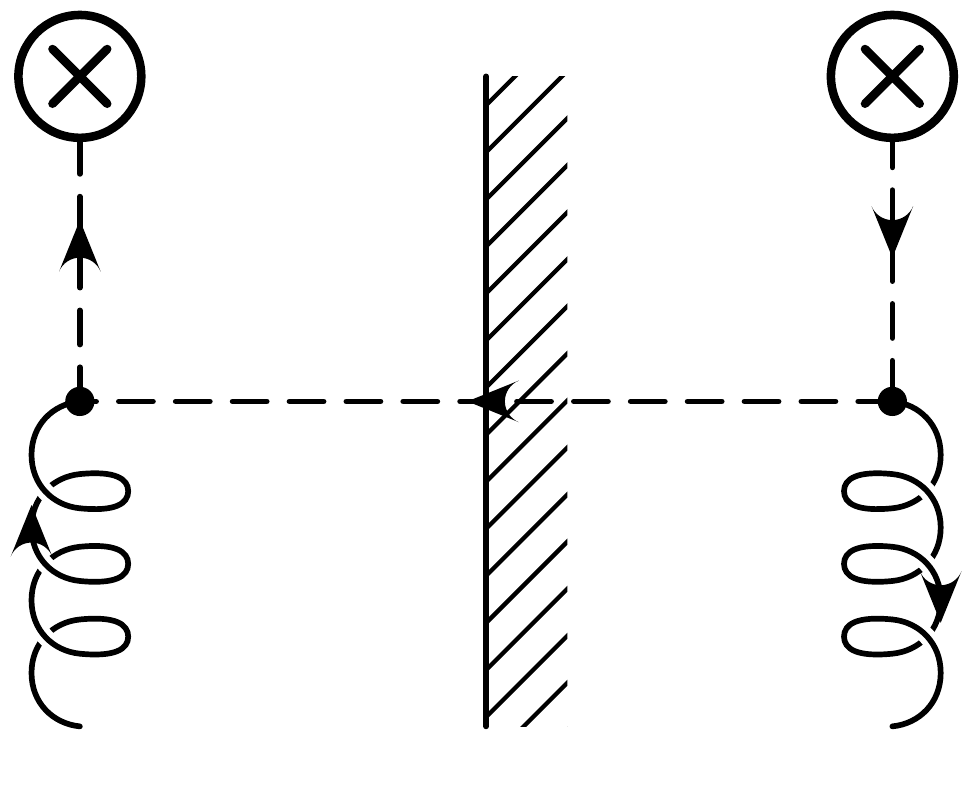}} \end{minipage} & z(1-z)  & 0 
\\ \hline\hline
\end{array}
\end{align*}
\end{minipage}
\begin{minipage}{0.4\textwidth}
\begin{align*}
\begin{array}{c|c|c}
\hline\hline
 \text{Graph} & \multicolumn{2}{c}{P_{G_+H}={P_{G_-H}}}  \\
\hline
& \hat \gamma_\mu & \hat \gamma_\nu  \\
\hline
\begin{minipage}{2cm} \mybox{\includegraphics[width=2cm]{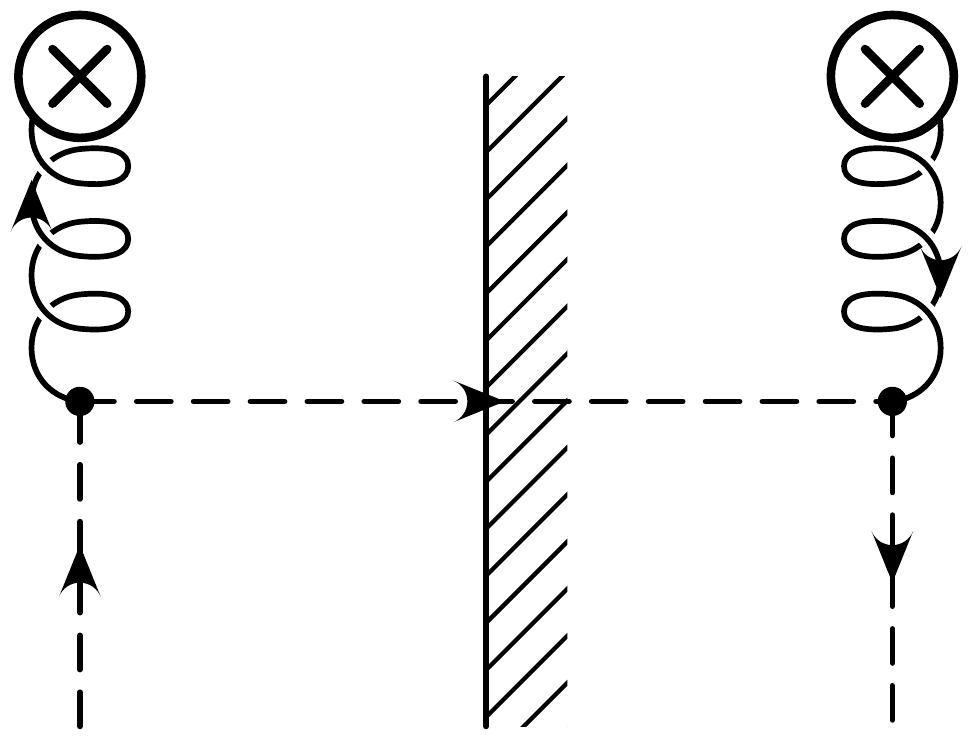}} \end{minipage} &  \frac{(1-z)}{z} & 0 \\
\hline\hline
\end{array}
\end{align*}
\end{minipage}
\caption{\label{tab:7} The one-loop diagrams contributing to $HG$ and $GH$ mixing in the anomalous dimension of collinear operators. The columns show the graph and contribution to the $\mu$ and $\nu$ anomalous dimension. The group theory factors are given in eqs.~\eqref{eq:3.48}, \eqref{eq:barQG} and \eqref{eq:3.52}.}
\end{table*}
\endgroup

The above analysis also applies to scalars, using the results in table~\ref{tab:7},
%%%
\begin{align}
 \hat \gamma_{\mu,HG_\pm }^{(R)}   &= c_{HG}(R)\, \widetilde P_{HG _\pm}(z) , &
\widetilde P_{HG_\pm}(z)  &=  z(1-z) , 
\end{align}
%%%
where the group theory factors $c_{HG}(R)$ are the same as for fermions in eqs.~\eqref{eq:3.48}--\eqref{eq:3.30}.

%===============================================================================
\subsection{$\gamma_{GQ}$ and $\gamma_{GH}$}
%===============================================================================

\begingroup
\begin{table*}
\renewcommand{\arraystretch}{1.5}
\setlength{\arraycolsep}{3pt}
\centering
\begin{align*}
\begin{array}{c|c|c|c|c}
\hline\hline
 \text{Graph} & \multicolumn{2}{c|}{P_{G_+ Q_+}} & \multicolumn{2}{c}{P_{G_+ Q_-}} \\
 \hline
 & \hat \gamma_\mu & \hat \gamma_\nu & \hat \gamma_\mu & \hat \gamma_\nu \\
\hline
\begin{minipage}{2cm} \mybox{\includegraphics[width=2cm]{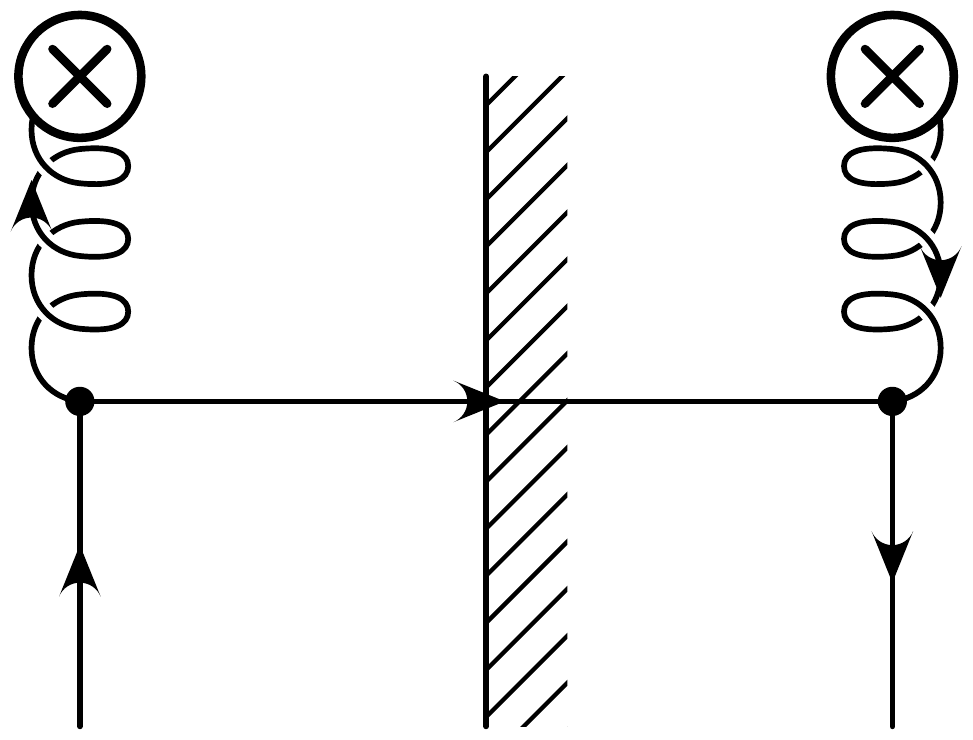}} \end{minipage} &  \frac{1}{z} & 0 &  \frac{(1-z)^2}{z} & 0 \\
\hline\hline
\end{array}
\end{align*}
\caption{\label{tab:GQ} One-loop diagrams contributing to $GQ$ mixing. The columns show the graph and contribution to the $\mu$ and $\nu$ anomalous dimension. The group theory factors are given in \eq{3.52}.}
\end{table*}
\endgroup

The $GQ$ element of the mixing matrix is due to the graph shown in table~\ref{tab:GQ}, and gives rise to the anomalous dimension 
%%%
\begin{align}\label{eq:3.51}
\hat \gamma_{\mu,G_\pm Q_\pm }^{(R)}   &= c_{GQ}(R)\, \widetilde P_{G_\pm Q_\pm}(z) \,, &
\widetilde P_{G_+Q_+}(z)    &=  \frac{1}{z}\,,  &
\widetilde P_{G_+ Q_-}(z)    &=  \frac{(1-z)^2}{z}\,, 
\end{align}
%%%
where $c_{GQ}(R)$ is the group theory factor,  and $\widetilde P_{G_{\pm} Q_{\pm}}(z)$  are the usual Altarelli-Parisi kernels.  At one-loop, $\widetilde P_{G_+Q_+}(z)=\widetilde P_{G_-Q_-}(z)$ and $\widetilde P_{G_+Q_-}(z)=\widetilde P_{G_-Q_+}(z)$, but the anomalous dimensions do not satisfy these equalities  because the group theory factors for $Q_+$ and $Q_-$ are
not equal. As a result,  gauge boson PDFs (including the gluon) develop a polarization asymmetry $f_{g_+}-f_{g_-}$ through parity-violating $\mu$ evolution.

 The only mixing is in the singlet and adjoint sectors, for which
%%%
\begin{align}\label{eq:3.52}
c_{GQ}(1) &= c_F , & c_{GQ}(\text{adj}_S) &= \frac{N^2-4}{2N} , & c_{GQ}(\text{adj}_A) &= \frac12 c_A ,\nn \\
c_{G \bar Q}(1) &=c_F, & c_{G \bar Q}(\text{adj}_S) &= -\frac{N^2-4}{2N}, & c_{G \bar Q}(\text{adj}_A) &= \frac12 c_A.
\end{align}
%%%
The $W\!B$ and $BW$ PDFs can mix with the triplet quark PDF, with anomalous dimensions
%%%
\begin{align}\label{eq:3.54}
\gamma_{\mu, W\!B\pm\, Q\pm}   &=  \gamma_{\mu, BW\pm\, Q\pm} = \frac{g_1 g_2}{4 \pi^2}\, \hyp_Q\, \widetilde P_{G\pm Q\pm}(z) , \nn \\
\gamma_{\mu, W\!B\pm\, \bar Q\pm}   &=  \gamma_{\mu, BW\pm\, \bar Q\pm} = -\frac{g_1 g_2}{4 \pi^2}\, \hyp_Q \,\widetilde P_{G\pm Q\pm}(z)  .
\end{align}
%%%

Similar results hold for the $GH$ entries using table~\ref{tab:7},
%%%
\begin{align}\label{eq:3.51a}
 \hat \gamma_{\mu,G_\pm H}^{(R)}   &= c_{GH}(R)\, \widetilde P_{GH}(z) \,, &
\widetilde P_{G_\pm H}(z)    &= \frac{(1-z)}{z}\,, 
\end{align}
%%%
where the group theory factors $c_{GH}(R)$ are the same as for fermions in \eqs{3.52}{3.54}.

%===============================================================================
\subsection{$\gamma_{HQ}$ and $\gamma_{QH}$}
%===============================================================================

\begingroup
\begin{table*}
\renewcommand{\arraystretch}{1.5}
\setlength{\arraycolsep}{3pt}
\centering
\begin{minipage}{0.4\textwidth}
\begin{align*}
\begin{array}{c|c|c}
\hline\hline
 \text{Graph} & \multicolumn{2}{c}{P_{HQ_+}={P_{HQ_-}}}  \\
 \hline
& \hat \gamma_\mu & \hat \gamma_\nu  \\
\hline
\begin{minipage}{2cm} \mybox{\includegraphics[width=2cm]{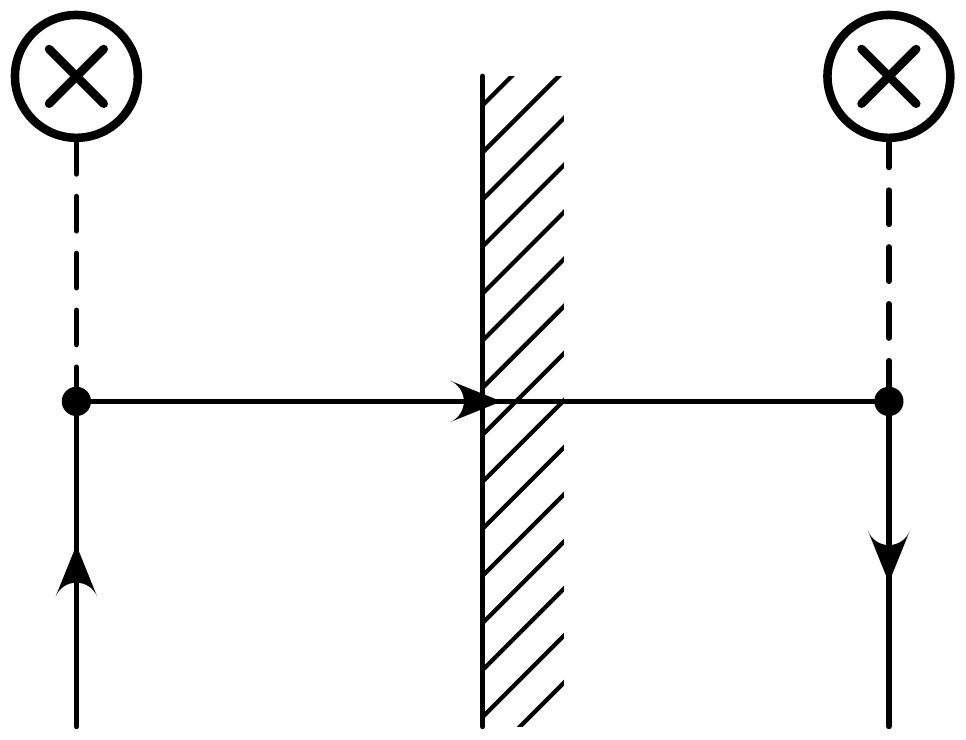}} \end{minipage} &\quad \frac{z}{2} \quad  & 0 
\\ \hline\hline
\end{array}
\end{align*}
\end{minipage}
\begin{minipage}{0.4\textwidth}
\begin{align*}
\begin{array}{c|c|c}
\hline\hline
 \text{Graph} & \multicolumn{2}{c}{P_{Q_+H}={P_{Q_-H}}}  \\
 \hline
& \hat \gamma_\mu & \hat \gamma_\nu  \\
\hline
\begin{minipage}{2cm} \mybox{\includegraphics[width=2cm]{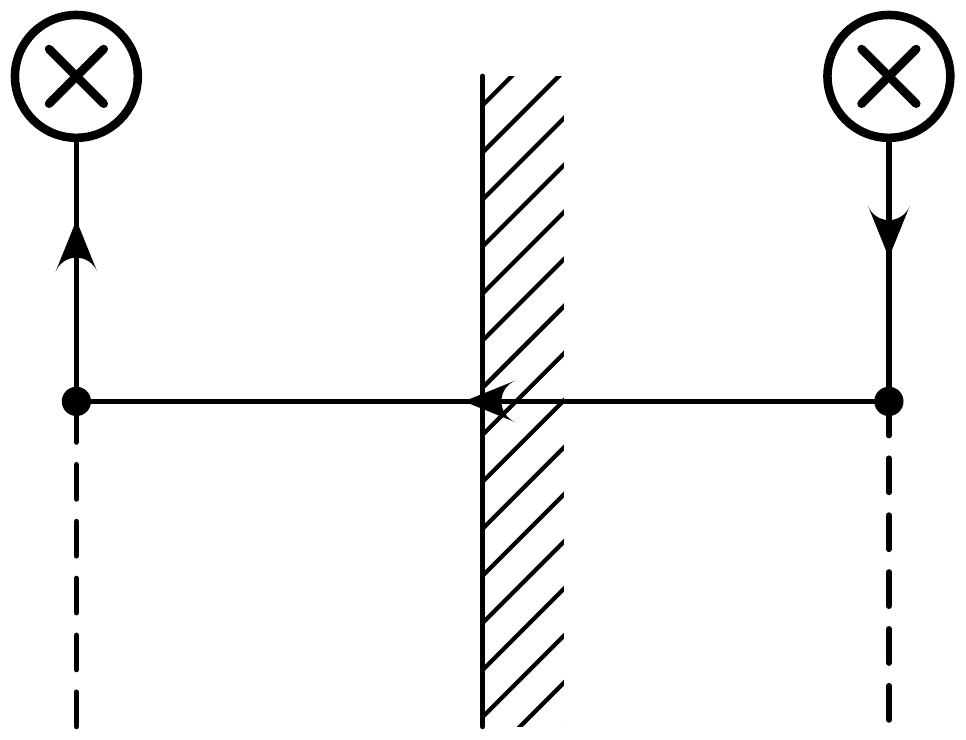}} \end{minipage} &\quad  \frac12 \quad & 0 \\
\hline\hline
\end{array}
\end{align*}
\end{minipage}
\caption{\label{tab:8} The one-loop diagrams contributing to $HQ$ and $QH$ mixing in the anomalous dimension of collinear operators. The columns show the graph and contribution to the $\mu$ and $\nu$ anomalous dimension. The Yukawa factors are give in \eqs{HQ}{QH}.}
\end{table*}
\endgroup

The mixing of fermion and scalar operators via Yukawa couplings is shown in table~\ref{tab:8}. The $HQ$ anomalous dimensions are 
%%%
\begin{align} \label{eq:HQ}
\mu \frac{\df}{\df \mu}\ O^{(I=0)}_{H} &= \frac{1}{4\pi^2} \frac{z}{2} \biggl[  [Y_d^\dagger Y_d]_{rs} \otimes O^{(I=0)}_{q,r,s}
+ [Y_u^\dagger Y_u]_{rs} \otimes O^{(I=0)}_{\bar q,s,r} + 2 [Y_d Y_d^\dagger]_{rs} \otimes O^{(I=0)}_{\bar d,s,r}  \nn \\
& +  2 [Y_u Y_u^\dagger]_{rs} \otimes O^{(I=0)}_{u,r,s} + [Y_e^\dagger Y_e]_{rs} \otimes O^{(I=0)}_{\ell,r,s}
 + 2 [Y_e Y_e^\dagger]_{rs} \otimes O^{(I=0)}_{\bar e,s,r}  \biggr] \,, \nn \\
\mu \frac{\df}{\df \mu}\ O^{(I=1)}_{H} &= \frac{1}{4\pi^2} \frac{z}{2} \biggl[  [Y_d^\dagger Y_d]_{rs} \otimes O^{(I=1)}_{q,r,s}
+ [Y_u^\dagger Y_u]_{rs} \otimes O^{(I=1)}_{\bar q,s,r} + [Y_e^\dagger Y_e]_{rs} \otimes O^{(I=1)}_{\ell,r,s}\biggr]\,,
\end{align}
%%%
and the $\bar H$ anomalous dimensions are given by charge conjugation, i.e.\ replacing $d,r,s \leftrightarrow \bar d,s,r$, etc.~on the right-hand side.

The $QH$ anomalous dimensions are
%%%
\begin{align} \label{eq:QH}
\mu \frac{\df}{\df \mu}\ O^{(I=0)}_{d,r,s} &=  \frac{N_c}{8\pi^2}  [Y_d Y_d^\dagger]_{sr} \otimes O^{(I=0)}_{\bar H} \,, \nn \\
\mu \frac{\df}{\df \mu}\ O^{(I=0)}_{u,r,s} &= \frac{N_c}{8\pi^2}  [Y_u Y_u^\dagger]_{sr} \otimes O^{(I=0)}_{H} \,, \nn \\
\mu \frac{\df}{\df \mu}\ O^{(I=0)}_{q,r,s} &= \frac{N_c}{8\pi^2}   \Bigl( [ Y_d^\dagger  Y_d]_{sr} \otimes O^{(I=0)}_{ H}
+[ Y_u^\dagger Y_u]_{sr} \otimes O^{(I=0)}_{\bar H} 
\Bigr)\,, \nn \\
\mu \frac{\df}{\df \mu}\ O^{(I=1)}_{q,r,s} &= \frac{N_c}{8\pi^2} \Bigl( [ Y_d^\dagger  Y_d]_{sr} \otimes O^{(I=1)}_{ H}
+[ Y_u^\dagger Y_u]_{sr} \otimes O^{(I=1)}_{\bar H} 
\Bigr)\,, \nn \\
\mu \frac{\df}{\df \mu}\ O^{(I=0)}_{e,r,s} &= \frac{1}{8\pi^2}  [Y_e Y_e^\dagger]_{sr} \otimes O^{(I=0)}_{\bar H} \,, \nn \\
\mu \frac{\df}{\df \mu}\ O^{(I=0)}_{\ell,r,s} &= \frac{1}{8 \pi^2}  [ Y_e^\dagger  Y_e]_{sr} \otimes O^{(I=0)}_{ H}\,, \nn \\
\mu \frac{\df}{\df \mu}\ O^{(I=1)}_{\ell,r,s} &= \frac{1}{8\pi^2}  [ Y_e^\dagger  Y_e]_{sr} \otimes O^{(I=1)}_{ H}\,,
\end{align}
%%%
and the $\bar QH$ anomalous dimensions are given by charge conjugation, i.e.\ by $d,r,s \to \bar d,s,r$ on the l.h.s.\ and $H \leftrightarrow \bar H$ on the r.h.s.\ These results greatly simplify when using \eq{approx}.

%===============================================================================
\subsection{$SU(2) \times U(1)$ mixing}
\label{sec:mixing}
%===============================================================================

$SU(2) \times U(1)$ mixing affects the $\nu$-anomalous dimension, since $\hat \ga_\nu$ contains an explicit gauge boson mass $M$, and $M_W \neq M_Z$ in the Standard Model. As we will now discuss, $M$ is equal to $M_W$ and not $M_Z$ (with one exception). In \sec{low_matching} we will see that the adjoint index in the $SU(2)$ triplet quark and gauge boson operator must be 3, which is a consequence of electric charge conservation. Direct inspection of the calculation in the previous subsections shows that for the exchange of a $W^3 = \cos \theta_W\, Z + \sin \theta_W\, A$ boson, the group theory factor for Total$_1$ and Total$_2$ are the same. Thus its contribution to $\ga_\nu$ drops out, leaving only contributions involving $M_W$. 
The one exception is $O_{\widetilde H H}^{(I=1)}$, where direct inspection of its corrections in the broken phase reveals that both $M_W$ and $M_Z$ enter, see \eq{tildeHH}. This is also obvious from the presence of $\al_1$ in $\ga_\nu$. 

%===============================================================================
\subsection{Fragmentation functions}
\label{sec:RGE_FF}
%===============================================================================

Next we consider the collinear operators for outgoing particles. We start with the case where a particle in the final state is identified, e.g.~the electron in DIS, and its momentum fraction is measured.
In this case, the matrix elements of collinear operators lead to fragmentation functions, which were defined in ref.~\cite{Collins:1981uw}. As is well known, the one-loop anomalous dimensions of fragmentation functions can be obtained from those for the PDF, and this holds for the gauge non-singlet case as well. The diagonal anomalous dimensions $QQ$ and $GG$ are the same for the PDF and fragmentation function, except that the helicity labels are interchanged because the role of the field and external state are swapped,
%%%
\begin{align}
\hat \gamma_{\mu, Q_\pm Q_\pm}^{(\text{frag})} (x,r^-,\mu,\nu) &= \hat \gamma_{\mu,Q_\pm Q_\pm}^{(\text{PDF})}  (x,r^-,\mu,\nu)
  \,, &
\hat \gamma_{\nu, Q_\pm}^{(\text{frag})} (\mu,\nu) &= \hat \gamma_{\nu,Q_\pm}^{(\text{PDF})}  (\mu,\nu)  \,,
\nn \\  
\hat \gamma_{\mu,G_+ G_\pm }^{(\text{frag})}  (x,r^-,\mu,\nu) &= \hat \gamma_{\mu,G_\pm G_+}^{(\text{PDF})} (x,r^-,\mu,\nu)
  \,, & 
\hat \gamma_{\nu, G_\pm}^{(\text{frag})} (\mu,\nu) &= \hat \gamma_{\nu,G_\pm}^{(\text{PDF})}  (\mu,\nu)  
\,,\end{align}
%%%
etc.
For the off-diagonal entries the flavor labels are also swapped,
%%%
\begin{align}
\hat \gamma_{\mu, Q_+G_\pm}^{(\text{frag})} (x,\mu) &= \hat \gamma_{\mu, G_\pm Q_+}^{(\text{PDF})} (x,\mu)
  \,, &
\hat \gamma_{\mu, G_+Q_\pm}^{(\text{frag})} (x,\mu) &= \hat \gamma_{\mu, Q_\pm G_+}^{(\text{PDF})} (x,\mu)
\,, \nn \\
\hat \gamma_{\mu, HG_\pm}^{(\text{frag})} (x,\mu) &= \hat \gamma_{\mu, G_\pm H}^{(\text{PDF})} (x,\mu)
  \,, &
\hat \gamma_{\mu, G_\pm H}^{(\text{frag})} (x,\mu) &= \hat \gamma_{\mu, HG_\pm}^{(\text{PDF})} (x,\mu)
\,,\end{align}
%%%
etc.

We also consider the case where no final state particle is detected, which we obtain by summing over all possible final states. If $D_{Q_\pm \to P}(x,\mu,\nu)$ is the fragmentation function for $Q_\pm$ to produce a particle $P$ with momentum fraction $x$, then not observing the final state gives the completeness relation\footnote{The factor of $x$ accounts for identical particles, as discussed in sec.~2.5 of ref.~\cite{Jain:2011xz}.} 
%%%
\begin{align}
  \sum_P \int_0^1 \df x\, x\, D_{Q_\pm \to P}(x,\mu,\nu)\,,
\end{align}
%%%
where the sum on $P$ runs over all final states, and the integral is over its momentum fraction.  The momentum sum rule for fragmentation functions then implies that
%%%
\begin{align}
\sum_P \int_0^1 \df x\, x\, D_{Q_\pm \to P}(x,\mu,\nu)&=1, & \sum_P \int_0^1 \df x\, x\, D_{G_\pm \to P}(x,\mu,\nu)&=1,
\end{align}
%%%
for the gauge singlet fragmentation functions, and
%%%
\begin{align}
\sum_P \int_0^1 \df x\, x\, D^{(R,\alpha)}_{Q_\pm \to P}(x,\mu,\nu)&=0, & \sum_P \int_0^1 \df x\, x\, D^{(R,\alpha)}_{G_\pm \to P}(x,\mu,\nu)&=0,
\end{align}
%%%
for the gauge non-singlet case.

%%%%%%%%%%%%%%%%%%%%%%%%%%%%%%%%%%%%%%%%%%%%%%%%%%%%%%%%%%%%%%%%%%%%%%%%%%%%%%%%
\section{Soft evolution}
\label{sec:RGE_S}
%%%%%%%%%%%%%%%%%%%%%%%%%%%%%%%%%%%%%%%%%%%%%%%%%%%%%%%%%%%%%%%%%%%%%%%%%%%%%%%%

We now move on to the soft operators, for which the RG equations are given by
%%%
\begin{align}
\frac{\df}{\df \ln \mu}\, \cS(\mu,\nu) &= \frac{\al(\mu)}{\pi}\, \hat \gamma_{\mu,\cS}(\mu,\nu)\, \cS(\mu,\nu)
\,,\nn \\
\frac{\df}{\df \ln \nu}\, \cS(\mu,\nu) &= \frac{\al(\mu)}{\pi}\,\hat \ga_{\nu,\cS}(\mu,\nu)\, \cS(\mu,\nu)
\,.\end{align}
%%%
Since non-Abelian gauge bosons carry gauge charge, soft operators can mix, turning $\hat \ga_\mu$ into a matrix. This first happens for soft functions with at least four gauge indices.

There is a single type of graph that contributes  at one-loop, shown in Table.~\ref{tab:soft_diagram}. Graphs where the gluon couples to a single line vanish since $n_i^2=0$. For the graph $\cS_i \cS_j$ in Table~\ref{tab:soft_diagram},  $\cS_i$ and $\cS_j$ are Wilson lines along the null vectors $n_i$ and $n_j$. We compute the graph in an Abelian theory, and put in the group theory factors later. The graph is
%%%
\begin{align} \label{eq:I_S}
 I_{\cS} &= -\img g^2 w^2 \Big(\frac{e^{\ga_E} \mu^2}{4\pi}\Big)^\eps \nu^\eta \int\! \frac{\df^d k}{(2\pi)^d} 
 \frac{n_i \sdt n_j\,|2k^0|^{-\eta}}{(-n_i \sdt k+\img 0)(k^2-M^2+\img 0)(n_j \sdt k +\img 0)} \nn \\
 &= -\img g^2 w^2 \Big(\frac{e^{\ga_E} \mu^2}{4\pi}\Big)^\eps \nu^\eta \int\! \frac{\df k^+\,\df k^-\,\df^{d-2} \mathbf{k}_\perp}{(2\pi)^d} 
 \frac{2\big|(k^+ - k^-)\sqrt{2/ |n_i \sdt n_j|}\big|^{-\eta}}{(-k^++ n_i^0\img 0)[k^- k^+  - \mathbf{k}_\perp^{\,2}-M^2+\img 0](-k^- + n_j^0 \img 0)}
  \nn \\
  &= \frac{\al w^2}{\pi}\, \bigg[  \frac{1}{\eta} \Big(\frac{1}{\eps} + \ln \frac{\mu^2}{M^2}\Big) - \frac{1}{2\eps^2} 
  + \frac{1}{2\eps} \ln \frac{(-n_i \sdt n_j-\img 0) \nu^2}{2\mu^2} + \ord{\eta^0,\eps^0}\bigg]
\,,\end{align}
%%%
when both $S_i$ and $S_j$ are time-ordered.
In contrast to refs.~\cite{Chiu:2011qc,Chiu:2012ir}, we use the gauge boson energy $k^0$ to regulate rapidity divergences, whis is more suitable for multiple collinear directions~\cite{Bertolini:2017efs} and does not affect the collinear calculation.
In the second line we employed the following light-cone coordinates
%%%
\begin{align}
  k^\mu = k^+ \frac{n_j^0 n_j^\mu}{\sqrt{2|n_i \sdt n_j|}} - k^- \frac{n_i^0 n_i^\mu}{\sqrt{2|n_i \sdt n_j|}} + k_\perp^\mu
  \,, \qquad
  \df^d k = \frac12\,\df k^+\, \df k^-\, \df^{d-2} \mathbf{k}_\perp
\,,\end{align}
%%%
where $n_{i,j}^0 = \pm 1$ is used to keep track of incoming vs.~outgoing directions. Strictly speaking, $k_\perp^\mu$ also enters in the rapidity regulator when $n_i$ and $n_j$ are not back-to-back, but this only contributes at $\ord{\eta}$. Due to this choice of integration variables, eq.~(\ref{eq:I_S}) is the same as the expression when $n_i \cdot n_j = 2$ (given in eq.~(95) of ref.~\cite{Manohar:2012jr}), apart from an additional factor of $|n_i \cdot n_j/2|^{\eta/2}$.\footnote{Depending on the sign of $\img 0$ in the eikonal propagators there are $\img \pi$ contributions~\cite{Rothstein:2016bsq}, which we account for in the branch-cut prescription of the logarithm. Contributions in the conjugate amplitude have the opposite prescription, i.e.~$\ln (-n_i \sdt n_j+\img 0)$, and for exchange between a Wilson line in the amplitude and conjugate amplitude we find $\ln |n_i \sdt n_j|$. Fortunately, these $\img \pi$ contributions do not enter in the final expression.} The $\mu$ and $\nu$ anomalous dimensions can be read off from the results in ref.~\cite{Manohar:2012jr}, and are shown in table \ref{tab:soft_diagram}.

\begin{table}
\centering
\begin{eqnarray*}
\begin{array}{c|c|c}
\hline\hline
\text{Graph} & \hat \gamma_\mu &\hat \gamma_\nu \\
 \hline
 \begin{minipage}{2cm} \mybox{\includegraphics[height=2cm]{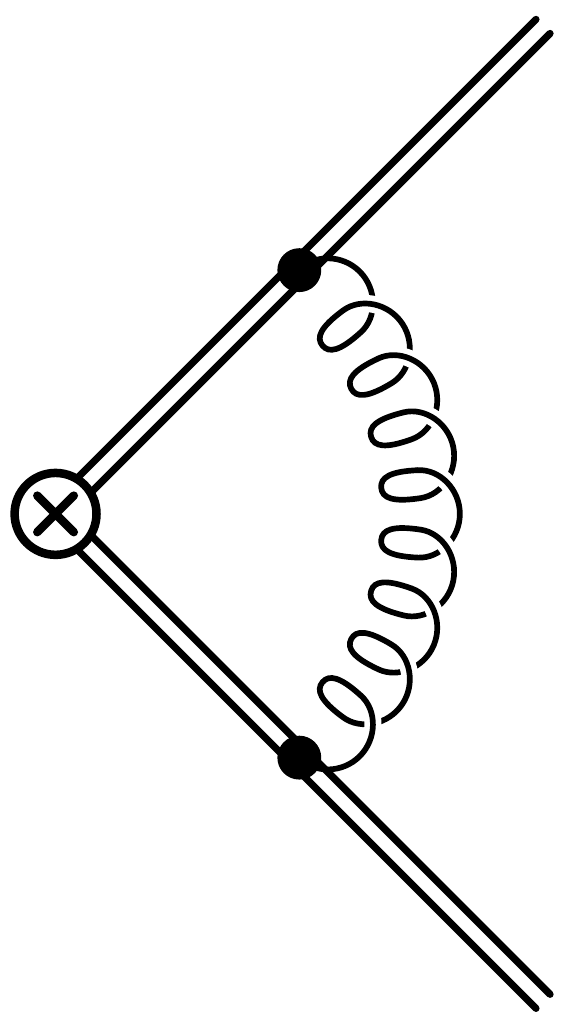}} \end{minipage}
& \ln \frac{(-n_i \cdot n_j-\img 0) \nu^2}{2\mu^2} & \ln \frac{\mu^2}{M^2} \\
\hline\hline
\end{array}
\end{eqnarray*}
\caption{The one-loop diagram for the soft operator $\cS_i \cS_j$. The double lines denote the Wilson lines $\cS_i$ and $\cS_j$. The columns show the graph and its contribution to the $\mu$ and $\nu$ anomalous dimension, apart from the group theory factors given in the text. 
\label{tab:soft_diagram}}
\end{table}

We now compute the anomalous dimensions for the soft operators in \eq{soft_ops} including the group theory factors. Write
%%%
\begin{align}\label{eq:3.62}
 \ln \frac{(-n_i \cdot n_j-\img 0) \nu^2}{2\mu^2} &=  \ln \frac{(-n_i \cdot n_j -\img 0)} {2} +  \ln \frac{ \nu^2}{\mu^2}\,.
\end{align}
%%%
First consider the $(n_i \cdot n_j)$-independent pieces in the soft anomalous dimension. For the soft operator
%%%
\begin{align}
\cS_{12\ldots n}^{ab}=\tr\big[(\cS_1 t^{a_1} \cS_1^\dagger)( \cS_2 t^{a_2} \cS_2^\dagger) \ldots (\cS_n t^{a_n} \cS_n^\dagger)\big],
\end{align}
%%%
one-loop graphs where gauge fields are exchanged between the same Wilson line or between a Wilson line $\cS_i$ and its conjugate $\cS_i^\dagger$ vanish, since $n_i^2=0$. 
Gauge boson exchange between $\cS_1$ or $\cS_1^\dagger$ and the other Wilson lines gives a group theory factor
%%%
\begin{align} \label{eq:casimir}
  &\tr\ [t^x,t^{a_1}] [t^x,t^{a_2}] t^{a_3} \ldots t^{a_n} + \tr\ [t^x,t^{a_1}] t^{a_2} [t^x,t^{a_3}] \ldots t^{a_n} +\ldots +  \tr\ [t^x,t^{a_1}] t^{a_2}  t^{a_3} \ldots [t^x,t^{a_n}]  \nn \\
&=-\tr\ [t^x, [t^x,t^{a_1}]] t^{a_2} t^{a_3} \ldots t^{a_n} =- c_A \tr\ t^{a_1} t^{a_2} t^{a_3} \ldots t^{a_n} 
  \,.\end{align}
%%%
Similarly, we add the exchanges between $\cS_2,\cS_2^\dagger$ and all the other Wilson lines, etc. The sum of all these contributions counts all exchanges twice,
so the overall group theory factor is $-n_I c_A/2 $, where $n_I$ is the number of indices, e.g.\ 2 for $\cS_{12}^{cd}$ and 4 for $\cS_{12}^{cd}\cS_{34}^{ef}$. 
The $\nu$ anomalous dimension has no $n_i \cdot n_j$ dependent terms, so
%%%
\begin{align}\label{eq:4.7}
  \hat \gamma_{\nu,\cS} = - \frac12 n_I\, c_A \ln \frac{\mu^2}{M^2}
\,.\end{align}
%%%
The $\nu$-anomalous dimensions of the soft and collinear operators cancel.

The $\mu$-anomalous dimensions do not cancel between the soft and collinear operators, as there is also a contribution from the hard matching coefficient $\mathcal{H}$ in \eq{hard_L}. For the $\mu$-anomalous dimension, the second term in \eq{3.62} can be treated in the same manner as $\gamma_\nu$, but the first term has to be computed explicitly, accouting for the imaginary part of $\ln (-n_i \cdot n_j)$ depending on whether the soft Wilson lines are time-ordered or anti-time-ordered. We find 
%%%
\begin{align}\label{eq:3.65}
  \hat \gamma_{\mu,\cS_{12}} &= c_A \Big[\ln \frac{\mu^2}{\nu^2}  - L_{12}\Big]
  \,, \nn \\
   \hat \gamma_{\mu,\cS_{123}} &= c_A \Big[\frac32 \ln \frac{\mu^2}{\nu^2} - \frac12 (L_{12}+L_{13}+L_{23}) \Big]
\,,\end{align}
%%%
where we use the abbreviation
%%%
\begin{align}
 L_{ij} \equiv \ln \left| \frac{n_i \sdt n_j}{2} \right|
\,.\end{align}
%%%
Eq.~\eqref{eq:3.65} holds irrespective of whether particles $1,2,3$ are incoming or outgoing particles. There are no $i\pi$ terms from the imaginary part of the logarithm.

In an $SU(2)$ gauge theory, there are three soft functions with four gauge indices, $\{\cS_{12}^{cd} \cS_{34}^{ef}$, $\cS_{13}^{ce} \cS_{24}^{df}$, $\cS_{14}^{cf} \cS_{23}^{de}\}$, which mix under renormalization. Their $\mu$-anomalous dimension is given in this basis by ($SU(2)$ only)
%%%
\begin{align} \label{eq:mix}
  \hat \gamma_{\mu,\cS_{1234}} &= c_A \left[2 \ln \frac{\mu^2}{\nu^2} -
 \begin{pmatrix}
  L_{12} + L_{34} & 0 & 0\\
  0 & L_{13} + L_{24} & 0 \\
  0 & 0 & L_{14} + L_{23} \\
  \end{pmatrix} \right]
  +
 \begin{pmatrix}
  0 & -w & w \\
  -v & 0 & v \\
  -u & u & 0 \\
  \end{pmatrix}
\,,\end{align}
%%%
using the conformal cross ratios 
%%%
\begin{align}
  u = \ln \frac{(n_1 \sdt n_2)\, (n_3 \sdt n_4) }{(n_1 \sdt n_3) \, (n_2 \sdt n_4)}
  \,, \qquad
  v = \ln \frac{(n_1 \sdt n_2)\, (n_3 \sdt n_4) }{(n_1 \sdt n_4)\,(  n_2 \sdt n_3)}  
  \,, \qquad
  w = \ln \frac{(n_1 \sdt n_3)\, (n_2 \sdt n_4) }{(n_1 \sdt n_4)\,  (n_2 \sdt n_3)} = v - u
\,.\end{align}
%%%
Again, there are no $i\pi$ terms from the imaginary part of the logarithm.  

We conclude this section by discussing the $U(1)$ soft operator that \emph{only} appears for $O_{\widetilde HH}$. Because $O_{\widetilde HH}$ violates hypercharge, it must be accompanied by $O_{H\widetilde H}$ to ensure that the hard scattering conserves hypercharge.\footnote{Of course hypercharge is spontaneously broken, but this is power suppressed in $M/Q$.} Assuming only a single pair of these collinear operators, the corresponding soft operator takes the form $\bar {\mathscr{S}}^\dagger_1 \mathscr{S}_1  \mathscr{S}_2^\dagger \bar {\mathscr{S}}_2$, where $\mathscr{S}$ and $\bar {\mathscr{S}}$ are $U(1)$ Wilson lines with hypercharge $\pm \hyp_H$. Its anomalous dimension is given by that for $S_{12}$ in \eqs{4.7}{3.65} with the replacement $c_A \to 4\hyp_H^2$, and with $M=M_Z$. Note that this $U(1)$ contribution should be expected because $M_Z$ shows up in the rapidity anomalous dimensions of these collinear operators, and the $Z$ boson has a $U(1)$ component. We conclude by noting that in this case the rapidity evolution of the $SU(2)$ soft operator is split equally in two terms involving $M_W$ and $M_Z$, respectively.

%%%%%%%%%%%%%%%%%%%%%%%%%%%%%%%%%%%%%%%%%%%%%%%%%%%%%%%%%%%%%%%%%%%%%%%%%%%%%%%%
\section{Matching onto the broken phase}
\label{sec:low_matching}
%%%%%%%%%%%%%%%%%%%%%%%%%%%%%%%%%%%%%%%%%%%%%%%%%%%%%%%%%%%%%%%%%%%%%%%%%%%%%%%%

We now switch to the broken phase of the theory, which involves matching at the scale $\mu \sim M$. Protons, electrons and neutrinos are now well-defined, and we can use them as external states. Tree-level matching suffices at NLL accuracy, and the matrix elements of collinear operators in proton states are the usual PDFs. For the transversely polarized gauge bosons with helicity $h=\pm 1$, using the Condon-Shortley phase convention,
%%%
\begin{align} \label{eq:PDF}
\langle T | O_{W_h}^{(I=0)} | T \rangle&=  f_{{W_h}^+/T} + f_{{W_h}^-/T} + \cos^2 \theta_W f_{{Z_h}/T} + \sin^2 \theta_W f_{{\ga_h}/T}
  \nn \\ & \quad
   +  \sin \theta_W \cos \theta_W \left( f_{{Z_h}{\ga_h}/T} + f_{{\ga_h}{ Z_h}/T}\right)
   \nn \\
&= \sin^2 \theta_W f_{{\ga_h}/T}
  \,, \nn \\  
\langle T | O_{W_h}^{(I=1,I_3=0)} | T \rangle &=  f_{{W_h}^+/T} - f_{{W_h}^-/T}
 \nn \\ &
  = 0
    \,, \nn \\   
\langle T | O_{W_h B_h}^{(I=1,I_3=0)} | T \rangle &= \cos \theta_W  \sin \theta_W \left(  f_{{\gamma_h} /T} - f_{{Z_h}/T} \right)
   +\cos^2 \theta_W f_{{Z_h}{\ga_h}/T} - \sin^2 \theta_W f_{{\ga_h} {Z_h}/T}    \nn \\   
&= \cos \theta_W  \sin \theta_W  f_{{\gamma_h} /T}  \,, \nn \\      
\langle T | O_{B_h W_h}^{(I=1,I_3=0)} | T \rangle &= \cos \theta_W  \sin \theta_W \left( f_{\gamma_h /T} - f_{Z_h/T} \right)
   +\cos^2 \theta_W f_{\ga_h Z_h/T} - \sin^2 \theta_W f_{Z_h \ga_h /T}     \nn \\   
  &= \cos \theta_W  \sin \theta_W  f_{\gamma_h /T}   \,, \nn \\      
\langle T | O_{W_h}^{(I=2,I_3=0)} | T \rangle &=  -\frac{1}{\sqrt 6} \big( f_{W_h^+/T} +  f_{W_h^-/T} \big) + 
   \frac2{\sqrt 6}\bigl[ \cos^2 \theta_W f_{Z_h/T} 
   +     \sin^2 \theta_W f_{\ga_h/T}
  \nn \\ & \quad
   +    \sin \theta_W \cos \theta_W  \left( f_{Z_h\ga_h/T} + f_{\ga_h Z_h/T} \right) \bigr]
   \nn \\ &
   =  \frac2{\sqrt 6}  \sin^2 \theta_W f_{\ga_h/T}
\end{align}
%%%
where we have omitted the arguments $z,\mu,\nu$ of the PDFs. We first rewrote each matrix element in terms of PDFs in the broken phase, and then dropped all terms except for the photon PDF, since they vanish below the electroweak scale. $f_{\gamma_h Z_h/p}$ and $f_{Z_h\gamma_h/p}$ are interference PDFs, and given by matrix elements of the form in \eq{3.36} where one field strength is that of the photon and the other is that of the $Z$ boson. Note that $W$ on the left-hand side refers to the full $SU(2)$ gauge field $W^1,W^2,W^3$, whereas on the right-side it denotes the $W^\pm$ bosons. Collinear operators in the adjoint representation with gauge indices $1,2$ change electric charge by one unit, and therefore have a vanishing matrix element between states of the same charge. 

Equation~(\ref{eq:PDF}) are initial conditions for DGLAP evolution, and the PDFs are evolved to the high scale $Q$ using the collinear and soft anomalous dimensions. The photon PDF has been computed recently in terms of the hadronic structure functions~\cite{Manohar:2016nzj,Manohar:2017eqh}. The other PDFs in \eq{PDF} can be computed similarly~\cite{Fornal:2018znf}, extending \eq{PDF} beyond tree level.

For right-handed fermions,
%%%
\begin{align}\label{eq:5.2}
\langle T | O_{u,r,s}^{(I=0)} | T \rangle &= \delta_{rs}\, f_{u_{+,r}/T}
\,, \qquad 
\langle T | O_{\bar u,r,s}^{(I=0)} | T \rangle = \delta_{rs}\,f_{\bar u_{-,r}/T}
\,,\end{align}
%%%
since $u$ annihilates positive helicity $u$ quarks and creates negative helicity $\bar u$ quarks. Here $r,s$ are generation indices, so $f_{u_{+,r}}$ for $r=1,2,3$ corresponds to the $u_+$, $c_+$ and $t_+$ PDFs (the top quark PDF vanishes below the electroweak scale). Similar expressions hold for $d$ and $e$.
The proton matrix element in \eq{5.2} is diagonal in flavor. Operators such as $O_{d,1,2}$ change flavor, and have vanishing matrix element in the proton to zeroth order in the weak interactions. Since the matrix elements in \eq{5.2} are taken at a scale of order $M_Z$, weak interaction corrections are not logarithmically enhanced and can be dropped at the accuracy we are working. Of course, electroweak interactions can not be neglected in evolving \eq{5.2} to high energies; this is after all the point of the paper.
Operators such as $O_{d,2,2}$, which measure the strange-quark content of the proton, can have non-zero matrix element~\cite{Kaplan:1988ku}.

Left-handed fields have to be converted from the weak eigenstate to the mass eigenstate basis
%%%
\begin{align}\label{eq:5.3}
\langle T | O^{(I=0)}_{q,r,s} | T \rangle &= \delta_{rs} f_{u_{-,r} /T} + \sum_{w} V_{rw}^* V_{sw} f _{d_{-,w} /T} 
\,, \nn \\
\langle T | O^{(I=0)}_{\bar q,r,s} | T \rangle &= \delta_{rs}  f_{\bar u_{+,r} /T} + \sum_{w} V_{rw}V_{sw}^* f_{\bar d_{+,w}/T}
\,, \nn \\
\langle T | O^{(I=1,I_3=0)}_{q,r,s} | T \rangle &= \frac12 \delta_{rs} f_{u_{-,r} /T} - \frac12 \sum_{w} V_{rw}^* V_{sw} f _{d_{-,w} /T} 
\,, \nn \\
\langle T | O^{(I=1,I_3=0)}_{\bar q,r,s} | T \rangle &= -\frac12 \delta_{rs}  f_{\bar u_{+,r} /T} +\frac12 \sum_{w} V_{rw}V_{sw}^* f_{\bar d_{+,w}/T}
\,, 
\end{align}
%%%
and similarly for $\ell$.

 The Higgs PDFs match on to ($W_L \equiv W_{h=0}$, etc.):
%%%
\begin{align}\label{eq:5.4b}
\langle T | O^{(I=0)}_H | T \rangle &= f_{W_L^+/T} + \frac12 (f_{h/T} + f_{Z_L/T} +  f_{Z_L h/T} + f_{h Z_L/T}) = 0
\,, \nn \\
\langle T | O^{(I=1,I_3=0)}_H | T \rangle &= \frac12 f_{W_L^+/T} - \frac14 (f_{h/T} + f_{Z_L/T} +  f_{Z_L h/T} + f_{h Z_L/T}) = 0
\,, \nn \\
\langle T | O^{(I=0)}_{\bar H} | T \rangle &= f_{W_L^-/T} + \frac12 (f_{h/T} + f_{Z_L/T} -  f_{Z_L h/T} -  f_{hZ_L/T}) = 0
\,, \nn \\
\langle T | O^{(I=1,I_3=0)}_{\bar H} | T \rangle &= -\frac12 f_{W_L^-/T} + \frac14 (f_{h/T} + f_{Z_L/T} -  f_{Z_Lh/T} - f_{hZ_L/T}) = 0
\,,\end{align}
%%%
at tree level, using the equivalence theorem to relate the scalar operators $\varphi$ to longitudinal gauge bosons~\cite{chanowitz,bohmbook}. Again, we first rewrote the the operators in terms of PDFs in the broken basis, which in this case all vanish below the electroweak scale.
Similarly, the $\widetilde H H$ PDFs give 
%%%
\begin{align}\label{eq:5.4a}
\langle T | O^{(I=0)}_{\widetilde H H} | T \rangle &= 0 \,, \nn \\
\langle T | O^{(I=1,I_3=1)}_{\widetilde H H} | T \rangle &= -\frac{1}{\sqrt 2} f_{\bar H^0 H^0/T}=-\frac1{2\sqrt 2} (f_{h/T} - f_{Z_L/T} -  f_{Z_L h/T} + f_{hZ_L/T}) = 0\,, \nn \\
\langle T | O^{(I=1,I_3=0)}_{\widetilde H H} | T \rangle &= 0\,, \nn \\
\langle T | O^{(I=1,I_3=-1)}_{\widetilde H H} | T \rangle &= 0
\,.\end{align}
%%%
Three of the matrix elements vanish because they are operators with non-zero electric charge, and have no diagonal matrix elements in a proton state.

Operators in non-singlet representations are the most interesting, since the corresponding matrix elements would have vanished in QCD (i.e.~the PDF for ``red", ``green" and ``blue" quarks are all equal). However, the proton is not an electroweak singlet, and $f_{u/p} \neq f_{d/p}$ and $f_{W^+/p} \neq f_{W^-/p}$. Assuming the proton beam is unpolarized, we can further simplify \eqs{5.2}{5.3} using
%%%
\begin{align}\label{eq:5.6}
  f_{u_-/p} = f_{u_+/p} = \tfrac12 f_{u/p}
  \,, \qquad
  f_{d_-/p} = f_{d_+/p} = \tfrac12 f_{d/p}
\,,\end{align}
%%%
etc.\
For an incoming neutrino we can simply take
%%%
\begin{align} \label{eq:f_nu}
  f_{\nu/\nu}(x,\mu) = \de(1-x) 
  \,, \qquad
  f_{e_-/\nu}(x,\mu) = 
  f_{e_+/\nu}(x,\mu) = 0
\,,\end{align}
%%%
at $\mu \lesssim M$, since the neutrino is neutral under electromagnetic and QCD effects and thus unaffected by  renormalization group evolution below the electroweak scale. For an incoming electron the initial condition corresponding to \eq{f_nu}  only holds at $\mu = m_e$, because the electron still interacts electromagnetically. Quark distributions, and thus gauge boson distributions at large $\mu$ will become polarized, even if we assume they are unpolarized at low $\mu$ as in \eq{5.6} since the electroweak evolution is chiral.

Similarly, the collinear operators for outgoing directions correspond to fragmentation functions. The tree-level matching relations at the scale $\mu \sim M_W$ are
%%%
\begin{align}
  D^{(I=0)}_{\ell \to e}(x,\mu,\nu) &= \tfrac12 [D_{\nu_- \to e}(x,\mu) + D_{e_- \to e}(x,\mu)]
  \,,\nn \\
  D_{\ell \to e}^{(I=1,I_3=0)}(x,\mu,\nu) &= \tfrac12 [ \tfrac12 (D_{\nu_- \to e}(x,\mu) - D_{e_- \to e}(x,\mu))]
  \,,\nn \\
  D^{(I=0)}_{e \to e}(x,\mu,\nu) &=  D_{e_+ \to e}(x,\mu)
  \,,\nn \\
  D^{(I=0)}_{W_h \to e}(x,\mu,\nu) &=  \tfrac13 \big[D_{W_h^+\to e}(x,\mu) + D_{W_h^-\to e}(x,\mu) + \cos^2 \theta_W D_{Z\to e}(x,\mu) + \sin^2 \theta_W D_{\ga\to e}(x,\mu)
  \nn \\ & \qquad
   +  \sin \theta_W \cos \theta_W \big(D_{Z\ga\to e}(x,\mu) + D_{\ga Z\to e}(x,\mu)\big)\big]
  \nn \\ &
  =  \tfrac13 \sin^2 \theta_W D_{\ga\to e}(x,\mu)
  \,,\nn \\
  D_{W_h\to e}^{(I=1,I_3=0)}(x,\mu,\nu) &= \tfrac13 [D_{W_h^+ \to e}(x,\mu) - D_{W_h^- \to e}(x,\mu)]
  \nn \\ &
  = 0
\,,\end{align}
%%%
etc., where $h$ is a helicity label. The extra factors of 1/2 for $\ell$ and 1/3 for $W$ compared to \eq{PDF} arise because the fragmentation functions in the symmetric phase are averaged over gauge configurations of the field from which the particle fragments.

Lastly, the matrix elements of soft operators at tree-level are
%%%
\begin{align}
    \langle 0 | \cS_{12}^{cd} |0 \rangle
  &\ = \tfrac12 \de^{cd}
 \,, \nn \\
   \langle 0 | \cS_{123}^{cde} |0 \rangle &= \tfrac{\img}{4} f^{cde} 
  \!\stackrel{N=2}{=} \tfrac{\img}{4} \eps^{cde} 
  \,, \nn \\
    \langle 0 |  \cS_{12}^{cd} \cS_{34}^{ef} |0 \rangle
 &\ = \tfrac{1}{4} \de^{cd} \de^{ef}
\,,\end{align}
%%%
From the collinear functions we know that the only non-vanishing contribution at tree-level requires all gauge indices to be 3,
%%%
\begin{align} \label{eq:S_tree}
    \langle 0 | \cS_{12}^{33} |0 \rangle
  = \tfrac12 
 \,, \qquad
   \langle 0 | \cS_{123}^{333} |0 \rangle
  = 0
  \,, \qquad
    \langle 0 |  \cS_{12}^{33} \cS_{34}^{33} |0 \rangle
 = \tfrac{1}{4}
\,.\end{align}
%%%
The one-loop soft matching has been computed in ref.~\cite{Chiu:2009mg}. The $\mu$ dependence of the one-loop matching converts the  double-logarithmic Sudakov evolution above the electroweak scale into the usual single-logarithmic DGLAP evolution below the electroweak scale.

%%%%%%%%%%%%%%%%%%%%%%%%%%%%%%%%%%%%%%%%%%%%%%%%%%%%%%%%%%%%%%%%%%%%%%%%%%%%%%%%
\section{Resummation}
\label{sec:resummation}
%%%%%%%%%%%%%%%%%%%%%%%%%%%%%%%%%%%%%%%%%%%%%%%%%%%%%%%%%%%%%%%%%%%%%%%%%%%%%%%%

\begin{figure}
\begin{center}
\includegraphics[]{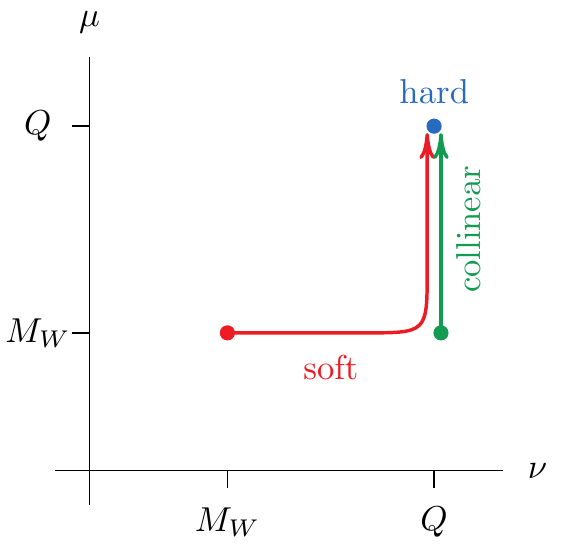}
\end{center}

\vspace{-3ex}

\caption{\label{fig:nu} Path in $(\nu,\mu)$ space for integrating the anomalous dimensions of collinear and soft operators.}
\end{figure}

The factorization in \eq{factor} enables the resummation of logarithms of $Q/M$, by separating the cross section into factors that each involve a single invariant mass and rapidity scale. Specifically, the resummation is accomplished by evaluating the hard matching coefficients $\mathcal{H}$, collinear operators $\C$ and soft operators $\cS$ at their natural scales\footnote{We did not specify a rapidity scale for the hard coefficient, because it does not contain rapidity divergences, $\gamma_{\nu,\cS}+\gamma_{\nu, \C}=0$.}
%%%
\begin{align}
   \mu_\mathcal{H} \sim Q
   \,, \qquad
   \mu_\C \sim \mu_\cS \sim M
   \,, \qquad
   \nu_\C \sim Q
   \,, \qquad
   \nu_\cS \sim M
\,,\end{align}
%%% 
where they do not contain large logarithms. RG evolving them to a common scale $(\mu,\nu)$ will exponentiate the logarithms. The $\mu$-anomalous dimension contains $\ln \nu$ terms, and the $\nu$-anomalous dimension contains $\ln \mu$ terms, which are related to each other and proportional to the cusp anomalous dimension.

We will evolve the collinear and soft operators to the hard scale. This avoids having to calculate (the evolution of) the process-dependent hard matching coefficients at one loop. The $\nu$-anomalous dimension contains $\ln \mu^2/M_W^2$ terms, so to avoid large logarithms the simplest strategy is to first do the $\nu$ evolution of the soft operator from $\nu = M_W$ to $\nu=Q$ at $\mu=M_W$, and then perform the $\mu$ evolution of the soft and collinear operators from $\mu=M_W$ to $\mu=Q$, as shown in \fig{nu} (see also the discussion above eq.~(4.30) in ref.~\cite{Chiu:2012ir}). 
Using \eq{4.7}, the $\nu$ evolution of the soft operator gives
%%%
\begin{align}
 U_\nu = 
 \exp\bigg[\int_{\nu_\cS}^{\nu_\C}\! \frac{\df \nu}{\nu}\, \gamma_{\nu, \cS}\bigg]
 =  \exp \bigg[- n_I\, \frac{\alpha_2(\mu)}{\pi} \ln \frac{Q}{M_W} \ln \frac{\mu^2}{M_W^2}\bigg]
\,,\end{align}
%%%
where $n_I$ is the number of gauge indices in the soft factor. When $\mu = M_W$ exactly,  $U_\nu = 1$ and can be ignored, but otherwise it must be kept to achieve NLL accuracy. In particular it should be kept when estimating the perturbative uncertainty from scale variations. Note that this analysis does not apply to the quintet contribution $O_{W_\pm}^{(I=2)}$ or the special case of $O_{\tilde H H}^{(I=1)}$, which will be discussed in \sec{PDF_SM}.

Moving on to the $\mu$-evolution, we first consider the terms in the collinear and soft $\mu$-evolution that give rise to double logarithms. They are described by the following multiplicative anomalous dimensions: For the soft anomalous dimension, the relevant terms in eqs.~\eqref{eq:3.65}--\eqref{eq:mix} are given by (using $c_A=2$)
%%%
\begin{align}\label{eq:6.3}
   \ga_{\mu,\cS}^{\rm DL} = n_I\, \frac{\al_2}{\pi}\, \ln \frac{\mu^2}{\nu^2}  
\,.\end{align}
%%%
For the collinear anomalous dimensions, the double logarithms arise from the $\ln \nu/(\bar n \cdot p)$ term, which vanishes for the off-diagonal elements. For the diagonal elements, it vanishes for the singlet, and for the triplet PDFs (and FFs) it is given by 
%%%
\begin{align}\label{eq:6.4}
  \ga_{\mu,qq}^{(I=1),{\rm DL}} &=   \ga_{\mu,WW}^{(I=1),{\rm DL}} =   \ga_{\mu,W\! B}^{(I=1),{\rm DL}} = \cdots =  \frac{2\alpha_2}{\pi}\, \ln \frac{\nu}{\bar n \cdot r}\ \delta(1-z) 
\,.\end{align}
%%%
Here $\bar n \cdot r = 2E$, with $E$ the energy of the parton. The triplet PDFs have a single gauge index which is contracted with a soft operator, and the soft operator anomalous dimension \eq{6.3} is proportional to $n_I$, the number of gauge indices.  Combining the collinear anomalous dimension in \eq{6.4} for a triplet operator with \eq{6.3} for $n_I=1$ gives
%%%
\begin{align}\label{eq:6.5}
   \ga_{\mu}^{(I=1),\rm DL} = \frac{2\alpha_2}{\pi}\, \ln \frac{\mu}{\bar n \cdot r}\ \delta(1-z)  
\,.\end{align}
%%%
The $\nu$ dependence has disappeared in the combined soft plus collinear anomalous dimensions, and the $\ln \mu/(\bar n \cdot r)$ anomalous dimensions is precisely the form of the anomalous dimension computed in refs.~\cite{Chiu:2009mg,Chiu:2009ft}, that gives rise to Sudakov double logarithms.
Integrating \eq{6.5} yields the evolution kernel
%%%
\begin{align}\label{eq:sudakov}
  U_\mu^{\rm DL} &=
  \exp\bigg[\int_{M_W}^{Q}\, \frac{\df \mu}{\mu}\, \frac{2\al_2 }{\pi}\, \ln \frac{\mu}{\bar n \cdot r}\bigg]
  \approx   \exp\bigg[ - \frac{\al_2}{\pi}\, \left(\ln^2 \frac{\bar n \cdot r}{M_W} - \ln^2 \frac{\bar n \cdot r}{Q}\right)\bigg]
\nn \\ &
 \approx   \exp\bigg[ - \frac{\al_2}{\pi}\, \ln^2 \frac{Q}{M_W} \bigg]
\,,\end{align}
%%%
to leading-logarithmic accuracy, since $\bar n \cdot r \sim Q$, the partonic center-of-mass energy of the collision. We recognize \eq{sudakov}
as an electroweak Sudakov factor. It suppresses the contribution from collinear operators in the triplet representation.

In addition to the double logarithms, there are single logarithms from the evolution of collinear and soft operators. The coefficients of the splitting functions for the collinear operators are modified for non-singlet representations. We give their explicit form for the Standard Model in \sec{PDF_SM}.

If there are no identified particles in the final state, then the only collinear functions are for the two incoming beam directions $n_1$ and $n_2$. In this case, the number of adjoint electroweak indices in the soft or collinear sector from the factorization theorem can be $n_I=0,1,2$, depending on whether the collinear operator for the two beams is a singlet or adjoint. The $n_I=1$ case does not occur, since the corresponding soft function vanishes. The beams are back-to-back, so $n_1 \cdot n_2=2$ and the electroweak logarithms have no angular dependence. When there is a single identified particle in the final state, there is a third direction $n_3$, and the number of adjoint indices can be $n_I=0,1,2,3$. There can be angular dependence due to electroweak logarithms from $\ln n_1 \cdot n_3$ and $\ln n_2 \cdot n_3 $ terms in the soft anomalous dimension. These arise from terms with $n_I=2$, since 
\eq{S_tree} prohibits contributions with $n_I=3$. The $n_I=0$ terms have no angular dependence, and there are no $n_I=1$ terms, as before.
With two identified particles in the final-state, as is the case for Drell-Yan, there are now four directions, and contributions with $n_I=4$ are also allowed. Mixing effects in the soft sector, described in \eq{mix}, enter for the first time.

%===============================================================================
\subsection{PDF evolution in the Standard Model}
\label{sec:PDF_SM}
%===============================================================================

In this section, we use the collinear evolution results of \sec{RGE_C} to give the PDF evolution equations in the Standard Model, keeping only the Yukawa of the top quark (see \eq{approx}). The splitting functions for $z < 1$ agree with those computed in ref.~\cite{Bauer:2017isx}. The $I=0$ sector gives the usual evolution of gauge-invariant PDFs,
%%%
\begin{align}
\mu \frac{\df}{\df \mu} f_{q,r,s}^{(I=0)}(z) &= \frac{\alpha_3}{\pi} \biggl[ \frac43  \widetilde P_{Q_-Q_-} \otimes f_{q,r,s}^{(I=0)} + \delta_{rs} \widetilde P_{Q_-G_+} \otimes f_{g_+}^{(I=0)} + \delta_{rs} \widetilde P_{Q_-G_-} \otimes f_{g_-}^{(I=0)} \biggr] \nn \\
& \quad + \frac{\alpha_2}{\pi}  \biggl[ \frac34  \widetilde P_{Q_-Q_-} \otimes f_{q,r,s}^{(I=0)}  
+  \frac{N_c}2 \delta_{rs} \widetilde P_{Q_-G_+} \otimes f_{W_+}^{(I=0)}+  \frac{N_c}2 \delta_{rs} \widetilde P_{Q_-G_-} \otimes f_{W_-}^{(I=0)}  \biggr] \nn \\ 
& \quad +  \frac{\alpha_1}{\pi} \left[ \hyp_q^2 \widetilde P_{Q_-Q_-} \!\otimes\! f_{q,r,s}^{(I=0)} \!+\! 2 N_c \hyp_q^2 \delta_{rs} \widetilde P_{Q_-G_+} \otimes f_{B_+}^{(I=0)} \!+\! 2 N_c \hyp_q^2 \delta_{rs} \widetilde P_{Q_-G_-} \!\otimes\! f_{B_-}^{(I=0)} \right] 
 \nn \\[5pt]
& \quad  + \frac{Y_t^2}{4\pi^2} \biggl[ 
 \delta_{r3}\delta_{s3} (1-z) \otimes f^{(I=0)}_{u,3,3}    -\frac18 \delta_{r3} f^{(I=0)}_{q,3,s}(z) 
 -\frac18 \delta_{s3} f^{(I=0)}_{q,r,3}(z) 
\nn \\[5pt] & \quad
+ \frac{N_c}{2} \delta_{r3}\delta_{s3}\, 1 \otimes f_{\bar H}^{(I=0)}  \biggr] 
\,,\end{align}
%%%
\vspace{-2ex}
\begin{align}
\mu \frac{\df}{\df \mu} f_{u,r,s}^{(I=0)} &= \frac{\alpha_3}{\pi} \left[ \frac43  \widetilde P_{Q_+Q_+} \otimes f_{u,r,s}^{(I=0)} +  \frac12  \delta_{rs} \widetilde P_{Q_+G_+} \otimes f_{g_+}^{(I=0)}+  \frac12  \delta_{rs} \widetilde P_{Q_+G_-} \otimes f_{g_-}^{(I=0)}  \right]  \nn \\
& \quad +  \frac{\alpha_1}{\pi} \Bigl[ \hyp_u^2 \widetilde P_{Q_+Q_+} \otimes f_{u,r,s}^{(I=0)} 
 +  N_c \hyp_u^2   \delta_{rs} \widetilde P_{Q_+G_+} \otimes f_{B_+}^{(I=0)} +   N_c \hyp_u^2   \delta_{rs} \widetilde P_{Q_+G_-} \otimes f_{B_-}^{(I=0)}  \Bigr]  \nn \\
& \quad + \frac{Y_t^2}{4\pi^2} \biggl[  \frac12  \delta_{r3}\delta_{s3} (1-z) \otimes f^{(I=0)}_{q,3,3}  
 -\frac14 \delta_{r3}  f^{(I=0)}_{u,3,s}(z) -\frac14  \delta_{s3} f^{(I=0)}_{u,r,3}(z)    \nn \\
& \quad + \frac{N_c}{2} \delta_{r3}\delta_{s3}\, 1 \otimes f_{H}^{(I=0)}  \biggr]
\,,\end{align}
%%%
\vspace{-2ex}
\begin{align}
\mu \frac{\df}{\df \mu} f_{d,r,s}^{(I=0)} &= \frac{\alpha_3}{\pi}  \left[ \frac43  \widetilde P_{Q_+Q_+} \otimes f_{d,r,s}^{(I=0)} +  \frac12  \delta_{rs} \widetilde P_{Q_+G_+} \otimes f_{g_+}^{(I=0)} +  \frac12  \delta_{rs} \widetilde P_{Q_+G_-} \otimes f_{g_-}^{(I=0)}  \right]  
\\
& \quad+  \frac{\alpha_1}{\pi}\left[ \hyp_d^2 \widetilde P_{Q_+Q_+} \otimes f_{d,r,s}^{(I=0)} + N_c  \hyp_d^2  \delta_{rs} \widetilde P_{Q_+G_+} \otimes f_{B_+}^{(I=0)} + N_c  \hyp_d^2  \delta_{rs} \widetilde P_{Q_+G_-} \otimes f_{B_-}^{(I=0)}  \right]  
,\nonumber\end{align}
%%%
\vspace{-2ex}
\begin{align}
\mu \frac{\df}{\df \mu} f_{\ell,r,s}^{(I=0)} &=   \frac{\alpha_2}{\pi} \left[ \frac34  \widetilde P_{Q_-Q_-} \otimes f_{\ell,r,s}^{(I=0)}+  \frac12  \delta_{rs} \widetilde P_{Q_-G_+} \otimes f_{W_+}^{(I=0)}+  \frac12  \delta_{rs} \widetilde P_{Q_-G_-} \otimes f_{W_-}^{(I=0)}  \right]   \\
& \quad +  \frac{\alpha_1}{\pi} \left[ \hyp_\ell^2 \widetilde P_{Q_-Q_-} \otimes f_{\ell,r,s}^{(I=0)} +  \hyp_\ell^2  \delta_{rs} \widetilde P_{Q_-G_+} \otimes f_{B_+}^{(I=0)}  +  \hyp_\ell^2  \delta_{rs} \widetilde P_{Q_-G_-} \otimes f_{B_-}^{(I=0)}  \right] 
,\nonumber\end{align}
%%%
\vspace{-3ex}
\begin{align}
\mu \frac{\df}{\df \mu} f_{e,r,s}^{(I=0)} &=  
  \frac{\alpha_1}{\pi}\left[ \hyp_e^2 \widetilde P_{Q_+Q_+} \otimes f_{e,r,s}^{(I=0)} +  \hyp_e^2  \delta_{rs} \widetilde P_{Q_+G_+} \otimes f_{B_+}^{(I=0)} +  \hyp_e^2  \delta_{rs} \widetilde P_{Q_+G_-} \otimes f_{B_-}^{(I=0)}  \right]
,\end{align}
%%% 
\vspace{-4ex}
\begin{align}\label{eq:6.12}
\mu \frac{\df}{\df \mu} f_{g_\pm}^{(I=0)} &= \frac{\alpha_3}{\pi} \bigg[  3 \widetilde P_{G_\pm G_+} \otimes f_{g_+}^{(I=0)} 
+3 \widetilde P_{G_\pm G_-} \otimes f_{g_-}^{(I=0)} 
+ \frac12 b_{0,3} f_{g_\pm}^{(I=0)}(z) \nn \\
&\quad + \frac43 \widetilde P_{G_\pm Q_+} \otimes  \sum_{\substack{ i=\bar q, u, d,  \\ r=1,\ldots,n_g}}   f_{i,r,r}^{(I=0)} 
+ \frac43  \widetilde P_{G_\pm Q_-}  \otimes \sum_{\substack{ i=q, \bar u, \bar d \\ r=1,\ldots,n_g}}   f_{i,r,r}^{(I=0)} \bigg]
,\end{align}
%%%
\vspace{-2ex}
\begin{align}
\mu \frac{\df}{\df \mu} f_{W_\pm}^{(I=0)} &=  \frac{\alpha_2}{\pi} \bigg[ 2 \widetilde P_{G_\pm G_+} \otimes f_{W_+} ^{(I=0)}
+ 2 \widetilde P_{G_\pm G_-} \otimes f_{W_-} ^{(I=0)}  + \frac12 b_{0,2} f_{W_\pm}^{(I=0)}(z) \\
&\quad + \frac34  \widetilde P_{G_\pm Q_+ } \otimes \!\sum_{\substack{i=\bar q, \bar \ell \\ r=1,\ldots,n_g}}\!\!   f_{i,r,r}^{(I=0)}
+ \frac34  \widetilde P_{G_\pm Q_-} \otimes \!\sum_{\substack{i=q, \ell,  \\ r=1,\ldots,n_g}}\!\!  f_{i,r,r}^{(I=0)} + \frac34  \widetilde P_{G_\pm H} \otimes  \! \sum_{i=H,\bar H} f_i^{(I=0)}   \biggr] 
,\nonumber\end{align}
%%%
\vspace{-1ex}
%%%
\begin{align}
\mu \frac{\df}{\df \mu} f_{B_\pm}^{(I=0)} &=   \frac{\alpha_1}{\pi} \bigg[  \frac12 b_{0,1} f_{B_\pm}^{(I=0)}(z) +   \widetilde P_{G_\pm Q_+} \otimes \sum_{\substack{i=\bar q, u, d,  \bar \ell, e  \\ r=1,\ldots,n_g}} \hyp_i^2 f_{i,r,r}^{(I=0)}  \nn \\
& \quad +   \widetilde P_{G_\pm Q_-} \otimes \sum_{\substack{i=q,  \bar u, \bar d,   \ell , \bar e \\ r=1,\ldots,n_g}} \hyp_i^2  f_{i,r,r}^{(I=0)} + \hyp_H^2  \widetilde P_{G_\pm H} \otimes \sum_{i=H,\bar H} f_i^{(I=0)} \bigg]
\,,\end{align}
\vspace{-2ex}
\begin{align}
\mu \frac{\df}{\df \mu} f_H^{(I=0)} &=    \frac{\alpha_2}{\pi} \left[ \frac34  \widetilde P_{HH} \otimes f_H^{(I=0)}+  \frac12 \widetilde P_{HG_+} \otimes f_{W_+}^{(I=0)} +  \frac12 \widetilde P_{HG_-} \otimes f_{W_-}^{(I=0)}  \right]  \nn \\
& \quad +  \frac{\alpha_1}{\pi} \Bigl[ \hyp_H^2 \widetilde P_{HH} \otimes f_H^{(I=0)} 
+  \hyp_H^2 \widetilde P_{HG_+} \otimes f_{B_+}^{(I=0)} +  \hyp_H^2 \widetilde P_{HG_-} \otimes f_{B_-}^{(I=0)}  \Bigr]  \nn \\
& \quad + \frac{Y_t^2}{8\pi^2} \left[ z  \otimes \left(  f_{\bar q,3,3}^{(I=0)} + 2  f_{\bar u,3,3}^{(I=0)}  \right) - N_c    f_H^{(I=0)}(z) \right]
\,,\end{align}
%%%
In addition, we also have the antiparticle equations given by $CP$ conjugation, $q_-,r,s \leftrightarrow \bar q_+,s,r$, $g_+ \leftrightarrow g_-$, $H \leftrightarrow \bar H$, etc.
Some terms have been simplified using $\delta(1-z) \otimes f = f(z)$.

In the $I=1$ sector,
%%%
\begin{align}
\mu \frac{\df}{\df \mu} f_{q,r,s}^{(I=1)} &= \frac{\alpha_3}{\pi}  \frac43  \widetilde P_{Q_-Q_-} \otimes f_{q,r,s}^{(I=1)} \nn \\
&\quad \!+\! \frac{\alpha_2}{\pi} \left[\!- \frac14  \widetilde P_{Q_-Q_-} \!\otimes\! f_{q,r,s}^{(I=1)} \!+\! \Gamma_1  f_{q,r,s}^{(I=1)}(z) \!+\!  \frac14 N_c  \delta_{rs} \widetilde P_{Q_-G_+} \otimes f_{W_+}^{(I=1)}\!+\!  \frac14 N_c  \delta_{rs} \widetilde P_{Q_-G_-} \!\otimes\! f_{W_-}^{(I=1)}  \right]
 \nn \\[5pt] & \quad 
 +  \frac{\alpha_1}{\pi} \hyp_q^2 \widetilde P_{Q_-Q_-} \otimes f_{q,r,s}^{(I=1)}  + \frac{g_1 g_2}{4\pi^2} \hyp_q N_c \delta_{rs} \widetilde P_{Q_-G_+} \otimes \Bigl( f_{W_+\!B_+}^{(I=1)} +
f_{B_+W_+}^{(I=1)} \Bigr)  \nn \\
&\quad + \frac{g_1 g_2}{4\pi^2} \hyp_q N_c \delta_{rs} \widetilde P_{Q_-G_-} \otimes \Bigl( f_{W_-\!B_-}^{(I=1)} +
f_{B_-W_-}^{(I=1)} \Bigr)  \nn \\ 
&\quad  + \frac{Y_t^2}{4\pi^2} \left[  -\frac18  \delta_{r3} f^{(I=1)}_{q,3,s}(z) -\frac18  \delta_{s3} f^{(I=1)}_{q,r,3}(z) + \frac{N_c}{2}    \delta_{r3}\delta_{s3}\, 1 \otimes f_{\bar H}^{(I=1)}  \right] 
\,,\end{align}
%%%
\vspace{-3ex}
\begin{align}
\mu \frac{\df}{\df \mu} f_{\ell,r,s}^{(I=1)} &=   \frac{\alpha_2}{\pi} \left[\!- \frac14  \widetilde P_{Q_-Q_-} \!\otimes\! f_{\ell,r,s}^{(I=1)} \!+\! \Gamma_1 f_{\ell,r,s}^{(I=1)}(z) \!+\!  \frac14  \delta_{rs}  \widetilde P_{Q_-G_+} \!\otimes\! f_{W_+}^{(I=1)}  \!+\!  \frac14  \delta_{rs}  \widetilde P_{Q_-G_-} \!\otimes\! f_{W_-}^{(I=1)}  \right] \nn \\[5pt] 
& \quad +  \frac{\alpha_1}{\pi} \hyp_\ell^2 \widetilde P_{Q_-Q_-} \otimes f_{\ell,r,s}^{(I=1)} + \frac{g_1 g_2}{4\pi^2} \hyp_\ell  \delta_{rs}  \widetilde P_{Q_-G_+} \otimes \Bigl( f_{W_+\!B_+}^{(I=1)} +
f_{B_+W_+}^{(I=1)}\Bigr)  \nn \\
& \quad + \frac{g_1 g_2}{4\pi^2} \hyp_\ell  \delta_{rs}  \widetilde P_{Q_-G_-} \otimes \Bigl( f_{W_-\!B_-}^{(I=1)} +
f_{B_-W_-}^{(I=1)}\Bigr)
\,,\end{align}
%%%
\vspace{-2ex}
\begin{align}
\mu \frac{\df}{\df \mu} f_{W_\pm}^{(I=1)} &=  \frac{\alpha_2}{\pi}\bigg[ \widetilde P_{G_\pm G_+}   \otimes f_{W_+}^{(I=1)} + \widetilde P_{G_\pm G_-}   \otimes f_{W_-}^{(I=1)}  + \Gamma_2 f_{W_\pm}^{(I=1)} (z)
\!+\! P_{G_\pm Q_+ } \otimes \!\!\! \sum_{\substack{i=\bar q, \bar \ell \\ r=1,\ldots,n_g}}   f_{i,r,r}^{(I=1)} \nn \\
&\quad \!+\! P_{G_\pm Q_-} \otimes \!\!\! \sum_{\substack{i=q, \ell, \\ r=1,\ldots,n_g}}   f_{i,r,r}^{(I=1)}  \!+\! P_{G_\pm H}(z) \otimes \!\! \sum_{i=H,\bar H} f_i^{(I=1)} \bigg]
,\end{align}
%%%
\vspace{-2ex}
\begin{align}
\mu \frac{\df}{\df \mu} f_{W_\pm B_\pm}^{(I=1)} &=  \left[\frac{\alpha_2}{\pi} \Gamma_3 +\frac{\alpha_1}{\pi} \frac14 b_{0,1} \right]  f_{W_\pm B_\pm}^{(I=1)} (z) 
+ \frac{g_1 g_2}{4\pi^2} \widetilde P_{G_\pm Q_-} \otimes  \sum_{i=q, \ell, r=1,\ldots,n_g}   \hyp_i f_{i,r,r}^{(I=1)}  \nn \\
& \quad -  \frac{g_1 g_2}{4\pi^2}  \widetilde P_{G_\pm Q_+} \otimes \sum_{i=\bar q, \bar \ell, r=1,\ldots,n_g} f_{i,r,r}^{(I=1)}  
\,,\end{align}
%%%
\vspace{-2ex}
\begin{align}
\mu \frac{\df}{\df \mu} f_H^{(I=1)} &= \frac{\alpha_2}{\pi} \left[ - \frac14  \widetilde P_{HH}  \otimes f_H^{(I=1)} + \Gamma_4  f_H^{(I=1)}(z) +  \frac14  \widetilde P_{HG_+} \otimes f_{W_+}^{(I=1)}  +  \frac14  \widetilde P_{HG_-} \otimes f_{W_-}^{(I=1)}  \right] \nn \\
& \quad +  \frac{\alpha_1}{\pi} \left[ \hyp_H^2 \widetilde P_{HH} \otimes f_H^{(I=1)}  \right]  + \frac{Y_t^2}{8\pi^2} \left[ z  \otimes f_{\bar q,3,3}^{(I=1)}  - N_c   f_H^{(I=1)} (z) \right]
\,,\end{align}
%%%
\vspace{-2ex}
\begin{align}
\mu \frac{\df}{\df \mu} f_{\widetilde H H}^{(I=1)} &= \frac{\alpha_2}{\pi} \left[ - \frac14  \widetilde P_{HH}  \otimes f_{\widetilde H H}^{(I=1)} + \Gamma_4  f_{\widetilde H H}^{(I=1)}(z) \right]  \nn \\[5pt]
&\quad +   \frac{\alpha_1}{\pi} \left[ -\hyp_H^2 \widetilde P_{HH} \otimes f_{\widetilde H H}^{(I=1)} + 2 \hyp_H^2 \Gamma_4 f_{\widetilde H H}^{(I=1)} (z)  \right] -\frac{Y_t^2}{8\pi^2} N_c   f_{\widetilde H H}^{(I=1)} (z) 
\,.\end{align}
%%%
The constants $\Gamma_i$ are
%%%
\begin{align}\label{eq:Ga_def}
\Gamma_1 &= \frac32 + 2 \ln \frac{\nu}{\bar n \cdot r}  \,,  &
\Gamma_2 &= \frac{b_{0,2}}{2}+2 \ln \frac{\nu}{\bar n \cdot r} \,, \nn \\
\Gamma_3 &= \frac{b_{0,2}}{4}+2 \ln \frac{\nu}{\bar n \cdot r}\,, &
\Gamma_4 &= 2 + 2 \ln \frac{\nu}{\bar n \cdot r}  \,.  
\end{align}
%%%
The antiparticle equations are given by $CP$ conjugation, $q,r,s \leftrightarrow \bar q,s,r$, $g_+ \leftrightarrow g_-$, $H \leftrightarrow \bar H$, $\hyp_{q,\ell} \to - \hyp_{q,\ell}$, etc.
With the sign convention between $q$ and $\bar q$ PDFs discussed below \eq{3.33}, $q + \bar q$ is $CP=+$ for the $I=0$ PDF, and $CP=-$ for the $I=1$ PDF. The $\ln \nu/(\bar n \cdot r)$ term, when combined with the soft anomalous dimension, turns in to $\ln \mu/(\bar n \cdot r)$ that yield electroweak Sudakov double logarithms, as discussed in the first part of \sec{resummation}. 

The evolution in the $I=2$ sector is given by
%%%
\begin{align}
\mu \frac{\df}{\df \mu} f_{W_\pm}^{(I=2)} &=  \frac{\alpha_2}{\pi} \left[ -\widetilde P_{G_\pm G_+} \otimes f_{W_+}^{(I=2)} -\widetilde P_{G_\pm G_-} \otimes f_{W_-}^{(I=2)}  +\biggl(\frac{b_{0,2}}{2}+6 \ln \frac{\nu}{\bar n \cdot r}\biggr)  f_{W_\pm} ^{(I=2)} (z) \right] , 
\end{align}
%%%
and only involves the transverse $W$ PDFs.

The chiral nature of the electroweak interactions implies that all parton distributions will become polarized. The polarized gluon distribution
$f_{\Delta g}=f_{g_+}-f_{g_-}$ will evolve to a non-zero value using \eq{6.12}, even if it vanishes at small values of $\mu$.

Finally, the $\nu$ anomalous dimensions are diagonal, and take the simple form
%%%
\begin{align}
\nu \frac{\df}{\df \nu} f_i^{(I=0)} &= 0 ,\nn \\
\nu \frac{\df}{\df \nu} f_i^{(I=1,I_3=0)} &= \frac{\alpha_2}{\pi} \ln \frac{\mu^2}{M_W^2} f_i^{(I=1,I_3=0)}, \nn \\
\nu \frac{\df}{\df \nu} f_i^{(I=2,I_3=0)} &= \frac{3\alpha_2}{\pi}  \ln \frac{\mu^2}{M_W^2} f_i^{(I=2,I_3=0)}
\,,
\end{align}
%%%
with the exception of  $\widetilde H H$, for which 
%%%
\begin{align} \label{eq:tildeHH}
\nu \frac{\df}{\df \nu} f_{\widetilde H H}^{(I=1,I_3=1)} &= 
\bigg[\frac{\alpha_2}{2\pi}  \ln \frac{\mu^2}{M_W^2}
+ \frac{(\alpha_2+ 4\hyp_H^2\alpha_1)}{2\pi} \ln \frac{\mu^2}{M_Z^2} \bigg]f_{\widetilde H H}^{(I=1,I_3=1)}
\nn \\
&= 
\bigg[\frac{\alpha_2}{2\pi}  \ln \frac{\mu^2}{M_W^2} +
\frac{\alpha_{\text{em}}}{2\pi \sin^2 \theta_W \cos^2 \theta_W} \ln \frac{\mu^2}{M_Z^2} \bigg]f_{\widetilde H H}^{(I=1,I_3=1)}
\,.
\end{align}
%%%

%%%%%%%%%%%%%%%%%%%%%%%%%%%%%%%%%%%%%%%%%%%%%%%%%%%%%%%%%%%%%%%%%%%%%%%%%%%%%%%%
\section{Comparison to literature}
\label{sec:comparison}
%%%%%%%%%%%%%%%%%%%%%%%%%%%%%%%%%%%%%%%%%%%%%%%%%%%%%%%%%%%%%%%%%%%%%%%%%%%%%%%%

We compare our results to those obtained for the (electroweak) PDF evolution in refs.~\cite{Ciafaloni:2001mu,Ciafaloni:2005fm,Bauer:2017isx}, which is based on splitting functions in the broken phase.
Their approach yields, for example,  for the $SU(2)$ running with $\mu \gg M$,
%%%
\begin{align} \label{eq:BFW}
  \frac{\df}{\df \ln \mu}\, f_q^{(I=1,I_3=0)}(x,\mu) &= \frac{\al_2}{\pi}\, \int_0^{1-M/\mu}\, \df z\, \bigg[-\frac14 \widetilde P_{QQ}(z)\, f_q^{(I=1,I_3=0)}\Big(\frac{x}{z},\mu\Big) 
  \nn \\ & \quad 
  + \frac18  N_c \widetilde P_{QG}(z)\, f_W^{(I=1,I_3=0)}\Big(\frac{x}{z},\mu\Big) + \dots \bigg]
\,,  \nn \\ 
  \frac{\df}{\df \ln \mu}\, f_W^{(I=1,I_3=0)}(x,\mu) &= \frac{\al_2}{\pi}\, \int_0^{1-M/\mu}\, \df z\, \bigg[\widetilde P_{GG}(z)\, f_W^{(I=1,I_3=0)}\Big(\frac{x}{z},\mu\Big) 
  \nn \\ & \quad
  + \widetilde P_{GQ}(z)\, \sum_{i=q,\bar q, \ell, \bar \ell} f_i^{(I=1,I_3=0)}\Big(\frac{x}{z},\mu\Big) + \dots \bigg]
\,.\end{align}
%%%
Here QCD is accounted for through the number of colors $N_c$,  the triplet PDFs are
%%%
\begin{align}\label{eq:3.81}
  f_q^{(I=1,I_3=0)}(x,\mu) &= \tfrac12[f_{u_-}(x,\mu) - f_{d_-}(x,\mu)]
  \,, \nn \\
  f_W^{(I=1,I_3=0)}(x,\mu) &= \sum_{h=\pm} f_{W_h^+}(x,\mu) - f_{W_h^-}(x,\mu)
\,,\end{align}
%%%
i.e.~we assume a single generation, and the (conventional) QCD splitting functions are given by
%%%
\begin{align}
  \widetilde P_{QQ} &=  \widetilde P_{Q_-Q_-}
  \,, &
  \widetilde P_{QG} &= \widetilde P_{Q_-G_-} + \widetilde P_{Q_-G_+}  
  \,, \nn \\
  \widetilde P_{GG} &=  \widetilde P_{G_-G_-} + \widetilde P_{G_-G_+}
  \,, &
  \widetilde P_{GQ} &=  \widetilde P_{G_-Q_-} + \widetilde P_{G_+Q_-}
\,.\end{align}
%%%
Before comparing this to our results, we want to stress that the polarized $f_{\Delta W}^{(I=1,I_3=0)}$ also mixes into $f_q^{(I=1,I_3=0)}$, and was not accounted for in these earlier calculations. Since $f_{\Delta W}^{(I=1,I_3=0)}$ does not vanish, this effect cannot be ignored~\cite{Fornal:2018znf}.

The splitting functions in \eq{BFW} agree with our results in \secs{RGE_C}{PDF_SM} for $z<1$. In writing \eq{BFW}, the soft singularity is cut off by hand through the upper bound on the $z$ integral, so the $\delta$-function terms in the splitting functions do not contribute. To compensate for this, they add ``virtual" contributions obtained from sum rules 
%%%
\begin{align} \label{eq:virtual}
  \frac{\df}{\df \ln \mu}\, f_q^{(I=1,I_3=0)}(x,\mu) &=   
  \frac{\al_2}{\pi} f_q^{(I=1,I_3=0)}(x,\mu) \, \int_0^{1-M/\mu}\, \df z\, z\,\Big[-\frac34 \widetilde P_{QQ}(z) - \frac34 \widetilde P_{GQ}(z) \Big]+ \dots
\nn \\ 
 &=   
  \frac{\al_2}{\pi}\, \Big(\frac32 \ln \frac{M}{\mu} + \frac98 \Big) f_q^{(I=1,I_3=0)}(x,\mu) + \dots  
 \,, \nn \\ 
  \frac{\df}{\df \ln \mu}\, f_W^{(I=1,I_3=0)}(x,\mu) &= 
  \frac{\al_2}{\pi} f_W^{(I=1,I_3=0)}(x,\mu)\, \int_0^{1-M/\mu}\, \df z\, z\,\Big[-2 \widetilde P_{GG}(z) - \frac12 (N_c+1) \widetilde P_{Q G}(z) \Big] + \dots
\nn \\ 
 &=   
  \frac{\al_2}{\pi}\, \Big(4 \ln \frac{M}{\mu} + \frac12 b_{0,2} \Big) f_q^{(I=1,I_3=0)}(x,\mu) + \dots  
\,,\end{align}
%%%
where $N_c+1$ is the number of quark plus lepton electroweak doublets.

To compare with our expressions, we need to remove the cutoff in \eq{BFW}, which only matters for the diagonal terms $P_{QQ}$ and $P_{GG}$. Specifically, 
%%%
\begin{align} \label{eq:cutoff}
  \int_{1-M/\mu}^1\, \df z\,  \widetilde P_{QQ}(z)\, f_q^{(I=1,I_3=0)}\Big(\frac{x}{z},\mu\Big) 
  &= \Big(2 \ln \frac{M}{\mu} + \frac32 \Big) \, f_q^{(I=1,I_3=0)}(x,\mu),
 \nn \\ 
  \int_{1-M/\mu}^1\, \df z\,  \widetilde P_{GG}(z)\, f_W^{(I=1,I_3=0)}\Big(\frac{x}{z},\mu\Big) 
  &= 2 \ln \frac{M}{\mu} \, f_W^{(I=1,I_3=0)}(x,\mu)
\,,\end{align}
%%%
up to power corrections in $M/Q$. Combining \eqs{virtual}{cutoff} gives, 
%%%
\begin{align} \label{eq:7.5}
  \frac{\df}{\df \ln \mu}\, f_q^{(I=1,I_3=0)}(x,\mu) &= 
    \frac{\al_2}{\pi}\, \bigg[\Big(\frac32 \ln \frac{M}{\mu} + \frac98 \Big)  + \frac14 \Big(2 \ln \frac{M}{\mu} + \frac32 \Big)\bigg] f_q^{(I=1,I_3=0)}(x,\mu) + \dots
    \,, \nn \\
    &= 
    \frac{\al_2}{\pi}\, \Big(2  \ln \frac{M}{\mu} + \frac32 \Big) f_q^{(I=1,I_3=0)}(x,\mu) + \dots
    \,, \nn \\
  \frac{\df}{\df \ln \mu}\, f_W^{(I=1,I_3=0)}(x,\mu) &= 
    \frac{\al_2}{\pi}\, \bigg[\Big(4 \ln \frac{M}{\mu} + \frac12 b_{0,2} \Big)  - 2 \ln \frac{M}{\mu} \bigg] f_W^{(I=1,I_3=0)}(x,\mu) + \dots \nn \\
    &= 
    \frac{\al_2}{\pi}\, \Big(2 \ln \frac{M}{\mu} + \frac12 b_{0,2}  \Big) f_W^{(I=1,I_3=0)}(x,\mu) + \dots
\,,\end{align}
%%%
which when evolved from $\mu=M$ to $\mu=Q$ leads to the same Sudakov factor as \eq{sudakov}. Interestingly, the constant terms in the anomalous dimensions in \eq{7.5} agree with those of $\Ga_1$ and $\Ga_2$ in \eq{Ga_def}.

The approach in refs.~\cite{Ciafaloni:2001mu,Ciafaloni:2005fm,Bauer:2017isx} was developed to obtain the LL corrections, and it does not capture the full NLL corrections. First of all, the anomalous dimensions are not the same as our result. Although they integrate to the same Sudakov factor at LL accuracy, \eq{sudakov} indicates that there are differences beyond LL, including from the $\mu$-dependence of the coupling $\al_2$ in the anomalous dimension. Secondly, the cut off $z \leq 1-M/Q$ in \eq{BFW}, though physically motivated, is somewhat arbitrary. In particular, changing $M$ between e.g.~$M_W$ and $M_Z$ leads to changes at NLL accuracy. 
Nevertheless it is interesting that so much of our calculation is reproduced by just considering the splitting functions in the broken phase of the theory and imposing the momentum sum rules. As discussed earlier, polarization effects must be taken into account because SU(2) is chiral. Finally, once particles are identified in the final state, there is a nontrivial soft function which cannot be reproduced by splitting functions (which is a purely collinear approximation) at NLL.

%%%%%%%%%%%%%%%%%%%%%%%%%%%%%%%%%%%%%%%%%%%%%%%%%%%%%%%%%%%%%%%%%%%%%%%%%%%%%%%%
\section{Generalizations and extensions}
\label{sec:extensions}
%%%%%%%%%%%%%%%%%%%%%%%%%%%%%%%%%%%%%%%%%%%%%%%%%%%%%%%%%%%%%%%%%%%%%%%%%%%%%%%%

In this section we provide a roadmap for a range of extensions, that will be presented in more detail in a forthcoming publication. Specifically, we will touch on extending our approach to higher orders, other processes, kinematic hierarchies and jets. Perhaps most interestingly, we consider a hybrid between inclusive and exclusive processes, which are fully exclusive in the central region of the collision (where detectors are located), but fully inclusive in the forward (beam) regions. These generalizations can of course be combined, depending on the specific process and measurement under consideration.

%===============================================================================
\subsection{Higher orders}
%===============================================================================

%%%
\begin{table}[t]
  \centering
  \begin{tabular}{l | c c c c c c}
  \hline \hline
  & Matching & Non-cusp & Cusp \\ \hline
  LL & tree & - & $1$-loop \\
  NLL & tree & $1$-loop & $2$-loop \\
  NLL$'$ & $1$-loop & $1$-loop & $2$-loop \\
  NNLL & $1$-loop & $2$-loop & $3$-loop \\
  \hline\hline
  \end{tabular}
  \caption{Perturbative ingredients needed at different orders in resummed perturbation theory. The columns correspond to 
  the loop order of the matching (both at the high scale $Q$ and low scale $M$), non-cusp and cusp anomalous dimensions.}
\label{tab:orders}
\end{table}
%%%

In this paper we limited ourselves to NLL accuracy, which allowed us to perform one-loop calculations of the anomalous dimensions and tree-level calculations of the matching at the hard scale $Q$ in \sec{high_matching} and at the low scale $M$ in \sec{low_matching}. The ingredients needed at different orders in perturbation theory are summarized in table \ref{tab:orders}, where the cusp anomalous dimension refers to the coefficient of the $\ln \nu$ and $\ln \mu$ terms in the anomalous dimensions. The cusp anomalous dimension~\cite{Korchemsky:1987wg} is universal and gives rise to double logarithms (see \sec{resummation}) which is why it is needed at one higher order than the rest of the anomalous dimension. 

To extend our approach to NLL$'$ requires carrying out the matching at one-loop order. The high-scale matching depends on the process, but is relatively easy because it can be carried out in the symmetric phase of $SU(2)$. The virtual corrections to the high-scale matching for $2 \to 2$ processes are known~\cite{Fuhrer:2010eu}.
The low-scale matching involves calculations in the broken phase (see e.g.~\cite{Chiu:2009mg,Chiu:2009ft}), but these can be carried out separately for each of the ingredients of the factorized cross section. Furthermore, these same ingredients appear for other processes and can thus be recycled.
Pushing on to NNLL requires in addition the three-loop cusp anomalous dimension~\cite{Korchemsky:1987wg,Moch:2004pa}, and the remainder of the anomalous dimension at two-loop order. Even for $z<1$ these are not simply a multiple of the known two-loop splitting functions~\cite{Furmanski:1980cm,Curci:1980uw,Floratos:1981hs,Ellis:1996nn}, because the group theory factors differ between the real-virtual and real-real contributions. However, the double virtual diagrams do not need to be calculated, because they are independent of the representation, and the sum of all diagrams for the gauge singlet case reproduces the known two-loop splitting functions.

%===============================================================================
\subsection{Other processes}
%===============================================================================

The examples we focussed on in this paper are $2 \to 2$ processes with one quark and one lepton current. For each new type of process that is considered, the hard matching in \sec{high_matching} has to be repeated. The anomalous dimensions of the collinear and soft functions do not depend on the process,  so the results of \secs{RGE_C}{RGE_S} are universal, and can be used again. New soft operators will appear for $2 \to N$ processes with $N>2$, or when collinear operators for SU(2) gauge bosons in the quintet representation contribute. This only requires determining the relevant group theory factors, since the basic diagram in table \ref{tab:soft_diagram} is the same.

%===============================================================================
\subsection{Kinematic hierarchies}
%===============================================================================

In our analysis we have assumed that there is a single hard scale $\hat s = Q^2$ describing the short distance process. However, it is possible that the Mandelstam invariants $\hat s_{ij} = (p_i + p_j)^2$ are hierarchical. For example, in a $2 \to 3$ process, two of the energetic final-state particles could be relatively close to each other, such that
%%%
\begin{align}
  q^2 \sim \hat s_{45} \ll |\hat s_{ij\neq 45}| \sim Q^2 
\,,\end{align}
%%% 
or one of the particles could be much less energetic,
%%%
\begin{align}
  q^2 \sim |\hat s_{i5}| \ll |\hat s_{i<j \neq 5}| \sim Q^2 
\,.\end{align}
%%%
This can be described using SCET$_+$~\cite{Bauer:2011uc,Procura:2014cba,Larkoski:2015zka}, by first matching onto SCET for a $2 \to 2$ process at the high scale $Q$, and then resolving the two nearby particles or resolving the soft particle at the lower scale $q$. If $q \lesssim M$, all this is irrelevant from the point of view of electroweak corrections, and these will simply be the same as for the corresponding $2 \to 2$ process. If $Q \gg q \gg M$, one first uses the evolution for the $2 \to 2$ process from the scale $\mu = Q$ to $\mu = q$, then matches onto the $2 \to 3$ process, and uses the evolution for the $2 \to 3$ process from $\mu = q$ down to $\mu = M$. Thus, instead of only $\ln Q/M$ we now also get $\ln q/M$, or equivalently $\ln Q/q$.

Lets make this a bit more concrete for the case of two energetic final-state particles that are close to each other (see refs.~\cite{Bauer:2011uc,Pietrulewicz:2016nwo} for additional details). The matching at the scale $\mu \sim q$ maps a single collinear function onto the two collinear functions for the nearby particles and a collinear-soft function describing the radiation between them. 
The tree-level matching coefficient is simply the appropriate collinear splitting function. The main difference with respect to refs.~\cite{Bauer:2011uc,Pietrulewicz:2016nwo}, is the gauge representation of collinear operators. For example, 
%%%
\begin{align}\label{eq:cmatch}
  \C_q &\to \frac14\, \C_q \C_W - \frac12\,\C_q^a \C_W^b \cS_{q W}^{ab}
\,.\end{align}
%%%
The soft operator $\cS_{q W}^{ab}$ in \eq{cmatch} is the collinear-soft function, where the subscript indicates that the Wilson lines are along $q$ and $W$. Thus instead of a soft function for the $2\to 3$ process, we have a soft function for the $2 \to 2$ process and a collinear-soft function. This soft function and collinear-soft function have the same invariant mass scale $\mu \sim M$, but different rapidity scales $\nu \sim M$ vs.~$QM/q$, and the resulting $\nu$-evolution sums single logarithms of $Q/q$. The matching relations for other collinear operators take a form similar to \eq{cmatch}.

%===============================================================================
\subsection{Jets}
%===============================================================================

For the processes we considered, only leptons were identified in the final state (the jet in DIS was not identified). However, one can also consider jets defined through an algorithm like anti-k$_T$~\cite{Cacciari:2008gp} and a jet radius parameter $R$. For $R \ll 1$\footnote{The collinear approximation holds surprisingly well, with smaller than 10\% corrections for values as large as $R = 0.7$~\cite{Jager:2004jh,Mukherjee:2012uz}.}, inclusive jet production can be described by a fragmentation function, which accounts for the jets produced by a parton~\cite{Dasgupta:2014yra,Kang:2016ehg,Dai:2016hzf}. The only difference with standard fragmentation functions is that the evolution stops at the jet scale $QR$, where this fragmentation function is perturbatively calculable. This introduces EW logarithms of $\ln QR/Q = \ln R$ in addition to $\ln Q/M$. The tree-level matching at the scale $\mu = QR$ yields 
%%%
\begin{align}
  D_{W_\pm \to {\rm jet}}^{(I=0)}(x,\mu,\nu) &= \de(1-x)
\,, \qquad
   D_{W_\pm \to {\rm jet}}^{(I=1,I_3=0)}(x,\mu,\nu) = 0
\,, \qquad
   D_{W_\pm \to {\rm jet}}^{(I=2,I_3=0)}(x,\mu,\nu) = 0
\,, \nn \\
  D_{q\to {\rm jet}}^{(I=0)}(x,\mu,\nu) &= \de(1-x)
\,, \qquad
   D_{q\to {\rm jet}}^{(I=1,I_3=0)}(x,\mu,\nu) = 0
\,, \nn \\
  D_{u\to {\rm jet}}(x,\mu,\nu) &= \de(1-x)
\,,\end{align}
%%%
etc. Because no particle is identified, we do not get a contribution from the gauge non-singlet fragmentation functions.

%===============================================================================
\subsection{Combining inclusive and exclusive processes}
\label{sec:incl_excl}
%===============================================================================

\begin{figure}
\centering
 \includegraphics[width=0.6\textwidth]{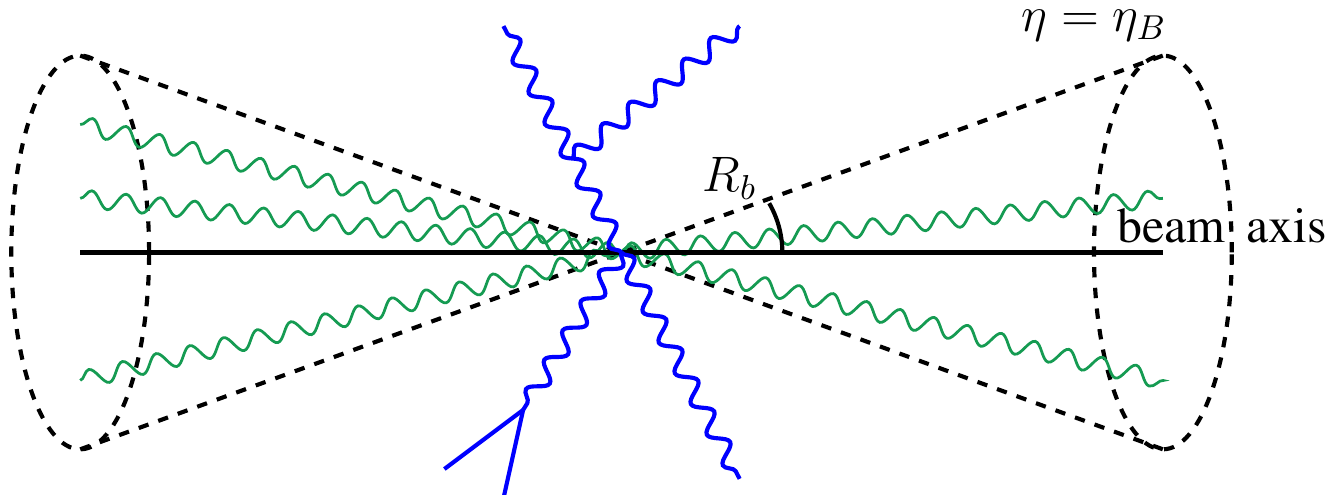}
 \caption{We treat the radiation in cones around the beam axis (green) inclusively, because it is unresolved. $W$ and $Z$ bosons emitted into the central region (blue) are detected, and can therefore be treated in an exclusive manner.}
\label{fig:beam_cones}
\end{figure}

The final generalization we consider mixes our resummation of EW logarithms for inclusive processes with the EW resummation for exclusive processes developed in refs.~\cite{Chiu:2007yn,Chiu:2007dg,Chiu:2008vv,Chiu:2009mg,Chiu:2009ft,Chiu:2009yx,Fuhrer:2010eu,Fuhrer:2010vi}. Specifically, we will assume that we are fully exclusive in the region $|\eta| < \eta_b$ covered by detectors, and fully inclusive near the beams, where $|\eta| > \eta_b$, see \fig{beam_cones}. These beam regions correspond to cones with half opening angle $R_b = 2 \arctan(e^{-\eta_b})$ around the beam axis. We will exploit that for the LHC we can safely assume that $R_b$ is small. Note that fully exclusive only refers to the electroweak corrections, which is justified because the $W$ and $Z$ boson are massive, and not to QCD or QED radiation. In particular, we will not discuss what happens below the scale $M$.

Before describing how to combine the inclusive and exclusive cases, it is useful to highlight the similarities and differences between the two. In the exclusive case only virtual diagrams contribute to the collinear functions. For example, from table~\ref{tab:coll_fermion} we read off that the fermion collinear operator has the following \emph{multiplicative} anomalous dimension ($SU(2)$ part only) 
%%%
\begin{align} \label{eq:C_ex}
\hat \ga_{\mu,Q}^{\rm ex} =
 c_F \Big( 2\ln \frac{\nu}{\bar n_i \cdot r_i} +\frac32 \Big) 
 \,, \qquad
\hat \ga_{\nu,Q}^{\rm ex} = c_F \ln \frac{\mu^2}{M^2} 
\,.\end{align}
%%%
Because we only include virtual diagrams, the anomalous dimension is independent of the representation (i.e.~singlet vs.~triplet), and of course there is no mixing. Including the corresponding $\ln \nu^2/\mu^2$ terms from the soft function anomalous dimension, we obtain the same $\mu$-anomalous dimension as in ref.~\cite{Chiu:2007yn}\footnote{The anomalous dimensions differ by an overall factor of -2. The factor of 2 arises because they consider amplitudes, and we consider squared amplitudes. The minus sign is due to the fact that they run their collinear functions from $Q$ to $M$, whereas we do the opposite, i.e.~they renormalize the coefficient functions whereas we renormalize the operators.}. Moving on to the soft function, we note that one cannot make the same simplifications as in \sec{gauge_spin} in the exclusive case. Specifically, the measurement restricting the real radiation sits between the $\cS_i$ and $\cS_i^\dagger$ in the amplitude and conjugate amplitude, prohibiting us from cancelling them against each other, even for the singlet. The one-loop diagram responsible for the soft anomalous dimension is the same, but in the exclusive case contributions where the gauge boson crosses the cut are of course not allowed.

We now describe how to combine  exclusive and inclusive resummation to calculate the EW corrections. For this we build on recent work~\cite{Chien:2015cka,Becher:2015hka,Hornig:2016ahz} that has addressed QCD corrections for a similar setup. The main difference is that here we restrict \emph{all} EW radiation in the central region, whereas they restrict the energy of soft radiation through some central jet veto. Instead we have to account for the gauge boson mass $M$. Our approach can be summarized as follows: we use the results from the exclusive case to evolve from $\mu=Q$ to $\mu=QR$,  match onto an inclusive descriptions of the two beams at that scale, and then evolve from $\mu=QR$ to $\mu=M$. This matching amounts to replacing the exclusive collinear functions for the beams by inclusive collinear functions and collinear-soft functions. Normalizing the collinear-soft function to 1, the tree-level matching coefficient is 1. The switch from exclusive to inclusive collinear functions at NLL thus simply involves changing the anomalous dimension from \eq{C_ex} to \eq{ga_C}, etc. The collinear-soft function consists of two back-to-back Wilson lines, and its one-loop anomalous dimension is therefore again the result of the diagram in table~\ref{tab:soft_diagram}. Since $n_i \cdot n_j/2 = 1$ (back-to-back Wilson lines), we are left with the following collinear-soft anomalous dimension
%%%
\begin{align}
 \C_{Q}:& \quad \hat \ga_{\mu,CS} = c_F \ln \frac{\nu^2}{\mu^2}
 \,, & 
 \hat \ga_{\nu,CS} &= c_F \ln \frac{\mu^2}{M^2}
  \nn \\
  \C_{Q}^a: & \quad \hat \ga_{\mu,CS} = (c_F - \tfrac12 c_A) \ln \frac{\nu^2}{\mu^2}
 \,, & 
 \hat \ga_{\nu,CS} &= (c_F - \tfrac12 c_A) \ln \frac{\mu^2}{M^2}  
\,.\end{align}
%%%
The overall normalization of the anomalous dimension was fixed using consistency, since the difference in $\nu$-anomalous dimension between the exclusive and inclusive collinear function must be cancelled by the collinear-soft function. The collinear-soft function has the same invariant mass scale $\mu_{CS} \sim M$ as the soft function, but its rapidity scale is $\nu_{CS} \sim M/R_b$. Thus the $\nu$-evolution will resum logarithms of $(M/R_b) / M = R_b$ in addition logarithms of $Q/M$. 

Finally, we note that there are electroweak nonglobal logarithms (NGLs)~\cite{Dasgupta:2001sh} of the form $\al^n \ln^n (QR/M)$ that arise because the radiation from collinear-soft function is unconstrained in the beam region and fully constrained in the central region. Although these logarithms formally enter at NLL, they are only visible in a two-loop calculation and are expected to be relatively small.

%%%%%%%%%%%%%%%%%%%%%%%%%%%%%%%%%%%%%%%%%%%%%%%%%%%%%%%%%%%%%%%%%%%%%%%%%%%%%%%%
\section{Conclusions}
\label{sec:conclusions}
%%%%%%%%%%%%%%%%%%%%%%%%%%%%%%%%%%%%%%%%%%%%%%%%%%%%%%%%%%%%%%%%%%%%%%%%%%%%%%%%

In this paper we considered an alternative to the usual paradigm in which the emissions of the massive $Z$ and $W$ bosons are treated as resolved, such that only virtual electroweak corrections need to be calculated. In fact we started from the opposite extreme, considering a rather inclusive setup in which only one or two particles are identified in the final state, and developing the framework to resum electroweak logarithms. The fact that incoming and outgoing particles are not electroweak singlets played an important role, introducing parton distribution functions and fragmentation functions in the corresponding representations that have a rather different evolution. Specifically, the resummation of nonsinglet distributions involves double logarithms, leading to a Sudakov suppression in the extreme high-energy limit. These contributions are also sensitive to the exchange of soft gauge bosons, and we performed our calculations using a separate ultraviolet, rapidity and infrared regulator to highlight the structure. Furthermore, we demonstrated the importance of  polarization effects for gauge bosons, due to the chiral nature of SU(2). For the user mainly interested in our results, we also provided an explicit recipe on how to incorporate these effects at next-to-leading logarithmic accuracy in the appendix.

Of course the experimental reality is probably somewhere between the fully exclusive and inclusive case. We therefore consider a mixed case where certain regions of phase space are treated exclusively (inclusively) due to presence (absence) of detectors. This involves a combination of our framework and that devised for exclusive processes, and we can furthermore lean heavily on related developments in QCD cross sections. We also consider the case where a jet (instead of a lepton) is identified in the final state, which is straightforward to incorporate. 

%%%%%%%%%%%%%%%%%%%%%%%%%%%%%%%%%%%%%%%%%%%%%%%%%%%%%%%%%%%%%%%%%%%%%%%%%%%%%%%%
\begin{acknowledgments}
We thank Bartosz Fornal for discussions, and for collaboration in initial stages of this project.
This work is supported by the DOE grant DE-SC0009919, the ERC grant ERC-STG-2015-677323,  the D-ITP consortium, a program of the Netherlands Organization for Scientific Research (NWO) that is funded by the Dutch Ministry of Education, Culture and Science (OCW),  and the Munich Institute for Astro- and Particle Physics (MIAPP) of the DFG cluster of excellence ``Origin and Structure of the Universe".
\end{acknowledgments}
%%%%%%%%%%%%%%%%%%%%%%%%%%%%%%%%%%%%%%%%%%%%%%%%%%%%%%%%%%%%%%%%%%%%%%%%%%%%%%%%

\appendix

%%%%%%%%%%%%%%%%%%%%%%%%%%%%%%%%%%%%%%%%%%%%%%%%%%%%%%%%%%%%%%%%%%%%%%%%%%%%%%%%
\section{Recipe for electroweak resummation in inclusive processes at NLL}
\label{app:recipe}
%%%%%%%%%%%%%%%%%%%%%%%%%%%%%%%%%%%%%%%%%%%%%%%%%%%%%%%%%%%%%%%%%%%%%%%%%%%%%%%%

In this section we show how to implement electroweak resummation in inclusive processes at NLL order using results in this paper. We provide a recipe, and use Drell-Yan as an example to illustrate it. \\

\noindent {\bf Step 1:} Determine the tree-level cross section, separating contributions by the helicity of external particles, as these receive different electroweak corrections.

For the Drell-Yan process, $pp \to \mu \bar \mu$, we will assume that the invariant mass $Q^2$ as well as the total rapidity $Y$ of the lepton pair is measured, and that we are otherwise inclusive in the final state. The corresponding cross section is given by 
%%%
\begin{align} \label{eq:DYtree}
  \frac{\df \si}{\df Q^2\, \df Y} &= \int\! \df x_1\, \df x_2\, \df x_3\, \df x_4\, 
  \int\! \df \eta\, \frac{1}{2\cosh^2(\eta-Y)}
  \nn \\ & \quad \times
  \sum_{i,j} f_{u_i}(x_1,\mu)\, f_{\bar {u_{-i}}}(x_2,\mu)\, \hat \si_{u\mu}^{ij}(x_1 x_2, \eta, Y,  \mu)  D_{\mu_j \to \mu}(x_3,\mu)\, D_{\bar \mu_{-j} \to \bar \mu}(x_4,\mu)
  \nn \\ & \quad \times
  \de(Q^2 - x_1 x_2 x_3 x_4 \Ecm^2)\, \de\Big(Y - \frac 12 \ln \frac{x_1}{x_2}\Big)
\,,\end{align}
%%%
where $Q$ is the invariant mass of the muon pair, $Y$ is the total rapidity of the muon pair, and $\eta$ is the rapidity of the $\mu^-$, and the sum on $i,j = \pm$ run over the helicity of the quark and lepton. For simplicity we restricted to a single channel, $u \bar u \to \mu \bar \mu$. The fragmentation function (FF) $D_{\mu_j \to \mu}(x_3,\mu)$ is included to account for the fact that muon $\mu$ produced in the hard interaction loses a fraction $1-x_3$ of its momentum due to collinear radiation (and similarly for $\bar \mu$). We have kept the integral over the pseudorapidity $\eta$ of the muon because electroweak corrections will depend on it. The partonic cross section in \eq{DYtree} is given by 
%%%
\begin{align}
\hat \si_{u\mu}^{ij} = \frac{\pi\al_{\rm em}^2}{3N_c\, x_1 x_2 \Ecm^2}\,
\big|Q_u Q_\mu + v_u^i v_\mu^j P_Z(x_1 x_2 \Ecm^2) \big|^2\, \frac{3 e^{2(\eta-Y)}}{4\cosh^2(\eta-Y)} 
\,,\end{align}
%%%
where the last factor corresponds to $s_{u\mu}^2/s_{u \bar u}^2$ in terms of rapidity coordinates.
Here $\al_{\rm em}$ is the electromagnetic coupling constant, $N_c= 3$ is the number of colors, $Q_A$ is the electric charge in units of $|e|$, and $v_A^i$ are the couplings to the $Z$ boson
%%%
\begin{align}
 v^-_A = \frac{2 t^3_A - 2Q_A \sin^2 \theta_W}{\sin(2\theta_W)}
 \,, \qquad
 v^+_A = - \frac{2Q_A \sin^2 \theta_W}{\sin(2\theta_W)}
\,,\end{align}
with $t^3_A$ the third component of weak isospin and $\theta_W$ the weak mixing angle. The factor
%%%
\begin{equation}
P_Z(s) = \frac{s}{s-M_Z^2 + \img \Ga_Z M_Z}
\,\end{equation}
%%%
encodes the difference between the photon and $Z$-boson propagator, where $M_Z$ and $\Gamma_Z$ are the mass and width of the $Z$ boson. Our resummation is only guaranteed to hold for the leading contribution in the $M^2/Q^2$ expansion, for which $P_Z(s) = 1$. Suppressed terms can of course be included through some matching procedure. \\

\noindent {\bf Step 2:} Rewrite the PDFs and FFs in terms of $SU(2)$ singlets and triplets. 

The resummation of electroweak logarithms will be carried out in the symmetric phase of $SU(2)$. This requires not only PDFs (and FFs) containing operators in the singlet representation, like $\bar q_L q_L = \bar u_L u_L + \bar d_L d_L$, but also in the triplet representation, such as $\bar q_L t^3 q_L = \tfrac12 \bar u_L u_L - \tfrac12 \bar d_L d_L$. In our notation we denote these by superscripts $(I=0)$ and $(I=1,I_3=0)$, respectively.
For Drell-Yan we need 
%%%
\begin{align} \label{eq:to_triplet}
  f_{u_+}(x,\mu) &= \tfrac12 f_{q}^{(I=0)}(x,\mu) + f_{q}^{(I=1,I_3=0)}(x,\mu)
  \,,\nn \\ 
  D_{e_+ \to e}(x,\mu) &= D^{(I=0)}_{\ell \to \mu}(x,\mu) -2 D_{\ell \to \mu}^{(I=1,I_3=0)}(x,\mu)
\,,\end{align}
%%%
which follows from inverting the equations in \sec{low_matching}. \\

\noindent {\bf Step 3:} Identify the relevant soft function for each term in the cross section. 

At tree-level the soft function is just a number, which we can conveniently normalize to 1. However, the different soft functions have different evolution equations. The level of complication depends on the number $n_I$ of PDFs and FFs in the triplet representation: 
\begin{itemize}
\item
$n_I$ is odd: the soft function vanishes, and these terms can therefore be dropped from the cross section.
\item
$n_I = 0$: the soft function is simply 1. For example, a term like $f_{q}^{(I=0)} f_{\bar q}^{(I=0)} D_{\ell}^{(I=0)} D_{\bar \ell}^{(I=0)}$ has no soft function. 
\item
$n_I=2$: the soft function consists of Wilson lines along the directions of the triplets. E.g. for the $f_{q}^{(I=1)} f_{\bar q}^{(I=1)} D_{\ell}^{(I=0)} D_{\bar \ell}^{(I=0)}$ the soft function is given by $2 \langle 0 | S_{q\bar q}^{33} |0 \rangle$. The overall factor of 2 ensures that it is normalized to 1 at tree-level, see \eq{S_tree}.
\item
$n_I=4$: the soft function consists of (correlated) pairs of Wilson lines, of which there are three linearly independent combinations. It depends on the gauge boson exchanged between the quark and lepton current:
\begin{align}
SU(2): & \quad -\tfrac{4}{3}\langle 0 | \cS_{q\bar q}^{33} \cS_{\ell \bar \ell}^{33} |0 \rangle
+\tfrac{8}{3} 
 \langle 0 |\cS_{q \ell}^{33} \cS_{\bar q \bar \ell}^{33} |0 \rangle
+ \tfrac{8}{3}  \langle 0 |\cS_{q \bar \ell}^{33} \cS_{\bar q \ell}^{33} |0 \rangle
\nn \\
 U(1): &  \quad 4 \langle 0 | \cS_{q\bar q}^{33} \cS_{\ell \bar \ell}^{33} |0 \rangle
\end{align}
The interference between $SU(2)$ and $U(1)$  can be dropped, because the corresponding soft function vanishes at tree-level. For Drell-Yan, $SU(2)$ only enters when $i,j=L$, for which these contributions can be identified by rewriting
%%%
\begin{align}
  e^2(Q_u Q_e + v_u^L v_e^L) = g_2^2 \,[t^3]_{11} [t^3]_{22}  + g_1^2\, \hyp_q \hyp_\ell
\,,\end{align}
%%%
where $\hyp$ denotes the hypercharge, and $[t^a]_{ij}$ are elements of the $SU(2)$ representation matrices. The first term on the right-hand side corresponds to $SU(2)$ and the second to $U(1)$. To be completely explicit,
%%%
\begin{align}
   f_{q}^{(I=1)} f_{\bar q}^{(I=1)} D_{\ell}^{(I=1)} D_{\bar \ell}^{(I=1)}
   & \Big[g_2^2 \,[t^3]_{11} [t^3]_{22}  
   \bigl(-\tfrac{4}{3}\langle 0 | \cS_{q\bar q}^{33} \cS_{\ell \bar \ell}^{33} |0 \rangle
\!+\! \tfrac{8}{3} 
 \langle 0 |\cS_{q \ell}^{33} \cS_{\bar q \bar \ell}^{33} |0 \rangle
\!+\! \tfrac{8}{3}  \langle 0 |\cS_{q \bar \ell}^{33} \cS_{\bar q \ell}^{33} |0 \rangle \bigr)
 \nn \\ &
 + g_1^2\, \hyp_q \hyp_\ell \bigl( 4 \langle 0 | \cS_{q\bar q}^{33} \cS_{\ell \bar \ell}^{33} |0 \rangle \bigr) \Big]
\,,\end{align}
%%%
where the soft function factors in brackets are 1 at tree-level but give a non-trivial contribution to the resummation in the next step.
\end{itemize}

\noindent {\bf Step 4:} Resum logarithms of $Q/M$ using the RG evolution.

The logarithms of $Q/M$ are resummed by evolving the PDFs, FFs and soft functions from their initial scale to the hard scale $\mu=Q$, as shown in \fig{nu}. Their RG equations were calculated in \secs{RGE_C}{RGE_S}, and fully explicit expressions for the PDFs were given in \sec{PDF_SM}. The soft function evolution introduces a dependence on the rapidity $\eta$ in \eq{DYtree} and involves mixing for $n_I = 4$, as described by \eq{mix}.

%%%%%%%%%%%%%%%%%%%%%%%%%%%%%%%%%%%%%%%%%%%%%%%%%%%%%%%%%%%%%%%%%%%%%%%%%%%%%%%%
\section{Examples of cross sections}
\label{app:xsec}
%%%%%%%%%%%%%%%%%%%%%%%%%%%%%%%%%%%%%%%%%%%%%%%%%%%%%%%%%%%%%%%%%%%%%%%%%%%%%%%%

We will illustrate the various combinations of PDFs that appear, by considering the inclusive production of a heavy particle $X$ in quark-antiquark annihilation. Depending on the quantum numbers of $X$ different interactions are possible. If $X$ is a vector, we can have
%%%
\begin{align}
\mathcal{L}_1 &=\bar q \gamma^\mu q X_\mu, & 
\mathcal{L}_2 &=\bar q \gamma^\mu t^a q X^a_\mu, & 
\mathcal{L}_3 &=\bar u \gamma^\mu u X_\mu, & 
\mathcal{L}_4 &=\bar d \gamma^\mu d X_\mu, & 
\mathcal{L}_5 &=\bar u \gamma^\mu d X_\mu + \text{h.c.}
\,,\end{align}
%%%
whereas if $X$ is a scalar,
%%%
\begin{align}
\mathcal{L}_6 &=\bar q_\alpha u X^\alpha + \text{h.c.}, & 
\mathcal{L}_7 &=\bar q_\alpha d X^\alpha + \text{h.c.} 
\,.\end{align}
%%%

We now describe the combinations of PDFs that enter in the factorization theorem for the corresponding cross sections. Working to NLL accuracy, we can restrict ourselves to tree-level matching. For brevity, we assume that the quark comes from the first proton and the antiquark from the second, but there is of course a contribution where these are swapped.

Starting with $\mathcal{L}_1$, $X$ must have quantum numbers $I=0$, $Y=0$, leading to
%%%
\begin{align}
\mathcal{L}_1&:  \bar q^\beta q_\alpha \otimes  \bar q^\alpha  q_\beta = \frac12\, \bar q q \otimes  \bar q  q +
2\, \bar q t^a q  \otimes    \bar q t^a q  \\
&\to \frac12 (f_{u_-} + f_{d_-}) \otimes (f_{\bar u_+}+f_{\bar d_+}) + \frac12 (f_{u_-} - f_{d_-}) \otimes (f_{\bar u_+} - f_{\bar d_+})  
= f_{u_-} \otimes f_{\bar u_+} + f_{d_-} \otimes f_{\bar d_+}
\,.\nonumber\end{align}
%%%
For $\mathcal{L}_2$: $X$ must be $I=1$, $Y=0$, and have 3 states, $X^+,X^0,X^-$, leading to
%%%
\begin{align} \label{eq:4}
\mathcal{L}_2&:  \bar q^\beta q_\alpha \otimes   \bar q^\lambda q_\sigma [t^a]_\lambda{}^\alpha [t^a]_\beta{}^\sigma = \frac38 \bar q q \otimes  \bar q q -
\frac12 \bar q t^a q  \otimes    \bar q t^a q \nn \\
&\to \frac38 (f_{u_-} + f_{d_-}) \otimes (f_{\bar u_+}+f_{\bar d_+}) - \frac18 (f_{u_-} - f_{d_-}) \otimes (f_{\bar u_+} - f_{\bar d_+})  \nn \\
&= \frac14 f_{u_-} \otimes f_{\bar u_+} +\frac14 f_{d_-} \otimes f_{\bar d_+} + \frac12 f_{u_-} \otimes f_{\bar d_+} +  \frac12 f_{d_-} \otimes f_{\bar u_+}
\end{align}
%%%
Alternatively, we can expand out $\mathcal{L}_2$ 
%%%
\begin{align}
\mathcal{L}_2 &= \frac1{\sqrt2} \bar u \gamma^\mu d X^+_\mu + \frac1{\sqrt2}\bar d \gamma^\mu u X^-_\mu +\frac12 (\bar u \gamma^\mu u - \bar d \gamma^\mu d ) X^0_\mu
\,,\end{align}
%%%
from which we read off that the ratios of contributions $u \bar d \to X^+$, $d \bar u \to X^-$, $u \bar u \to X^0$, $d \bar d \to X^0$ is given by $1/2:1/2:1/4:1/4$, in agreement with \eq{4}.

Moving on, $\mathcal{L}_3$ gives $f_{u+} \otimes f_{\bar u-}$, $\mathcal{L}_4$ gives $f_{d+} \otimes f_{\bar d-}$ for the production of $X^0$. $\mathcal{L}_5$ gives $f_{u+} \otimes f_{\bar d-}$ for the production of $X^+$ and $f_{d-} \otimes f_{\bar u_+}$ for the production of $X^-$
For $\mathcal{L}_6$ we find
%%%
\begin{align}
\mathcal{L}_6 &:  \bar q^\alpha q_\alpha \otimes  \bar u u + \bar u u \otimes \bar q^\alpha q_\alpha \to 2f_{u_-} \otimes f_{\bar u_+}  + f_{d_-} \otimes f_{\bar u_+}  + f_{u_-} \otimes f_{\bar d_+}
\end{align}
%%%
which give rise to the production of $X^0$, $X^-$ and $X^+$, and similarly for $\mathcal{L}_7$. 

\phantomsection
\bibliographystyle{jhep}
\bibliography{ewpdf}

\end{document}